# Graphene Lattices with Embedded Transition-Metal Atoms and Tunable Magnetic Anisotropy Energy: Implications for Spintronic Devices

Rostislav Langer[1,2], Kimmo Mustonen[3], Alexander Markevich[3], Michal Otyepka[1,4], Toma Susi*[3], Piotr Błoński*[1]

[1] Regional Centre of Advanced Technologies and Materials, Czech Advanced Technology and Research Institute (CATRIN), Palacký University Olomouc, 779 00 Olomouc, Czech Republic

[2] Department of Physical Chemistry, Faculty of Science, Palacký University Olomouc, 779 00 Olomouc, Czech Republic

[3] University of Vienna, Faculty of Physics, Boltzmanngasse 5, A-1090 Vienna, Austria

[4] IT4Innovations, Technical University of Ostrava, 708 00 Ostrava-Poruba, Czech Republic

*e-mail: piotr.blonski@upol.cz; toma.susi@univie.ac.at

## Abstract

Doping of the graphene lattice with transition metal atoms resulting in high magnetic anisotropy energy (MAE) is an important goal of materials research owing to its potential application in spintronics. In this article, by using spin-polarized density functional theory including spin-orbit coupling, we examined magnetic properties of graphene with vacancy defects, both bare and nitrogen-decorated, and doped by Cr, Mn and Fe transition metal single atom (TM-SA) and two different TM atoms simultaneously. Adsorption of a second TM atom on an already embedded TM atom, i.e., the formation of upright TM dimers, was also considered. It is found that the graphene-mediated coupling between TM dopants can significantly increase MAE compared to that of SA impurities. While the MAE of TM-SA did not exceed 2 meV, it was enhanced to –23 meV for Cr and Fe simultaneously embedded into two separated double-vacancy (DV) defects and to a remarkably high value of 119.7 meV for two upright Fe–Mn dimers bound to two separate DVs, considerably exceeding the sum for individual TM-SAs. The latter MAE corresponds to a blocking temperature of 34 K assuming a relaxation time of 10 years. The origin of the enhanced MAE is discussed in relation to the spin excitations at the Fermi level and changes in d-derived states accompanying the rotation of the magnetization between in-plane and out-of-plane direction. We demonstrate that the presence of partially occupied degenerate states at the Fermi level favors its formation. The stability of the systems is also discussed. The computational findings are supplemented by an atomic-resolution characterization of an incidental Mn impurity bonded to four carbon atoms, whose localized spin matches expectations as measured using core-level electron energy-loss spectroscopy. Conducting TM-doped graphene with robust magnetic features offers prospects for the design of graphene-based spintronic devices.

## Introduction

Since the isolation of graphene in 2004 [1], a great deal of research effort has been devoted to endowing it with robust magnetic features that are lacking in its pristine form in order to enhance its potential in spintronic applications [2] ranging from quantum computing [3] to storing data in magnets the size of single atoms [4]. One of the great challenges in this field is preventing thermally induced reorientation of the magnetic moments between the easy and hard magnetization axis or, in other words, increasing the blocking temperatures ($T_b$), which is enforced by MAE, *i.e.*, the energy barrier for magnetic moments to flip their directions. A high MAE necessitates large spin and orbital magnetic moments per atom and a large spin-orbit coupling (SOC). Further, magnetic anisotropy is highly dependent on symmetry, dimensionality and atomic composition and is often found to be much larger in low-symmetry nanostructures than in highly symmetric bulk materials [5–8]. Surface-supported nanoparticles offer additional degrees of freedom to tune the MAE by the particle shape and size - down to single atoms and coupling with the substrate [9–11].

To this end, transition metal (TM) atoms and their nanoclusters adsorbed onto graphene have been extensively studied [7,12,13]. While they can be mobile over structurally perfect regions of the graphene sheet, lattice imperfections create spots that can firmly anchor TM atoms or clusters, preventing their diffusion and conserving their size, symmetry and, accordingly, the desired properties [12,14–16].

As revealed by density functional theory calculations (DFT), Fe clusters can promote vacancy formation, *i.e.*, it becomes easier to remove C atoms from the graphene lattice in the presence of Fe [17]. A high diffusion barrier of 6.8 eV has been calculated for a Fe atom in a single vacancy (SV) defect in the graphene lattice using Perdew-Burke-Ernzerhof (PBE) functional, sufficient to prevent the migration of Fe at room temperature [18]. The strong adsorption of Fe in vacancy defects in graphene was found to give rise to a large anisotropy in geometries and MAE, which, however, did not exceed 1 meV [17]. Whereas theory predicted ferromagnetic (FM) alignment for four Fe atoms substituting C atoms in a 6×6 supercell of graphene that initially contains 72 C atoms (corresponding to 5.55 at. % dopant concentration) [19], a single Fe atom embedded in SV was found to be non-magnetic. In contrast, a single Fe atom in a double vacancy (DV) defect in graphene acquired a high magnetic moment of 3.5 $\mu_B$ [14]. The high spin state of Fe@DV was further experimentally verified via core-level electron energy-loss spectroscopy of its $L$ edge white-line intensity ratio [20].

Theoretical calculations based on DFT have also revealed strong binding of Sc-Zn, Pt and Au atoms in vacancy defects in graphene due to strong interaction between $sp^2$-carbon and $spd$ orbitals of a metal, with migration barriers around 2–4 eV for metals at SV and slightly higher for DV. The total magnetic moments of Cr and Mn were 2 and 3 $\mu_B$, respectively, for both types of defects [18]. The non-bonding $d$-orbitals were partially occupied for Cr and Mn, while the antibonding states were filled for Fe [21]. Further, the Curie temperature of Cr@SV was theoretically estimated to be 498.2 K [19].

DFT calculations have revealed that graphene doped by a single Mn atom undergoes a transition from non-magnetic semi-metal into FM half-metal [22]. A second Mn atom binds preferentially in the immediate vicinity of the Mn already present in the graphene lattice and they interact antiferromagnetically (AFM) in the ground state (GS) [19,22]. The magnetic moments of Mn@SV vary with the Mn concentration and originate from the $p$-$d$ exchange mechanism [23].

Such calculations have also indicated that doping a Cr atom into the SV defect in graphene tuned its Fermi level ($E_F$) from semi-metal to half-metal. The $p$-$d$ exchange mechanism between the C-$p$ and Cr-$d$ states was responsible for the induced magnetism in the otherwise non-magnetic graphene [24]. The doping of two Cr atoms into the graphene monolayer generated a magnetic moment of ~4 $\mu_B$. Depending on the distance between the Cr dopants, the GS was found to be FM, AFM or paramagnetic (PM). The origin of the particular magnetic state was described on the basis of RKKY indirect exchange interactions [25]. Experimentally, a Cr atom was inserted into e-beam induced vacancies in a graphene lattice *in situ* in STEM, though unfortunately its electron energy loss spectrum was not reported [26].

Despite the abundant computational literature on the structural, electronic and magnetic properties of graphene doped with Cr, Mn and Fe, calculations of their MAEs are more elusive [14,17] and, to the best of our knowledge, MAE of Cr- and Mn-doped graphene has not been calculated until now. This is surprising because the estimation of the magnetic anisotropy caused by SOC is of fundamental importance for assessing the material's applicability for spintronic applications. Further, information can be stored and processed in the atomic scale, if the atomic spins are coupled [3,27,28]. Thus, for the application of graphene in spintronic devices, the graphene lattice will contain multiple TM atoms rather than a single-atom dopant. In addition, the easy axis must be oriented perpendicular to the surface of the substrate to reduce the dipolar magnetic interactions between neighboring magnetic moments. Hence, examining the interaction of magnetic atoms in the graphene lattice is of both fundamental and practical interest.

In this work, systematic spin-polarized (SP) DFT calculations including SOC were carried out to investigate structural, electronic and magnetic properties of defective graphene containing SV and DV defects, optionally also N-decorated [29], doped with Cr, Fe and Mn. In light of recent advances in experimental manipulation of foreign atoms in the graphene lattice [30,31] and the theoretical prediction for manipulation of an Fe atom [32], these elements were selected as examples from Groups VI-VIII.B to elucidate changes in electronic and magnetic properties of doped graphene, including the MAE, with respect to the distances between TM atoms, both of the same element and two different TM atoms. The formation of upright TM dimers was further considered. Our main finding is that doping graphene with two different TM atoms and the formation of TM dimers leads to a tremendous increase in MAE compared to SA-doped graphene.

## Computational details

First-principle calculations were performed by using SP-DFT as implemented in the Vienna Ab-initio Simulation Package (VASP) [33–35]. The electron-ion interactions were treated by projector-augmented wave (PAW) method [36,37]. The basis set contained plane waves with a maximum kinetic energy of 500 eV. The electronic exchange and correlation effects were treated by the Perdew, Burke and Ernzerhof (PBE) [38] functional in the generalized gradient approximation (GGA).

The graphene monolayer was represented by 3×3 and 6×3 orthorhombic cells containing 48 and 96 atoms, respectively, and a vacuum layer of 15 Å deployed along the $z$-direction to avoid spurious interactions between the images of the graphene layers due to the periodic boundary conditions. To construct the SV and DV defects, one and two carbon atoms were removed from the graphene lattice, respectively, which resulted in the reconstruction of carbon bonds after optimization [39] (**Figure 1a,b**). In the SV five- and nine-membered rings were formed, and due to the unsaturated C center (2-fold-coordinated C atom), a local magnetic moment of about 1.2 $\mu_B$ was observed [39]. In the DV five- and eight-membered rings were formed and remained non-magnetic as reconstruction of the atomic positions facilitates an overlap of unpaired $sp^2$ electron clouds. Density of states (DOS) of SV showed spin-polarized metallic behavior (**Figure S1a**), while DV exhibited non-magnetic DOS with a bandgap of 0.1 eV (**Figure S1e**). Nitrogen-containing vacancy defects were also considered; the quadruple-N-decorated double vacancy (NDV) was non-magnetic, while the corresponding triple-N-decorated single vacancy (NSV) possessed magnetic moment of 0.3 $\mu_B$ [40]. To include TM atoms, we initially placed the Cr, Mn and Fe atoms ~1 Å above the center of the optimized vacancy because it is the most preferred binding site [14], and re-optimized the structures.

We further introduced two defects in the 6×3 cell to examine possible changes in the structural, electronic, and magnetic properties of graphene doped with two TM atoms of the same/different elements. We also considered changes in the distance between TM atoms and their effect on the properties of the doped graphene. Finally, we adsorbed another TM atom on top of TM@(N)SV and TM@(N)DV to study the properties of dimers $TM_2$@$TM_1$@(N)SV and $TM_2$@$TM_1$@(N)DV.

All structures were optimized until forces acting on all atoms were reduced to less than 10 meV/Å and the electronic and magnetic degrees of freedom were relaxed until the change in total energy between successive iteration steps was smaller than $10^{-6}$ eV. A Gaussian smearing of width 0.02 eV was used for partial occupancies of orbitals. For sampling of the Brillouin zone, convergence tests demonstrated the sufficiency of a $\Gamma$-centered $6 \times 6 \times 1$ $k$-point mesh ($3 \times 6 \times 1$ mesh for the 6×3 cell). Static calculations were performed with the tetrahedron method with Blöchl correction [41], while keeping the remaining computational parameters unchanged.

The binding energies, $E_{bind}$, of TM atoms to defective graphene were evaluated as

$$E_{bind} = E_{DG+TM} - E_{DF} - E_{TM},$$

where $E_{DG+TM}$, $E_{DF}$ and $E_{TM}$ are energies of defective graphene doped with TM, defective graphene, and a single TM atom in the gas phase, respectively. The binding energy of a second (upper) TM atom with the lower TM atom already embedded into the graphene lattice to form an upright TM dimer was evaluated as

$$E_{bind} = E_{DG+TM1+TM2} - E_{DF+TM1} - E_{TM2}.$$

To further assess the stability of TM atoms and dimers in their ground-state configurations, we evaluated their migration barriers from imperfect regions of the graphene lattice. In addition, the Hessian matrix was determined to exclude the presence of imaginary modes in the spectrum of phonon frequencies. The finite difference method was used, *i.e.*, each ion was displaced in the direction of each Cartesian coordinate, and from the forces the Hessian matrix was determined. Only symmetry inequivalent displacements were considered with a step size of 0.02 Å and the self-consistency cycle converged until the energy difference reached $10^{-7}$ eV.

Bader charge analysis [42,43] was performed to evaluate the charge located on each atom ($q_{Bader}$)

$$q_{Bader} = V_{val} - N_{Bader},$$

where, respectively, $V_{val}$ and $N_{Bader}$ denotes the number of valence electrons in a free atom and the computed number of valence electrons in the atom in the system.

Spin orbit coupling has been implemented in VASP by Kresse and Lebacq following the approach of Kleinman and Bylander [44] and MacDonald *et al.* [45]. Calculations including SOC were performed in the non-collinear mode as implemented in VASP by Hobbs *et al.* [46]. The magnetic anisotropy energy (MAE) per computational cell was evaluated as the difference in total energies (TE) between different orientations of the magnetization,

$$\text{MAE} = E_x - E_z.$$

In this convention, a positive MAE corresponds to an easy axis perpendicular to the graphene plane. For selected systems and the denser $11 \times 11 \times 1$ *k*-point mesh, the MAE changes were about 1 meV [14].

Since DFT+$U$ methods depend on semi-empirical values of $U$–$J$ [47] whose transferability between different systems is questionable, and if moderate values of the on-site Coulomb repulsion are admitted, they introduce only modest changes with respect to conventional DFT calculations [27], whereas large values of the on-site repulsion led to unrealistic eigenvalue spectra of TM dimers [48], we used DFT+$U$ method with $U$–$J$ of 2 eV.

In addition to fully self-consistent calculations of MAE, the magnetic force theorem [49], MAE$_{(FT)}$, was applied. The FT permits to approximate the MAE by the difference in the band energies from non-self-consistent calculations thus significantly reducing the computational efforts, though it can also lead to errors that are difficult to control [14].

The energy barrier against thermal agitations can be represented by the MAE via a simple equation, MAE = $k_B T$. Thus, MAE of 32 meV would be sufficient to allow the system withstanding a single thermal excitation up to 373 K. However, given the stochastic nature of the thermally induced reversal of the magnetization direction, the MAE corresponds to the blocking temperature $T_b$ by an equation based on Néel's theory [50],

$$\text{MAE} = k_B T_b \left( \ln \frac{\tau_N}{\tau_0} \right),$$

where $\tau_N$ represents the aimed relaxation time (10 years), and $\tau_0$ is the attempt period specific for every magnetic material (usually in the order of $10^{-10}$ s, which is also assumed here).

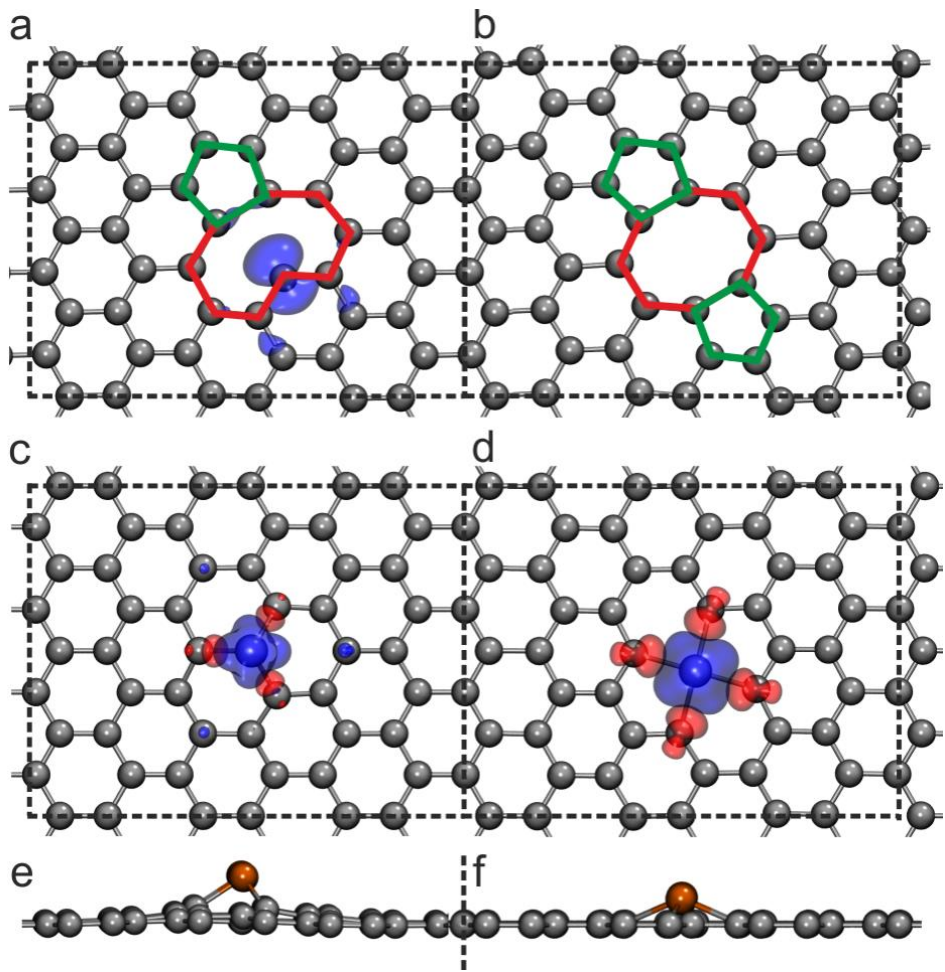

**Figure 1** Top view of **a)** single vacancy defect in graphene, **b)** double vacancy defect in graphene, **c)** TM@SV, **d)** TM@DV (where TM stands for Cr, Mn, Fe). Side view of **e)** TM@SV and **f)** TM@DV. Carbon atoms are depicted in grey, TM atoms in ochre. Five-membered rings are highlighted in green, eight- and nine-membered rings in red (*cf.* the text). Spin densities corresponding to positive/negative magnetic moments are displayed in blue/red at isosurface values of $\pm 0.035$ e$^-$Å$^{-3}$. Supercells are marked by dashed lines. After the cell was doubled in the *x*-direction, a second defect was introduced into the lattice.

## Experimental details

Scanning transmission electron microscopy images were recorded using an aberration-corrected Nion UltraSTEM100 instrument operated at 60 kV in near-ultrahigh vacuum using the MAADF (medium-angle annular dark-field) detector with an angular range of 80–200 mrad. As a sample we used chemically synthesized commercial monolayer graphene (Graphenea), which occasionally contains incidental lattice imperfections. Localized magnetic moment analysis was further conducted via electron energy-loss spectroscopy using a Gatan PEELS 666 spectrometer retrofitted with an Andor iXon 897 electron-multiplying charge-coupled device camera, with an energy spread of approximately 0.35 eV and a dispersion of 0.5 eV px$^{-1}$. The Mn $L$ edge fine structure white-line ratio $L_3/L_2$, which represents dipole transitions from the Mn $2p^{3/2}$ and Mn $2p^{1/2}$ to unoccupied $3d$ levels [20], was calculated after smoothing the spectrum with a 3.5 eV moving average, and then subtracting the pre-edge background to estimate the $L_3$ peak area, followed by subtracting the background from the tail of that peak to determine the $L_2$ peak area.

## Results

### 1) Single-Atom Doped Graphene

First, we focused on the binding energies and structural properties of TM@SV and TM@DV (**Figure 1**). All TM atoms were strongly bound to defective graphene in line with previous reports [14,18,21]. Binding in the DV defect was more favored than in the SV defect in line with a larger charge transfer from TM atoms to graphene lattice in TM@DV (**Table 1**). TM atoms were lifted out of the graphene plane during optimization due to the steric effect of their larger size. For the TM@SV, the length of the TM–C bond varied between 1.81–1.88 Å. The bonds of TM@DV were longer, in the range 1.90–2.07 Å (**Table 1**). Generally, the length of TM–C bond decreased in the order from Cr to Fe as the size of the atoms decreases. TM@SV caused buckling of the graphene layer which was more pronounced in the vicinity to the TM atom, while for TM@DV almost no buckling was observed (**Figure S2).**

Bader analysis showed charge transfer from TM atoms to their carbon neighbors (0.7–1.2 e$^-$, *cf.* **Table 1**) consistent with the higher electronegativity of C than the TM atoms. We also considered the charge states of TM (common oxidation states II and III) in SV and DV defects by using the finite model of ovalene (**Figure S3**), B3LYP [51,52] functional and def2SVP basis set as implemented in Gaussian 09 [53]. The natural bond orbital (NBO) analysis [54] was employed to evaluate TM's partial charges. The analysis revealed a charge transfer mainly from the *s* and *p* orbitals of TM atoms, while the C atoms of graphene lattice back-donate charge to the *d*-orbitals of TM atoms (**Table S1 and Table S2**).

All TM atoms studied here induced magnetic moments in graphene (**Figure 1c,d**). In TM@SV, three metal valence electrons participate in the formation of TM–C $\sigma$-bonds, one electron in a TM–C $\pi$-bond and remaining unpaired valence electrons (two of Cr and three of Mn) induce magnetic moments respectively of 2.0 $\mu_B$ and 3.0 $\mu_B$ (**Table 1**). In TM@DV, metal valence electrons are involved in the formation of four TM–C $\sigma$-bonds and one to create the $\pi$-bond. The $\pi$-electrons on the atoms in the immediate vicinity to the DV defect have opposite spins cancelling each other out; the remaining valence electrons of TM atoms induce magnetic moment in graphene (**Table 1**). The magnetic moments on Fe were 1.8 $\mu_B$ and 3.0 $\mu_B$ for Fe@SV and Fe@DV, respectively. An earlier study at the GGA-PBE level revealed zero magnetic moment for Fe@SV because the Fe impurity has doubly occupied states and an even number of electrons [18]; however, the on-site Coulomb correction led to the non-magnetic solution higher in energy than the magnetic one by 0.03 eV in agreement with the study by Santos *et al.* [35]

Doping with TM atoms induced changes in the electronic structure of graphene from semi-metal to metal (Mn@DV and Fe@DV), half-metal (Fe@SV [17,19]) and/or semiconductor (Cr@SV [19,26], Mn@SV [22,55], and Cr@DV [26]), in good agreement with the literature, as depicted in **Figure S1 and Table S3**. However, the atom- and orbital-decomposed partial densities of states (PDOS) revealed an even more complex picture of changes in the electronic structure of graphene upon doping. Whereas the $C_{3v}$ symmetry of TM@SV broke the degeneracy of *d*-orbitals into $d_{z^2}$, $d_{xy}/d_{x^2-y^2}$ and $d_{xz}/d_{yz}$ that coupled with the *p*-orbitals of the carbon atoms, the coupling of states with energies above and below the Fermi level ($E_F$) through spin-orbit interaction, which promotes either perpendicular or in-plane magnetization, depending on the spins and symmetries of the interacting states [56,57], as schematically shown in **Figure S4**. PDOS of Cr@SV demonstrated predominant contribution of the occupied $d_\delta$ (and in a smaller extent $d_{z^2}$) orbitals and empty $d_{z^2}$ orbital to the intragap states (**Figure S5**) favoring in-plane MAE, which is, however, dominated by the large energy denominator and, thus, very small. PDOS of Mn@SV and Fe@SV (**Figures S6** and **S7**) showed that the TM $d_{z^2}$ orbital contributes to the spin-up valence band (VB) edge and the TM $d_\delta$ and $d_\pi$ orbitals to the spin-down conduction band (CB) edge that provides competitive negative/positive contributions to the MAE. The spin-up VB(CB) edge of Cr@DV (**Figure S8**) exhibited sharp peaks of $d_{xy}/d_{x^2-y^2}$ ($d_{xz}/d_{yz}$), thus the electron hopping between the nearest occupied and unoccupied states favors the in-plane anisotropy, which is counterbalanced by another mechanism, *i.e.*, lowering of the total energy by a downshift of $d_{x^2-y^2}$ derived bands. Since the degenerate $d_{xy}/d_{x^2-y^2}$ states are far from $E_F$, the effect is small, and altogether, the MAE for Cr@DV does not exceed 0.05 meV (**Table 1**). Both Mn@DV (**Figure S9**) and Fe@DV (**Figure S10**) showed half-metallic DOS originating respectively from spin-up and spin-down of $d_{xz}/d_{yz}$, which provides the primary contribution to the positive MAE.

SP-DFT calculations including SOC revealed quite small MAEs ranging between ±0.1–1.1 meV and increasing within the Cr-Mn-Fe order (**Table 1**) in line with the low orbital moment anisotropy at isotropic spin moments (**Table S4**). The magnetic force theorem provided even lower values of MAE (**Table 1**).

**Table 1** The binding energy $E_{bind}$ (eV), TM–C bond length (Å), total magnetic moment $\mu_{tot}$ ($\mu_B$), magnetic moment of TM $\mu_{TM}$ ($\mu_B$), magnetic moment of nearest carbon atoms $\mu_C$ ($\mu_B$), Bader charges $q_{Bader}$ of TM (e⁻), MAE(TE) and MAE(FT) (both in meV) of TM@SV and TM@DV.

| system | $E_{bind}$ | TM–C | $\mu_{tot}$ | $\mu_{TM}$ | $\mu_C$ | $q_{Bader}$ | MAE(TE) | MAE(FT) |
|--------|-----------|------|-------------|------------|---------|-------------|---------|---------|
| **Cr@SV** | -6.47 | 1.88 | 2.00 | 2.44 | -0.12 | 1.04 | -0.15 | -0.11 |
| **Mn@SV** | -6.39 | 1.85 | 3.00 | 2.75 | -0.09 | 0.91 | -0.26 | -0.25 |
| **Fe@SV** | -7.13 | 1.81 | 1.95 | 1.81 | -0.03 | 0.74 | 0.83 | 0.25 |
| **Cr@DV** | -7.24 | 2.07 | 2.00 | 2.44 | -0.10 | 1.18 | 0.05 | 0.02 |
| **Mn@DV** | -7.21 | 1.98 | 3.25 | 3.42 | -0.08 | 1.17 | 0.64 | 0.49 |
| **Fe@DV** | -6.48 | 1.90 | 3.16 | 2.98 | -0.03 | 1.17 | 1.11 | 0.21 |

In our scanning transmission electron microscopy observations of chemically synthesized monolayer graphene, we were able to find an incidental Mn impurity bonded to four carbon neighbors in a defected area of the lattice (**Figure 2a**). The identity of the impurity was confirmed by electron energy-loss spectroscopy, which also allowed us to determine its Mn $L_3/L_2$ white-line ratio as 1.74±0.05 (**Figure 2b**). This value contrasts with the almost twice as high value of 3.38±0.05 reported by Lin *et al.* for Fe@DV [20]. While it is difficult to precisely quantify the values of their magnetic moment based on these ratios [58], such a large difference is surprising considering that our simulation results shown in **Table 1** would lead to expect that both Fe and Mn exhibit similar localized magnetic moments.

The reason for this discrepancy could be the local disorder here: while the Mn is bonded to four carbon neighbors, the nearby lattice contains pentagonal and heptagonal rings. Furthermore, other impurities that are presumably silicon are bound to vacancies in the nearby lattice, which might further disturb the localized properties of the Mn site. Indeed, simulations with simultaneous Mn and Si dopants in the 6×3 cell indicated the reduction of the magnetic moment on Mn because of nearby Si (**Table S5** and **Table S6**). Bader charge analysis revealed a charge transfer of 2.2–2.5 e$^-$ from Si to the graphene lattice and the DOS plots of Mn₁Si@SV and Mn₁Si@DV shown in **Figure S11** exhibit small electronic gaps, overlap of $p_z$- and $d_z{}^2$-derived states and, thus, the reduced Mn magnetic moment. There are multiple Si atoms in the experimental sample, which further reduces the Mn magnetic moment. These findings highlight the sensitivity of the experimental magnetic moment to sample quality, and the need for further systematic studies.

Very recently, the incorporation of substitutional Mn atoms into SV defects in graphene on Cu(111) has been reported [59]. While the $L_3/L_2$ Mn white-line ratio has not been studied, the reported magnetic moment on Mn from DFT calculations is ~3 $\mu_B$, which agrees with our results (**Table 1**).

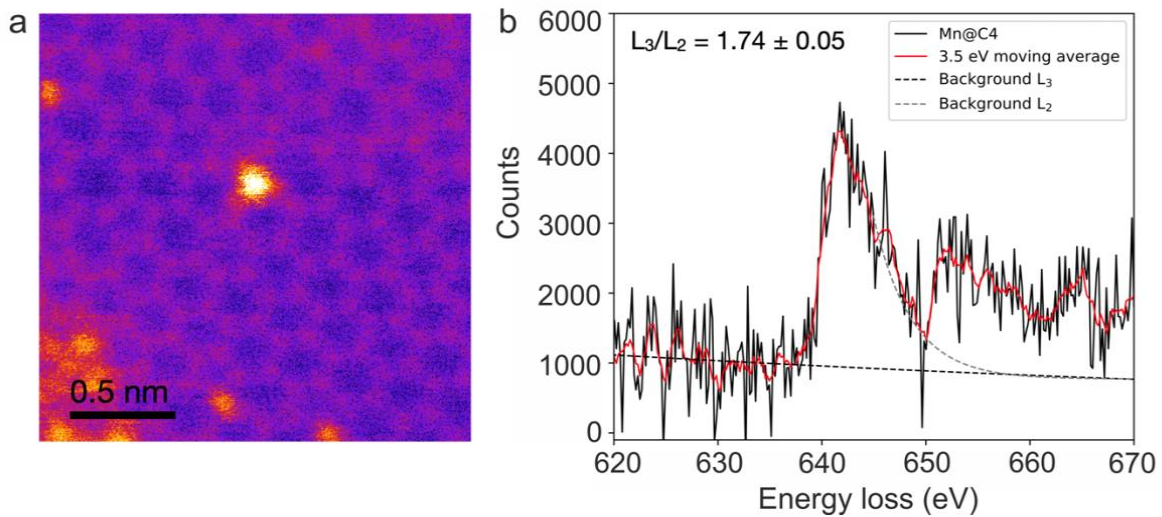

**Figure 2** Experimental observation of a Mn@DV impurity. **a)** STEM/MAADF image of the Mn atom (bright contrast) bonded to four carbon atoms in a disordered patch of graphene. The slight irregular variation in the contrast of some scan lines before the impurity is likely due to mechanical or electronic scan noise. **b)** EEL spectrum of the Mn $L$ core edge, confirming the identity of the impurity and allowing us to determine its localized magnetic moment via the $L_3/L_2$ white-line ratio.

### 2) Graphene simultaneously doped by two TM atoms

Recent experimental manipulation of impurity atoms in monolayer graphene [30,31,60] has raised our interest in studying the influence of the mutual arrangement and distance of TM atoms (thereafter denoted as $TM_1TM_2$@SV and $TM_1TM_2$@DV) on the electronic and magnetic properties of graphene. We used a larger 6×3 orthorhombic cell containing 96-carbon atoms to introduce a second SV or DV defect in the graphene lattice as shown in **Figure 1**, and by changing the distance between the vacancy defects (**Figure S12** and **Figure S13**), we simulated the mutual interaction of TM atoms at different separations within the graphene lattice. Both cis/trans position of TM atoms in graphene lattice (*i.e.*, both TM were above the graphene lattice vs. one TM above and one below the lattice) were considered, but only the most energetically stable structures are discussed in the following.

While virtually no differences in the structural features of $TM_1TM_2$@SV and $TM_1TM_2$@DV were observed compared to TM@SV and TM@DV, the magnetic moments significantly differed from those of single TM-atom doped graphene (**Table S7** and **Table S8**). One could expect that the magnetic moments of the larger 6×3 cells should correspond to the sum of the magnetic moments of TM-SAs (this is indeed true for CrCr@SV and MnMn@SV), however, CrCr@DV and MnMn@DV exhibited even higher magnetic moments indicating synergistic effect between the TM-atoms, *i.e.,* maximum value of 8.0 µ$_B$ per computational cell vs. the expected 4.0 µ$_B$ for Cr and 6.0 µ$_B$ for Mn. At the same time, the values of magnetic moments varied between 6.1 and 8.0 µ$_B$ (CrCr@DV) and 7.3 and 8.0 µ$_B$ (MnMn@DV) depending on the mutual TM-TM distance, and between 2.5–6.0 µ$_B$ and 5.6–6.9 µ$_B$ respectively for FeFe@SV and FeFe@DV (**Table S7** and **Table S8**). No systematic change of magnetic moments with the TM-TM distance was however observed. In agreement with the previous reports [19,22,25,61] depending on the distribution of the TM atoms in the graphene lattice, the AFM order was more energetically preferred than the FM one; for instance, MnMn@SV (**Figure S12-20**) and CrCr@DV (**Figure S13-4**) were AFM in the GS and lower in energy by 69.8 and 30.4 meV, respectively, as compared to the FM order.

As compared to TM@SV and TM@DV, DOS of $TM_1TM_2$@SV and $TM_1TM_2$@DV (**Figure S14-S23** and **Figure S24-S31**) changed significantly due to the introduction of a second TM atom into the graphene lattice. For the CrCr@SV GS structure (**Figure S14**; see also **Figure S15**), the degeneracy of the $d_\delta$-derived states was broken, and they were broadened due to the largely increased binding energy of the second Cr atom. Further, an up-shift towards lower binding energies was observed for occupied $d_{xz}/d_{yz}$, which were also broadened. Importantly, while the bandgap value (**Table S9**) for the GS structure was 0.5 (0.3) eV for the spin up (down) channel, the less stable FM structure (**Figure S15-4**) exhibited metallic DOS (**Figure S16**), while in the AFM alignment (more stable than the FM one by 99.1 meV) it exhibited an electronic gap of 0.3 eV (**Figure S17**).

For MnMn@SV, a broadening and up-shift of the occupied $d$ states brought about the metallic character to the DOS of the GS structure (**Figure S18, Table S9**) with FM alignment of the magnetic moments. For the GS structure ordered AFM (**Figure S19**), and for the less stable system (**Figure S20**), the electronic gap was shrunk by 0.1–0.2 eV comparing to the SA counterpart. Similarly, the electronic gap of FeFe@SV GS was as low as 0.1 eV (**Table S9, Figure S21**), or even disappeared for the GS–1 FM structure (**Figure S22**). The bandgap of the GS-1 AFM structure (more stable than the FM one by 98.4 meV) was 0.2 eV (**Figure S23**).

For $TM_1TM_2$@DV, we observed half-metallic (metallic) character of CrCr@DV in the GS (GS–1) (**Figure S24-S26**) and a small bandgap (up to 0.3 eV) opening for MnMn@DV (**Figure S27-S28**) and FeFe@DV in the spin up channel (**Figure S29-S30**) (**Table S9**). The AFM FeFe@DV (**Figure S31**) was lower in energy than the FM analogue by 29.7 meV and exhibited a tiny bandgap of 0.1 eV.

Whereas the MAE changed within the range of 0.4 meV (calculated per computational cell) for the TM pairs, no systematic dependence of the MAE on the TM–TM distance was observed. While the MAE of both MnMn@SV and MnMn@DV roughly reached twice the MAE value of Mn@SV and Mn@DV, the MAE of CrCr@SV and CrCr@DV increased and the MAE of FeFe@SV and FeFe@DV decreased compared to the SA counterparts (**Table S7 and Table S8**). It should be noted, however, that for the graphene-doping homoatomic pairs, the MAEs are in the sub-meV range, and the highest MAE found for CrCr@DV is only 1.6 meV. We also note that the sign of the MAE changed in some cases.

Since the value of MAE for homoatomic doped graphene did not significantly change with the mutual distance between TM atoms, for graphene doped with two different atoms we only examined MAE for a fixed distance of 12.87–13.24 Å between $TM_1$ and $TM_2$. The total magnetic moment of such a system increased to approximately the sum of the magnetic moments of the individual TM atoms (**Table 2** and **Table S10**). The electronic bandgaps reached roughly intermediate values obtained for the SA counterpart (**Table S11 and Figures**

**S32-S38**). FeMn@SV and FeMn@DV exhibited metallic DOS. Apart from CrFe@DV, whose GS was AFM and higher in energy than the FM one by 15.4 meV, all other systems were FM in the GS.

**Table 2** The binding energy $E_{bind}$ (eV), TM$_1$–TM$_2$ distance (Å), total magnetic moment $\mu_{tot}$ ($\mu_B$), magnetic moments of TM$_1$ and TM$_2$ $\mu_{TM1}$, $\mu_{TM2}$ ($\mu_B$), and MAE$_{(TE)}$ and MAE$_{(FT)}$ (per computational cell, both in meV) of TM$_1$TM$_2$@SV and TM$_1$TM$_2$@DV depicted in **Figures S32-S38**.

| Structure | $E_{bind}$ | TM$_1$-TM$_2$ | $\mu_{tot}$ | $\mu_{TM1}, \mu_{TM2}$ | MAE$_{(TE)}$ | MAE$_{(FT)}$ |
|---|---|---|---|---|---|---|
| **CrFe@SV** | -12.16 | 13.24 | 4.00 | 2.49, 1.55 | 8.80 | 0.15 |
| **CrMn@SV** | -11.73 | 12.86 | 5.00 | 2.48, 2.73 | 0.40 | -0.34 |
| **FeMn@SV** | -11.25 | 13.04 | 4.89 | 1.58, 2.84 | 2.90 | -0.16 |
| **CrFe@DV (AFM)** | -5.53 | 12.95 | 0.00 | 0.23, -0.04 | -23.20 | 0.61 |
| **CrFe@DV** | -5.52 | 12.95 | 6.00 | 2.43, 3.12 | -23.00 | 0.49 |
| **CrMn@DV** | -5.17 | 13.00 | 5.09 | 2.45, 3.42 | 16.00 | 0.59 |
| **FeMn@DV** | -3.97 | 12.87 | 6.70 | 3.02, 3.48 | -15.80 | 1.08 |

Noticeably, the MAE (calculated per computational cell) significantly increased as compared to the single TM-atoms, reaching –23.0 meV for FM CrFe@DV, 16.0 meV for CrMn@DV and –15.8 meV for FeMn@DV. The character of spin excitations near $E_F$ remains similar for Cr in Cr@DV (**Figure S8**) and in mixed systems CrTM@DV (**Figures S35, S36** and **S39**), *i.e.*, the electron hopping between the nearest occupied and unoccupied states, $d_{xy}$ and $d_{yz}$, favors the in-plane anisotropy. For the Mn atom in CrMn@DV, spin excitations involving spin-up occupied/empty $d_{xz/yz}$ and spin-up occupied $d_{xz}$ and spin-down empty $d_{z^2}$ favors positive MAE (**Figures S37** and **S40**). For Mn in FeMn@DV (**Figures S38** and **S41**), spin excitations in the immediate vicinity to $E_F$ involving $d_{yz/xz}$ favor positive MAE, which is counterbalanced by the negative contribution from the coupling between majority-spin occupied $d_{xy}$ and empty $d_{xz/yz}$. Similarly, the coupling between spin-majority occupied $d_{yz}$ and spin-minority empty $d_{xz}$, and between spin-majority occupied $d_{xy}$ and spin-minority unoccupied $d_{xz}$ favors, respectively, negative and positive MAE.

While the occupation scheme of Fe in FeMn@DV is like in Fe@DV – the negative contribution to MAE is due to the coupling of majority-spin and minority-spin $d_{yz/xz}$, which compete with the positive contribution as described above for Fe@DV – it changed in CrFe@DV (**Figure**

**S35**) so that now easy-plane is also favored due to the net negative contribution from the coupling between majority-spin/minority-spin occupied/unoccupied $d_{xz/yz}$, and between minority-spin occupied $d_{z^2}$ and unoccupied $d_{xz/yz}$; however, the coupling between majority-spin occupied $d_{xz/yz}$ and minority-spin unoccupied $d_{z^2}$ provides a positive contribution to the MAE. One shall also note that the energy denominators in the mixed systems and thus the corresponding contributions to the MAE can be different than in SA doped graphene.

The visible effect of the enhanced MAE in DV-graphene doped with two different TM atoms, which contrasts with SA doped graphene, is a changed splitting of energy bands that were degenerate in the absence of spin-orbit interaction, specifically the $d_{xz/yz}$ states, and the downshift in binding energies of the $d_{z^2}$ states, both leading to a lowering of the total energy (**Figures S39-S44**) upon the reorientation of the magnetization direction: The easy-plane of the CrFe@DV and FeMn@DV systems seems associated with the downshift in binding energies, respectively the $d_{yz}$, $d_{xz}$, $d_{z^2}$ states (**Figure S39-S40**) and the $d_{xz}$ and $d_{yz}$ states (**Figure S43-S44**). As for CrMn@DV, the changes in $d$-DOS of Cr at about –1.5 to –1.0 eV and of Mn at about –1.5 eV brought about the positive MAE (**Figure S41-S42**). Surprisingly, the spin and orbit anisotropy for both CrFe@DV and CrMn@DV was ~0.0 µ$_B$, while for FeMn@DV a substantial spin anisotropy at almost isotropic orbital moments was found (**Table S10**). Interestingly, when Cr and Mn atoms of CrMn@DV were separated by 6.55 Å, the MAE increased up to 47.5 meV due to a strong anisotropy of the spin moments (–0.55 µ$_B$).

Note that the MAE obtained by using the magnetic force theorem, MAE$_{(FT)}$, were much lower than MAE$_{(TE)}$. In principle, FT calculations can be used to decipher the atomic and orbital partial contribution to MAE; however, this was not possible for the systems studied here due to the negligible MAE$_{(FT)}$.

The presence of TM atoms induces magnetic moments in the surrounding carbon atoms, as revealed by asymmetry in electronic DOS (**Figures S32-S44**), *i.e.*, the sharing of electrons between the magnetic atoms and the lattice leads to a net magnetic moment on the carbon sites (**Figures S45-S47**). Further, a (half)metallic DOS at $E_F$ (CrMn@DV) and FeMn@DV can enable the conduction electrons to couple to the magnetic moments indicating the presence of an RKKY (Rudermann–Kittel–Kasuya–Yoshida) [62–64] exchange through the graphene lattice (**Figure S48**). For the CrFe@DV system with small energy gaps at $E_F$, the RKKY interaction is suppressed and the coupling between TM atoms is due to a super-exchange network of interactions throughout the carbon atoms. Thus, the origin of the greatly enhanced MAE up to –23 meV for DV-graphene doped simultaneously by two different TM atoms can be sought in graphene-mediated coupling due to the RKKY-like super-exchange interactions between TM-atoms. Electrons predominantly hop between the occupied and unoccupied states in the immediate vicinity to $E_F$, which leads to the switch of magnetic anisotropy from the easy-plane direction for CrFe@DV (**Figure S35**) and FeMn@DV (**Figure S38**) to the out-of-plane direction for CrMn@DV (**Figure S37**). Although the RKKY-like super-exchange interactions between Cr and Mn pairs and trimers doping the graphene lattice [65,66] have also been reported, only pairs of different TM atoms in DV appear to influence their electronic structures in favor of large MAEs.

Notably, the graphene-doping heteroatomic TM pairs should exhibit a high stability in the vacancy defects (**Table S12** and **Figure S49**). This contrasts the behavior of $3p/2p$ or $4p/2p$ co-doping elements, which in almost all cases prefer to replace a C–C bond [67,68].

For the sake of completeness, the simultaneous adsorption of two TM atoms in the centers of two different defect types separated by about 7.8 Å and 9.5 Å was considered: $Cr_1Cr_2$@SVDV and CrFe@SVDV (**Table S13** and **Figure S50**). The magnetic moments were close to the sum of the individual Cr@SV and Cr@DV systems; only the magnetic moment on Fe was lower as in Fe@DV. The MAEs were low, about 1 meV. However, it was significantly enhanced from 0.1 meV to 8.1 meV in $Cr_1Cr_2$@SVDV when the inter-vacancy distance changed from 7.8 to 9.4 Å, indicating that in experimental samples the precise implementation of TM atoms may play a significant role for achieving a large MAE.

### 3) Adsorption of a TM atom onto TM Doped Graphene

Theoretical calculations revealed that transition metals dimers can exhibit an enhanced MAE than the SA-doped graphene [7,14,48]. Therefore, we considered TM–TM dimers embedded into defective graphene as shown in **Figure S51-S62**. The binding energy of a second TM atom on top of TM@DV and TM@SV was exothermic in energy ranging from –4.05 eV to –0.04 eV with a $TM_1$–$TM_2$ bond length of 1.98-2.85 Å (**Table 3**). It must be noted that while more stable configurations may be obtained by binding TM atoms to graphene on both sides, the resulting MAEs are low, in sub-meV range (*cf.* **Table S14** and **Figure S63**), as are the MAEs of the graphene-doping TM-SAs.

**Table 3** The binding energy $E_{bind}$ (eV), $TM_1$–$TM_2$ distance (Å), total magnetic moment $\mu_{tot}$ ($\mu_B$), magnetic moments of $TM_1$ and $TM_2$ $\mu_{TM1}$, $\mu_{TM2}$ ($\mu_B$), and $MAE_{(TE)}$ and $MAE_{(FT)}$ (per computational cell, both in meV) of $TM_2$@$TM_1$@SV and dimers $TM_2$@$TM_1$@DV depicted in **Figures S51-S63**.

| Structure | $E_{bind}$ | $TM_1–TM_2$ | $\mu_{tot}$ | $\mu_{TM1}, \mu_{TM2}$ | $MAE_{(TE)}$ | $MAE_{(FT)}$ |
|---|---|---|---|---|---|---|
| **Cr@Cr@SV** | -0.39 | 2.85 | 3.72 | 2.32, 1.28 | -0.73 | -0.43 |
| **Fe@Cr@SV** | -1.66 | 2.19 | 2.97 | 1.89, 1.80 | -6.85 | -1.26 |
| **Mn@Cr@SV** | -2.70 | 2.19 | 6.86 | 1.90, 4.38 | -1.13 | -0.22 |
| **Fe@Fe@SV** | -0.09 | 2.27 | 1.42 | -0.03, 1.37 | -12.40 | 0.77 |
| **Fe@Mn@SV** | -3.61 | 2.31 | 5.00 | 1.65, 3.44 | -5.05 | -1.79 |
| **Mn@Mn@SV** | -0.19 | 2.47 | 7.98 | 2.48, 4.61 | 14.90 | -1.96 |
| **Cr@Cr@DV** | -0.04 | 2.48 | 6.00 | 0.75, 4.56 | -0.42 | -0.20 |
| **Fe@Cr@DV** | -3.39 | 2.19 | 4.00 | 0.76, 3.09 | -2.31 | -1.39 |
| **Mn@Cr@DV** | -4.05 | 2.19 | 3.00 | -1.60, 3.84 | 0.35 | 0.27 |
| **Fe@Fe@DV** | -1.58 | 2.13 | 3.96 | -0.96, 3.75 | -0.86 | 1.29 |
| **Fe@Mn@DV** | -1.25 | 1.98 | 3.03 | 0.46, 2.52 | 9.79 | 0.61 |
| **Mn@Mn@DV** | -2.64 | 2.37 | 8.04 | 3.30, 4.28 | 0.16 | 0.11 |

The interaction between the TM atoms in a dimer led to lower or higher total (1.4–8.0 $\mu_B$) and local magnetic moments as compared to the SA-doped defective graphene (**Table 3** and **Table S15**). The magnetic moment of the lower TM atom was quenched by the adsorption of a second TM atom.

The relatively strong binding of a second $TM_2$ atom above the $TM_1$ in $TM_2@TM_1@SV$ and $TM_2@TM_1@DV$ significantly altered the electronic structure of the $TM_1$ doped graphene (**Figures S51-S62**, **Table S16**). Particularly, broadening and/or upshift of the states led to the metallic DOS for several systems (Cr@Cr@SV, Fe@Cr@SV, Fe@Fe@SV, *cf.* **Figures S51, S52, S54**, and **Figures S64-S67**). For the metallic Fe@Cr@DV and Fe@Mn@DV, the reordering of states in the vicinity to $E_F$ was observed (**Figures S58, S61** and **S67**). In the remaining systems the band gap in the range of 0.1 and 0.5 eV was calculated (**Table S16**).

Among the TM dimers considered, a large positive MAE was calculated for Fe@Mn@DV (9.8 meV) and Mn@Mn@SV (14.9 meV), and a large negative MAE for Fe@Fe@SV (−12.4 meV) (**Table 3**). The systems with the largest MAE exhibited the largest anisotropies of the magnetic moments (**Table S15**). The shape anisotropy contribution [69] to MAE is negligible, below 1 meV, increasing from 0.04 meV for the mixed-atom system CrFe@DV to 0.91 meV for Fe@Mn@DV. Since only Fe@Mn@DV has a big $E_{bind}$ among dimers with a large MAE, we will focus only on it in the following.

**Figure S61** shows the change of occupation of the Mn atom in the Fe@Mn@DV system compared to SA Mn@DV (**Figure S9**). The $d_{xy/x^2-y^2}$ states of the lower atom are pushed down to –0.75 eV, the $d_{xz/yz}$ states above $E_F$, and $d_{z^2}$ split into two components, one below occupied $d_{xy/x^2-y^2}$ and another just above $E_F$. The states of Fe lie in the immediate vicinity to $E_F$. The $d_{z^2}$ states of both TM atoms split and coincide indicating $\sigma$ bonding between two metal atoms. The $d_{xz/yz}$ states of both TM atoms are broadened and overlap, indicating $\pi$ bonding.

The spin excitations near $E_F$ for the Mn atoms (**Figure S61**) favor a negative MAE. This, however, due to the large energy denominator is very small. Previous reports [13,14,16] have demonstrated that the leading contribution to the MAE is imposed by the upper TM atom despite the opposite sign of MAE of the lower TM atom. For the upper Fe atom, the coupling between minority-spin occupied $d_{xy}$ and majority-spin empty $d_{xz/yz}$ favors the positive MAE. Further positive contribution to the MAE is associated with the downshift of the peaks in the $d_{yz}$ spectra to about –1.4 to –1.2 eV for the perpendicular magnetization, which was located just below $E_F$ for the in-plane direction. Altogether, the downshift in binding energies of the $d$-states of the upper TM atom favoring the positive MAE can be seen in the relativistic PDOS (**Figure 3** and **Figure S67**).

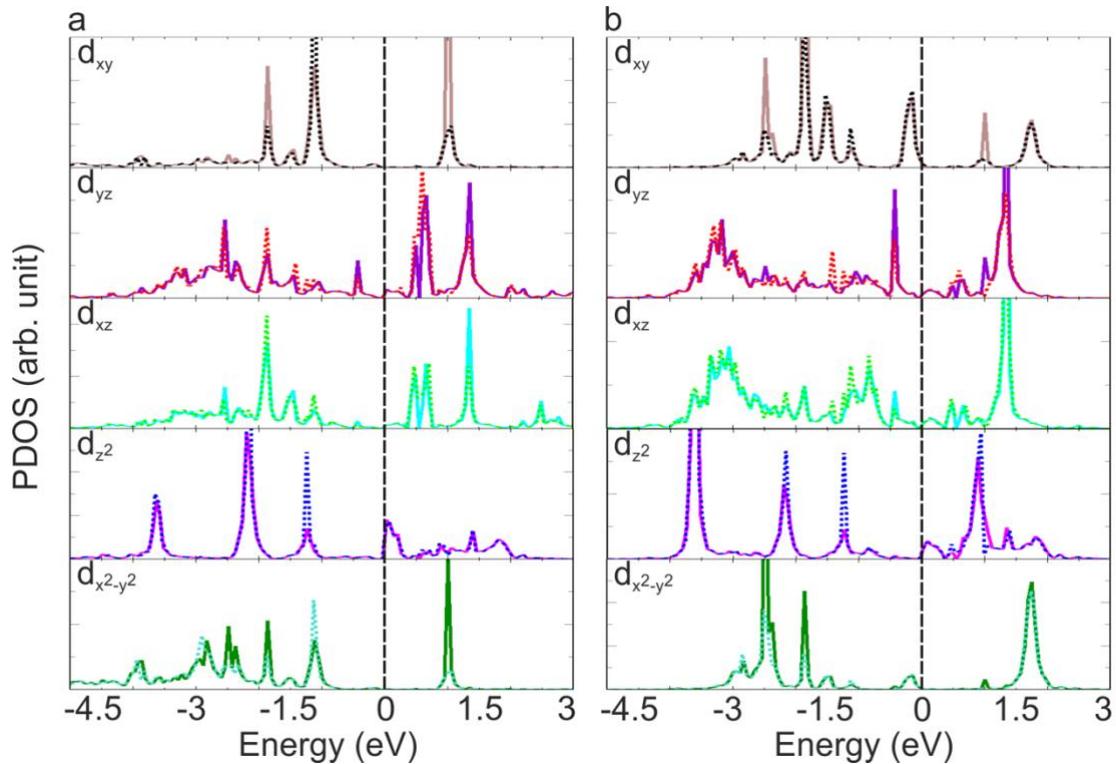

**Figure 3** Relativistic partial atom/orbital-resolved densities of states for Fe@Mn@DV (**Figure S61h**) for in-plane (solid lines) and perpendicular magnetization (dashed lines). **a)** Mn atom and **b)** Fe atom. (*Cf.* **Figure S67**).

Interestingly, the introduction of second dimer separated by 12.9 Å (6×3 cell, **Figure 4**) caused the increase of MAE up to 119.7 meV in Fe@Mn@DV due to the spin transition as indicated in spin anisotropy of –1.6 $\mu_B$ and orbital anisotropy of 0.1 $\mu_B$. The scalar-relativistic PDOS (**Figure S68**) for the two Fe–Mn dimers bound to separate DV defects indicates on competing contributions to positive/negative MAE due to excitations near $E_F$. However, the gapless PDOS facilitates graphene-mediated coupling between the dimers due to the RKKY exchange interactions (see above and **Figure S69**), and furthermore, the quasi-degeneracy of partially occupied states at $E_F$ for in-plane magnetization is lifted for axial magnetization, as shown in **Figure 5**, bringing about the dominant positive contribution to the enormous MAE of ~120 meV. Similarly, the existence of a partially occupied quasi-degenerate state at $E_F$ has been shown to favor the formation of a large magnetic anisotropy in an Ir–Co dimer supported on an ideal graphene layer [13].

The giant MAE of ~120 meV corresponds to the blocking temperature of 34 K based on Néel's theory [50].

It is noteworthy that the Fe–Mn dimers with the high MAE are expected to be stable in their GS configurations in the vacancy defects (**Table S17** and **Figure S70**), as also shown by the phonon frequency calculations (**Table S18**) and the calculated Gibbs energy [70] of –1.36 eV at 373 K.

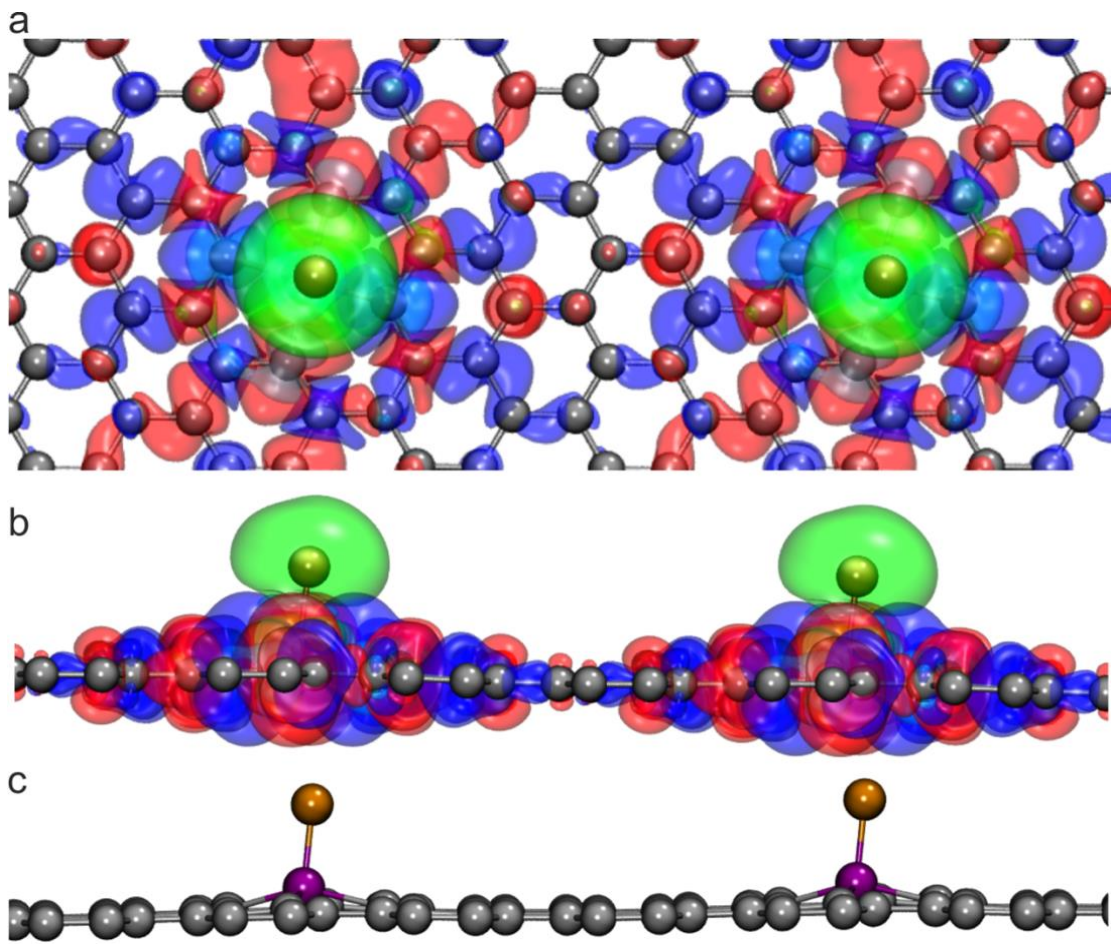

**Figure 4 a)** Top and **b)** side view of spin densities plotted at ±0.01 e⁻·Å⁻³ isovalues for Fe@Mn (displayed in green/cyan for spin densities corresponding to positive/negative magnetic moments) and ±0.001 e⁻·Å⁻³ (shown in blue/red) for DV-graphene for the system of two Fe— Mn dimers bound to separate DV defects. **c)** Side view of the corresponding structure. Gray, purple and ochre colors represent carbon, manganese and iron atoms, respectively.

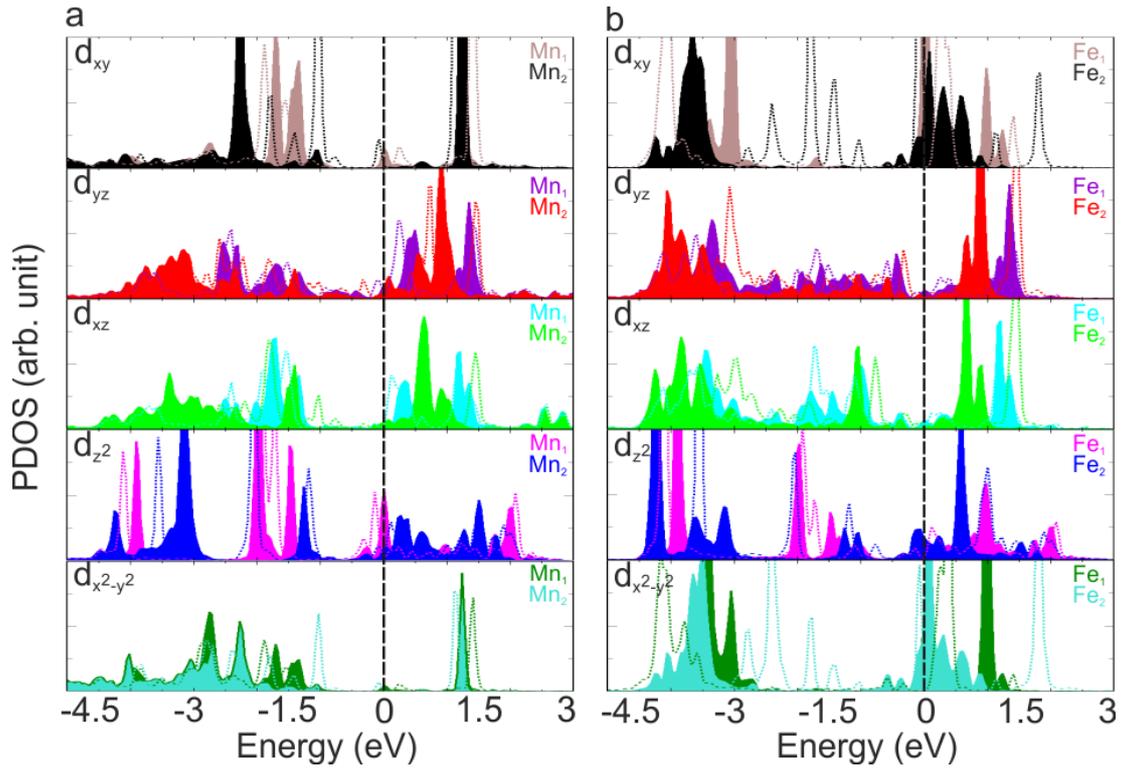

**Figure 5** Relativistic partial atom/orbital-resolved densities of states for two Fe—Mn dimers bound to separate DV defects (**Figure 4**) for in-plane (solid lines) and perpendicular magnetization (dashed lines). **a)** Mn atoms and **b)** Fe atoms.

### 4) TM doping of N-decorated Defective Graphene

The presence of N atoms at the edges of the defects in the graphene layer (**Figure S71**) led to higher binding energies of TM@NSV and TM@NDV, smaller TM–C/N distances and higher charge transfers in comparison with TM@SV and TM@DV (**Table S19**). Magnetic moments of TM atoms in TM@NSV and TM@NDV can either increase or decrease depending on the system and they varied in the range 0.3–1.9 $\mu_B$ (**Table S19** and **Table S20**). The lower spin state of Fe@NDV compared to Fe@DV is in line with experimental observations [20]. In general, N-impurities act as $n$-dopants and shift the density of states towards lower energies (**Figures S72-S77**). This also led to the semiconductor (Cr@SV and Mn@SV) to metal transition (Cr@NSV and Mn@NSV), half-metal (Fe@SV) to semiconductor transition (Fe@NSV) and metal (Mn@DV and Fe@DV) to semiconductor transition (Mn@NDV and Fe@NDV), **Table S21**. No significant changes of MAE (by 0.1–1.8 meV) were observed as compared to TM@DV and TM@SV. The magnetic force theorem revealed much higher MAE than the fully self-consistent calculations (**Table S19**), which indicates that MAE of atom-sized magnets calculated as the difference in the band energies at a fixed potential and charge density should be validated against MAE calculated as the difference in the total energies from self-consistent calculations for different orientations of the magnetic moments.

The mixed element dopants, $TM_1TM_2$@NSV and $TM_1TM_2$@NDV, showed a reversed stability trend as compared with pristine analogs, *i.e.*, the $TM_1TM_2$@NDV possessed lower binding energies than $TM_1TM_2$@NSV (**Table S22**). Their magnetic moments increased due to the combination of individual TM atoms ranging between 2.6–8.7 μB (**Table S22**). The lowest-in-energy structures of single and double vacancies demonstrated half-metallic DOS (**Figures S78** and **S79**, **Table S21**).

The FeMn@NSV system, with the largest magnetic moment of 8.7 μB, exhibited the biggest MAE (5.2 meV) among the $TM_1TM_2$@NSV and $TM_1TM_2$@NDV systems (with the MAE in the sub-meV range or about 1 meV) with an eight-fold increase of MAE in comparison to TM@NSV and TM@NDV (**Table S22** and **Table S23**), due to the favorable graphene-mediated coupling between TM atoms doping graphene (see above). However, the presence of nitrogen led to the significantly smaller values of MAE comparing to $TM_1TM_2$@SV and $TM_1TM_2$@DV, *i.e.*, 5.2 vs. 16.0 meV.

Finally, the formation of upright dimers $TM_2$@$TM_1$@NSV and $TM_2$@$TM_1$@NDV were energetically favorable (**Table S24**). The lowest-in-energy structure of single and double vacancies showed metallic and half-metallic states (**Figures S80-S81**, **Table S21**), differing from the pristine analogs. The MAE was lower than for the corresponding dimers bound to the pristine vacancy defects, except Cr@Cr@NSV, which possess MAE of –17.6 meV (**Table S24** and **Table S25**).

## Conclusion

In this work, we theoretically examined the stability as well as the electronic and magnetic properties of TM atoms and dimers of the elements Cr, Mn and Fe bound to graphene single or double vacancies. We studied the dependence of magnetic anisotropy energy (MAE) on (*i*) the distances between the TM atoms within the monolayer graphene, (*ii*) co-doping of graphene with two atoms of the same and different elements, (*iii*) formation of upright TM dimers, both homo- and hetero-atomic and, finally, (*iv*) the presence of nitrogen atoms within the vacancy defects.

While the MAE of TM single atoms in defective graphene did not exceed 2 meV, the binding of a second TM atom to another defect in the graphene lattice led to the significant increase of MAE to –23 meV for CrFe@DV and a remarkably high value of 119.7 meV for two upright

Fe–Mn dimers bound to two separate DVs, with the easy plane and perpendicular easy axis, respectively. This giant MAE corresponds to the blocking temperature of 34 K. The origin of the greatly enhanced MAE can be sought in graphene-mediated coupling due to the RKKY-like super-exchange interactions between TM-dopants that modifies the occupations of states near $E_F$, such that a changed splitting of energy bands that were degenerate in absence of spin-orbit interaction lowers the total energy. Specifically, the presence of partially occupied degenerate states at $E_F$ favors the formation of a giant MAE, which appears to be a general feature of the mechanism favoring the large MAE formation [5]. It is worth noticing that the value and even the sign of MAE as well as the electronic DOS are sensitive to both the distribution of TM atoms in the graphene lattice and composition of the graphene-doping atoms.

Finally, we provided the atomic-scale observation of a Mn substitution in graphene, including a spectroscopic measurement of its $L$-edge core energy-loss white-line intensity ratio. Mn dopants bonded to four carbon atoms may create spots for the Fe—Mn dimers formation.

The conductive TM-doped graphene with robust magnetic features offers new vista to the design of graphene-based spintronic devices. For practical applications of TM-doped graphene, the carbon sheet must be deposited on a solid substrate. It remains to be studied the effect of substrates on the MAE of TM atoms and dimers anchored by the defects in graphene.

## Supporting Information

Density of states of TM@SV and TM@DV; Finite model calculations of defective graphene; Scheme showing the coupling of eigenstates; Atom/orbital decomposed DOS of TM@SV and TM@DV; Properties of $Mn_1Si$@SV and $Mn_1Si$@DV; Properties and density of states of $TM_1TM_2$@SV and $TM_1TM_2$@DV with variable $TM_1$–$TM_2$ distance; Relativistic partial atom/orbital-resolved densities of states of $TM_1TM_2$@SV and $TM_1TM_2$@DV; Spin density distribution plots of $TM_1TM_2$@SV and $TM_1TM_2$@DV; Variation in the properties of CrFe@DV with the distance between TM atoms; Barrier for diffusion of $TM_1TM_2$@SV and $TM_1TM_2$@DV; Structure and density of states for a system of two TM atoms, one embedded into SV and the other into DV defect; Atom/orbital decomposed DOS of $TM_2$@$TM_1$@SV and $TM_2$@$TM_1$@DV; Relativistic partial atom/orbital-resolved densities of states of $TM_2$@$TM_1$@SV and $TM_2$@$TM_1$@DV; Variation in the properties of Fe@Mn@DV with the distance between TM atoms; Barrier for diffusion of $TM_2$@$TM_1$@SV and $TM_2$@$TM_1$@DV;

Structures, properties and density of states of TM@NSV, TM@NDV, $TM_1TM_2$@NSV, $TM_1TM_2$@NDV, $TM_2$@$TM_1$@NSV and $TM_2$@$TM_1$@NDV

## Acknowledgements

Work in Olomouc has been supported by the Operational Programme for Research, Development and Education of the European Regional Development Fund (project no. CZ.02.1.01/0.0/0.0/16_019/0000754). RL acknowledges support from the Internal Student Grant Agency of the Palacký University in Olomouc, Czech Republic (IGA_PrF_2021_031) and the internship at the University of Vienna. AM and TS acknowledge funding from the European Research Council (ERC) under the European Union's Horizon 2020 research and innovation programme (Grant agreement No. 756277-ATMEN). We thank Ulrich Kentsch, Mukesh Tripathi and Andreas Postl for experimental assistance.

## References

1.  Novoselov, K. S.; Geim, A. K.; Morozov, S. V.; Jiang, D.; Zhang, Y.; Dubonos, S. V.; Grigorieva, I. V.; Firsov, A. A. Electric Field Effect in Atomically Thin Carbon Films. *Scienc*e **2004**, *306*, 666-669, DOI: 10.1126/science.1102896.

2.  Tuček, J.; Błoński, P.; Ugolotti, J.; Swain, A. K.; Enoki, T.; Zbořil, R. Emerging Chemical Strategies for Imprinting Magnetism in Graphene and Related 2D Materials for Spintronic and Biomedical Applications. *Chem. Soc. Rev.* **2018**, *47*, 3899–3990, DOI: 10.1039/C7CS00288B.

3.  Khajetoorians, A. A.; Wiebe, J.; Chilian, B.. Wiesendanger, R. Realizing All-Spin-Based Logic Operations Atom by Atom. *Science* **2011**, 332, 1062–1064, DOI: 10.1126/science.1201725.

4.  Natterer, F. D.; Yang, K.; Paul, W.; Willke, P.; Choi, T.; Greber, T.; Heinrich, A. J.; Lutz, C. P. Reading and Writing single-Atom Magnets. *Nature* **2017**, 543, 226–228, DOI: 10.1038/nature21371.

5.  Strandberg, T. O.; Canali, C. M.; MacDonald, A. H. Transition-Metal Dimers and Physical Limits on Magnetic Anisotropy. *Nat. Mater.* **2007**, *6*, 648-51, DOI: 10.1038/nmat1968.

6.  Błoński, P.; Lehnert, A.; Dennler, S.; Rusponi, S.; Etzkorn, M.; Moulas, G.; Bencok, P.; Gambardella, P.; Brune, H.; Hafner, H. Magnetocrystalline Anisotropy Energy of Co and Fe Adatoms on the (111) Surfaces of Pd and Rh. *J. Phys. Rev. B* **2010**, *81*, 104426, DOI: 10.1103/PhysRevB.81.104426.

7.  Błoński, P.; Hafner, J. Magnetic Anisotropy of Heteronuclear Dimers in the Gas Phase and Supported on Graphene: Relativistic Density-Functional Calculations. *J. Phys. Condens. Matter* **2014**, *26*, 146002, DOI: 10.1088/0953-8984/26/14/146002.

8.  Lehnert, A.; Dennler, S.; Błoński, P; Rusponi, S.; Etzkorn, M.; Moulas, G.; Bencok, P.; Gambardella, P.; Brune, H.; Hafner, J. Magnetic Anisotropy of Fe and Co Ultrathin Films Deposited on Rh(111) and Pt(111) Substrates: An Experimental and First-Principles Investigation. *Phys. Rev. B* **2010**, *82*, 094409, DOI: 10.1103/PhysRevB.82.094409.

9.  Sun, S.; Murray, C. B.; Weller, D.; Folks, L.; Moser, A. Monodisperse FePt Nanoparticles and Ferromagnetic FePt Nanocrystal Superlattices. *Science* **2000**, *287*, 1989-1992, DOI: 10.1126/science.287.5460.1989.

10. Gambardella, P.; Rusponi, S.; Veronese, M.; Dhesi, S.; Grazioli, C.; Dallmeyer, A.; Cabria, I.; Zeller, R.; Dederichs, P.; Kern, K.; Carbone, C.; Brune, H. Giant Magnetic Anisotropy of Single Cobalt Atoms and Nanoparticles. *Science*, **2003**, *300*, 1130-1133, DOI: 10.1126/science.1082857.

11. Błoński, P.; Hafner, J. Density-Functional Theory of the Magnetic Anisotropy of Nanostructures: An Assessment of Different Approximations. *J. Phys. Condens. Matter*, **2009**, *21*, 426001, DOI: 10.1088/0953-8984/21/42/426001.

12. Hu, J.; Wu, R. Giant Magnetic Anisotropy of Transition-Metal Dimers on Defected Graphene. *Nano Lett.* **2014**, *14*, 1853-1858, DOI: 10.1021/nl404627h.

13. Błoński, P.; Hafner, J. Cu(1 1 1) Supported Graphene as a Substrate for Magnetic Dimers with a Large Magnetic Anisotropy: Relativistic Density-Functional Calculations. *J. Phys. Condens. Matter* **2014**, *26*, 256001, DOI: 10.1088/0953-8984/26/25/256001.

14. Navrátil, J.; Błoński, P.; Otyepka, M. Large Magnetic Anisotropy in an OsIr Dimer Anchored in Defective Graphene. *Nanotechnology* **2021**, *32*, 230001, DOI: 10.1088/1361-6528/abe966.

15. Zhang, Y.; Wang, Z.; Cao, J. Prediction of Magnetic Anisotropy of 5d Transition Metal-Doped g-$C_3N_4$. *J. Mater. Chem. C* **2014**, *2*, 8817–8821, DOI: 10.1039/C4TC01239A.

16.    Zhang, K. C.; Li, Y. F.; Liu, Y.; Zhu, Y. Protecting Quantum Anomalous Hall State from Thermal Fluctuation: Via the Giant Magnetic Anisotropy of Os-based Dimers. *Phys. Chem. Chem. Phys.* **2018**, *20*, 28169–28175, DOI: 10.1039/C8CP05407J.

17.    Haldar, S.; Pujari, B. S.; Bhandary, S.; Sanyal, B. Fe$_n$ (n = 1–6) Clusters Chemisorbed on Vacancy Defects in Graphene: Stability, Spin-dipole moment, and Magnetic Anisotropy. *Phys. Rev. B* **2014**, *89*, 20541, DOI: 10.1103/PhysRevB.89.205411.

18.    Krasheninnikov, A.; Lehtinen, P.; Foster, A.; Pyykkö, P.; Nieminen, R. Embedding Transition-Metal atoms in Graphene: Structure, Bonding, and Magnetism, *Phys. Rev. Lett.* **2009**, *102*, 126807, DOI: 10.1103/PhysRevLett.102.126807.

19.    Durajski, A.; Auguscik, A.; Szczęśniak, R. Tunable Electronic and Magnetic Properties of Substitutionally Doped Graphene. *Phys. E 2020, Low-Dimensional Syst. Nanostructures* **2020**, *119*, 113985, DOI: 10.1016/j.physe.2020.113985.

20.    Lin, C.; Teng, P.; Chiu, P.; Suenaga, K. Exploring the Single Atom Spin State by Electron Spectroscopy. *Phys. Rev. Lett.* **2015**, *115*, 206803, DOI: 10.1103/PhysRevLett.115.206803.

21.    Santos, E. J. G.; Ayuela, A.; Sánchez-Portal, D. First-Principles Study of Substitutional Metal Impurities in Graphene: Structural, Electronic and Magnetic Properties. *New J. Phys.* **2010**, *12*, 053012, DOI: 10.1088/1367-2630/12/5/053012.

22.    Wu, M.; Cao, C.; Jiang, J. Electronic Structure of Substitutionally Mn-Doped Graphene. *New J. Phys.* **2010**, *12*, 063020, DOI: 10.1088/1367-2630/12/6/063020.

23.    Liu, J.; Lei, T.; Zhang, Y.; Ma, P.; Zhang, Z. First-Principle Calculations for Magnetism of Mn-Doped Graphene. *Adv. Mater. Res.* **2013**, *709*, 184, DOI: 10.4028/AMR.709.184.

24.    Thakur, J,;  Saini, H.; Kashyap, M. Electronic and Magnetic Properties of Cr Doped Graphene; Full Potential Approach. *AIP Conf. Proc.* **2015**, *1675*, 030032, DOI: 10.1063/1.4929248.

25.    Thakur, J.; Kashyap, M.; Taya, A.; Rani, P.; Saini, H. Sublattice Dependent Magnetic Response of Dual Cr Doped Graphene Monolayer: A Full Potential Approach. *Indian J. Phys.* **2017**, *91*, 43-51, DOI: 10.1007/s12648-016-0899-5.

26.    Dyck, O.; Yoon, M.; Zhang, L.; Lupini, A.; Swett, J.; Jesse, S. Doping of Cr in Graphene Using Electron Beam Manipulation for Functional Defect Engineering. *ACS Appl. Nano Mater.* **2020**, *3*, 10855-10863, DOI: 10.1021/acsanm.0c02118.

27.     Meier, F.; Zhou, L.; Wiebe, J.; Wiesendanger, R. Revealing Magnetic Interactions from Single-Atom Magnetization Curves. *Science* **2008**, *320*, 82–86, DOI: 10.1126/science.1154415.

28.     Hermenau, J.; Brinker, S.; Marciani, M.; Steinbrecher, M.; dos Santos Dias, M.; Wiesendanger, R. Lounis, S.; Wiebe, J. Stabilizing Spin Systems via Symmetrically Tailored RKKY Interactions. *Nat. Commun.* **2019**, *10*, 2565, DOI: 10.1038/s41467-019-10516-2.

29.     Lin, Y.; Teng, P.; Yeh, C.; Koshino, M.; Chiu, O.; Suenaga, K. Structural and Chemical Dynamics of Pyridinic-Nitrogen Defects in Graphene. *Nano Lett.* **2015**, *15*, 7408-7413, DOI: 10.1021/acs.nanolett.5b02831.

30.     Susi, T.; Meyer, J.; Kotakoski, J. Manipulating Low-Dimensional Materials Down to The Level of Single Atoms with Electron Irradiation. *Ultramicroscopy* **2017**, *180*, 163-172, DOI: 10.1016/j.ultramic.2017.03.005.

31.     Tripathi, M.; Mittelberger, A.; Pike, N. A.; Mangler, C.; Meyer, J. C.; Verstraete, M. J.; Kotakoski, J.; Susi, T. Electron-Beam Manipulation of Silicon Dopants in Graphene. *Nano Lett.* **2018**, *18*, 5319-5323, DOI: 10.1021/acs.nanolett.8b02406.

32.     Markevich, A. V.; Baldoni, M.; Warner, J. H.; Kirkland, A. I.; Besley, E.; Dynamic Behavior of Single Fe Atoms Embedded in Graphene. J. *Phys. Chem. C* **2016**, *120*, 21998-22003, DOI: 10.1021/acs.jpcc.6b06554.

33.     Kresse, G.; Hafner, J. Ab Initio Molecular Dynamics for Liquid Metals. *Phys. Rev. B* **1993**, *47*, 558, DOI: 10.1103/physrevb.47.558.

34.     Kresse, G.; Furthmüller, J. Efficient Iterative Schemes for Ab Initio Total-Energy Calculations Using a Plane-Wave Basis Set. *Phys. Rev. B* **1996**, *54*, 11169, DOI: 10.1103/physrevb.54.11169.

35.     Kresse, G.; Furthmüller, J. Efficiency of Ab-Initio Total Energy Calculations for Metals and Semiconductors Using a Plane-Wave Basis Set. *Comput. Mat. Sci.* **1996**, *6*, 15-50, DOI: 10.1016/0927-0256(96)00008-0.

36.     Blöchl, P. E. Projector Augmented-Wave Method. *Phys. Rev. B* **1994**, *50*, 17953, DOI: 10.1103/PhysRevB.50.17953.

37.     Kresse, G.; Joubert, D. From Ultrasoft Pseudopotentials to the Projector Augmented-Wave Method. *Phys. Rev. B* **1999**, *59*, 1758, DOI: 10.1103/PhysRevB.59.1758.

38.    Perdew, J. P.; Burke, K.; Ernzerhof, M. Generalized Gradient Approximation Made Simple. *Phys. Rev. Lett.* **1996**, *77*, 3865, DOI: 10.1103/PhysRevLett.77.3865.

39.    Robertson, A.; Montanari, B.; He, K.; Allen, C.; Wu, Y.; Harrison, N.; Kirkland, A.; Warner, J. Structural Reconstruction of the Graphene Monovacancy. *ACS Nano* **2013**, *7*, 4495-4502, DOI: 10.1021/nn401113r.

40.    Babar, R.; Kabir, M. Ferromagnetism in Nitrogen-Doped Graphene. *Phys. Rev. B* **2018**, *99*, 115422, DOI: 10.1103/PhysRevB.99.115442.

41.    Blöchl, P. E.; Jepsen, O.; Andersen, O. Improved Tetrahedron Method for Brillouin-Zone Integrations. *Phys. Rev. B* **1994**, *49*, 16223, DOI: 10.1103/PhysRevB.49.16223.

42.    Bader, R. F. W. Atoms in Molecules. *Acc. Chem. Res.* **1985**, *18*, 9-15, DOI: 10.1021/ar00109a003.

43.    Henkelman, G.; Arnaldsson, A.; Jónsson, H. A Fast and Robust Algorithm for Bader Decomposition of Charge Density. *Comput. Mater. Sci.* **2006**, *36*, 354-360, DOI: 10.1016/j.commatsci.2005.04.010.

44.    Kleinman L., Relativistic Norm-Conserving Pseudopotential. *Phys. Rev. B* **1980**, *21*, 2630, DOI: 10.1103/PhysRevB.25.2103.

45.    MacDonald, A.; Picket, W.; Koelling, D. A Linearised Relativistic Augmented-Plane-Wave Method Utilising Approximate Pure Spin Basis Functions. *J. Phys. C Solid State Phys.* **1980**, *13*, 2675, DOI: 10.1088/0022-3719/13/14/009.

46.    Hobbs, D.; Kresse, G.; Hafner, J. Fully Unconstrained Noncollinear Magnetism within the Projector Augmented-Wave Method. *Phys. Rev. B* **2000**, *62*, 11556, DOI: 10.1103/PhysRevB.62.11556.

47.    Dudarev, S.; Botton, G.; Savrasov, S.; Humphreys, C.; Sutton, A. Electron-Energy-Loss Spectra and the Structural Stability of Nickel Oxide: An LSDA+U Study. *Phys. Rev. B* **1998**, *57*, 1505, DOI: 10.1103/PhysRevB.57.1505.

48.    Błoński, P.; Hafner, J. Magnetic Anisotropy of Transition-Metal Dimers: Density Functional Calculations. *Phys. Rev. B* **2009**, *79*, 224418, DOI: 10.1103/PhysRevB.79.224418.

49.    Heine, V. Electronic Structure from the Point of View of the Local Atomic Environment In Solid States Phys. - *Adv. Res. Appl. C* **1980**, *35*, 1-127, DOI: 10.1016/S0081-1947(08)60503-2.

50. Néel, L. Théorie du Traînage Magnétique des Ferromagnétiques en Grains Fins avec Application aux Terres Cuites. *Annales de Géophysique* **1949**, *5*, 99–136, DOI: 10.1007/978-1-4020-4423-6_220.

51. Becke, A. Density-Functional Thermochemistry. III. The Role of Exact Exchange. *J. Chem. Phys.* **1993**, *98*, 5648–5652, DOI: 10.1063/1.464913.

52. Lee, C.; Hill, C.; Carolina, N. Development of the Colle-Salvetti Correlation-Energy Formula into a Functional of the Electron Density. *Phys. Rev. B* **1998**, *37*, 785, DOI: 10.1103/PhysRevB.37.785.

53. Frisch, M. J.; Trucks, G. W.; Schlegel, H. B.; Scuseria, G. E.; Robb, M. A.; Cheeseman, J. R.; Scalmani, G.; Barone, V.; Petersson, G. A.; Nakatsuji, H.; Li, X.; Caricato, M.; Marenich, A. V.; Bloino, J.; Janesko, B. G.; Gomperts, R.; Mennucci, B.; Hratchian, H. P.; Ortiz, J. V.; Izmaylov, A. F.; Sonnenberg, J. L.; Williams-Young, D.; Ding, F.; Lipparini, F.; Egidi, F.; Goings, J.; Peng, B.; Petrone, A.; Henderson, T.; Ranasinghe, D.; Zakrzewski, V. G.; Gao, J.; Rega, N.; Zheng, G.; Liang, W.; Hada, M.; Ehara, M.; Toyota, K.; Fukuda, R.; Hasegawa, J.; Ishida, M.; Nakajima, T.; Honda, Y.; Kitao, O.; Nakai, H.; Vreven, T.; Throssell, K.; Montgomery, J. A., Jr.; Peralta, J. E.; Ogliaro, F.; Bearpark, M. J.; Heyd, J. J.; Brothers, E. N.; Kudin, K. N.; Staroverov, V. N.; Keith, T. A.; Kobayashi, R.; Normand, J.; Raghavachari, K.; Rendell, A. P.; Burant, J. C.; Iyengar, S. S.; Tomasi, J.; Cossi, M.; Millam, J. M.; Klene, M.; Adamo, C.; Cammi, R.; Ochterski, J. W.; Martin, R. L.; Morokuma, K.; Farkas, O.; Foresman, J. B.; Fox, D. J. Gaussian, Inc., Wallingford CT, **2016**.

54. Glendening, E. D.; Badenhoop, J. K.; Reed, A. E.; Carpenter, J. E.; Bohmann, J. A.; Morales, C. M.; Weinhol, F. NBO, Theoretical Chemistry Institute, University Of Wisconsin, Madison, WI, **2001**.

55. Huan, M.; Liang, M.; Liang-Cai, M. Tuning the Electronic and Magnetic Properties of Mn-Doped Graphene by Gas Adsorption and Effect of External Electric Field: First-Principles Study. *Int. J. Mod. Phys. B* **2019**, *33*, 1950166, DOI: 10.1142/S0217979219501662.

56. Daalderop, G. H. O.; Kelly, P. J.; Schuurmans, M. F. H. Magnetic Anisotropy of a Free-Standing Co Monolayer and of Multilayers which Contain Co Monolayers. *Phys. Rev. B* **1994**, *50*, 9989, DOI: 10.1103/PhysRevB.50.9989.

57. Whangbo, M. H.; Gordon, E. E.; Xiang, H.; Koo, H. J.; Lee, C. Prediction of Spin

Orientations in Terms of HOMO–LUMO Interactions Using Spin–Orbit Coupling as Perturbation. *Acc. Chem. Res.* **2015**, *48*, 12, 3080–3087, DOI: 10.1021/acs.accounts.5b00408.

58. Pease, D.; Fasihuddin, A.; Daniel, M.; Budnick, J. Method of Linearizing the 3d $L_3/L_2$ White Line Ratio as a Function of Magnetic Moment. *Ultramicroscopy* **2001**, *88*, 1-16, DOI: 10.1016/s0304-3991(00)00116-9.

59. Lin, P. C.; Villarreal, R.; Achilli, S.; Bana, H.; Nair, M. N.; Tejeda, A.; Verguts, K.; De Gendt, S.; Auge, M.; Hofsäss, H.; De Feyter, S.; Di Santo, G.; Petaccia, L.; Brems, S.; Fratesi, G.; Pereira, L. M. C. Doping Graphene with Substitutional Mn. *ACS Nano* **2021**, *15*, 5449-5458, DOI: 10.1021/acsnano.1c00139.

60. Tripathi, M.; Markevich, A.; Böttger, R.; Facsko, S.; Besley, E.; Kotakovski, J.; Susi, T. Implanting Germanium into Graphene. *ACS Nano* **2018**, *12*, 4641-4647, DOI: 10.1021/acsnano.8b01191.

61. Anagaw, M.; Mekonnen, S. The Effect of Manganese(Mn) Substitutional Doping on Structural, Electronic and Magnetic Properties of Pristine Hexagonal Graphene: Using Spin Polarized Density Functional Theory. *J. At. Mol. Condens. Nano Phys.* **2020**, *7*, 95, DOI: 10.26713/jamcnp.v7i2.1422.

62. Ruderman, M. A.; Kittel, C. Indirect Exchange Coupling of Nuclear Magnetic Moments by Conduction Electrons. *Phys. Rev.* **1954**, *96*, 99–102, DOI: 10.1103/PhysRev.96.99.

63. Kasuya, T. A Theory of Metallic Ferro- and Antiferromagnetism on Zener's Model. *Prog. Theor. Phys.* **1956**, *16*, 45–57, DOI: 10.1143/PTP.16.45.

64. Yosida, K. Magnetic Properties of Cu-Mn Alloys. *Phys. Rev.* **1957**, *106*, 893–898, DOI: 10.1103/PhysRev.106.893

65. Crook, C. B.; Constatntin, C.; Ahmed, T.; Zhu, J. X.; Balatsky, A. V.; Haraldsen, J. T. Proximity-Induced Magnetism in Transition-Metal Substituted Graphene. *Sci. Rep.* **2015**, *5*, 1–11, DOI: 10.1038/srep12322.

66. Crook, C. B.; Houchins, G.; Zhu, J. X., Balatsky, A. V.; Constantin, C.; Haraldsen, J. T. Spatial Dependence of the Super-Exchange Interactions for Transition-Metal Trimers in Graphene. *J. Appl. Phys.* **2018**, *123*, 0–6, DOI: 10.1063/1.5007274.

67. Denis, P. A.; Pereyra Huelmo, C. Structural characterization and chemical reactivity of dual doped graphene. *Carbon* **2015,** *87*, 106–115, DOI: 10.1016/j.carbon.2015.01.049.

68.  Denis, P. A.; Iribarne, F. Dual doped monolayer and bilayer graphene: The case of 4p and 2p elements. *Chem. Phys. Lett.* **2016**, *658*, 152–157, DOI: 10.1016/j.cplett.2016.06.032.

69.  Tejedor, M.; García, J. A.; Carrizo, J.; Elbaile, L.; Santos, J. D.; Mira, J.; Rivas, J. In-Plane Magnetic Anisotropy along the Width in Amorphous Magnetic Ribbons. *J. Magn.* **2004**, *227-276*, 1362-1364, DOI: 10.1016/j.jmmm.2003.12.089.

70.  Cramer, C. J. Essentials of Computational Chemistry: Theories and Models, 2nd Edition. *Wiley* **2004**, ISBN: 978-0-470-09182-1.

**Supporting information**

# Graphene Lattices with Embedded Transition-Metal Atoms and Tunable Magnetic Anisotropy Energy: Implications for Spintronic Devices

Rostislav Langer[1,2], Kimmo Mustonen[3], Alexander Markevich[3], Michal Otyepka[1,4], Toma Susi*[3], Piotr Błoński*[1]

[1] Regional Centre of Advanced Technologies and Materials, Czech Advanced Technology and Research Institute (CATRIN), Palacký University Olomouc, 779 00 Olomouc, Czech Republic

[2] Department of Physical Chemistry, Faculty of Science, Palacký University Olomouc, 779 00 Olomouc, Czech Republic

[3] Faculty of Physics, University of Vienna, Boltzmanngasse 5, A-1090 Vienna, Austria

[4] IT4Innovations, Technical University of Ostrava, 708 00 Ostrava-Poruba, Czech Republic

*e-mail: piotr.blonski@upol.cz; toma.susi@univie.ac.at

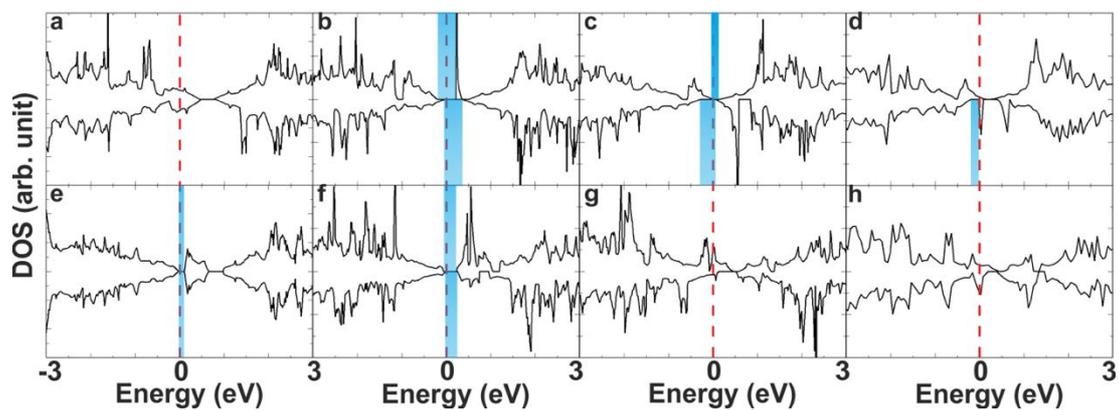

**Figure S1** Density of state plots of **a)** SV, **b)** Cr@SV, **c)** Mn@SV, **d)** Fe@SV, **e)** DV **f)** Cr@DV, **g)** Mn@DV, **h)** Fe@DV. The blue bars depict the bandgaps.

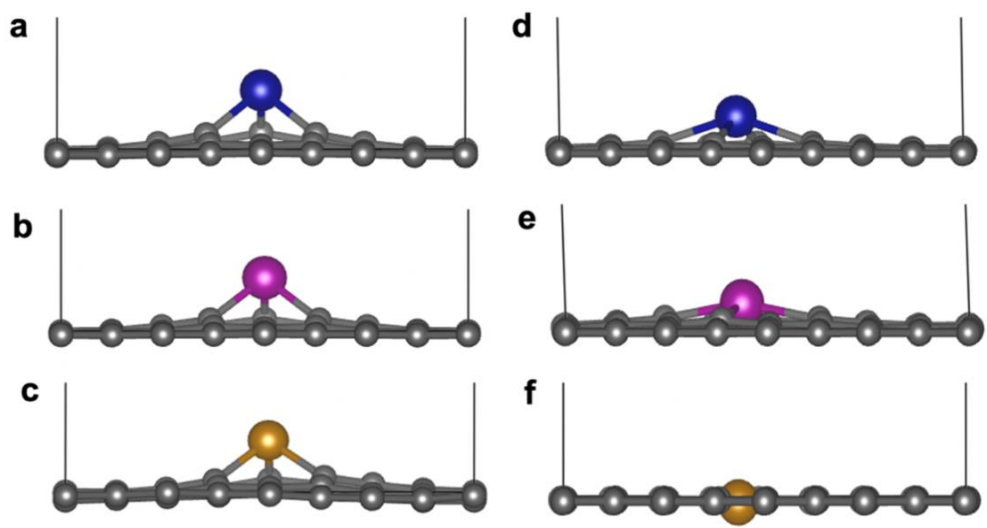

**Figure S2** Side-view of **a)** Cr@SV, **b)** Mn@SV, **c)** Fe@SV, **d)** Cr@DV, **e)** Mn@DV, **f)** Fe@DV.

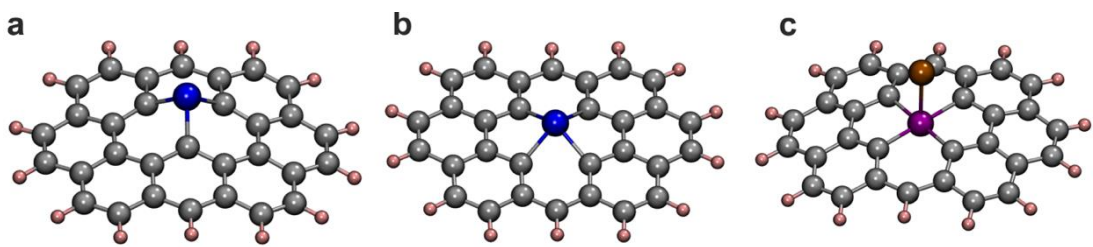

**Figure S3** The finite model of defective graphene, **a)** TM@SV, **b)** TM@DV and **c)** TM$_2$@TM$_1$@DV. Carbon atoms are in gray, hydrogen atoms are in pink and TM atoms are in blue/purple/ochre.

**Table S1** The binding energy $E_{bind}$ (eV), TM–C distance (Å), partial charge (e) and electronic configuration of TM@SV, TM@DV and TM$_2$@TM$_1$@DV in the oxidation state II.

| System (II) | TM–C | Charge | Electronic configuration of TM@S(D)V | Electronic configuration of TM |
|---|---|---|---|---|
| **Cr@SV** | 1.90 | 0.59 | [core] 4s(0.19) 3d(5.10) 4p(0.18) 4d(0.01) | [core] 4s(1.00) 3d(5.00) |
| **Mn@SV** | 1.84 | 0.99 | [core] 4s(0.23) 3d(5.69) 4p(0.12) 4d(0.01) | [core] 4s(1.00) 3d(5.00) |
| **Fe@SV** | 1.79 | 0.63 | [core] 4s(0.21) 3d(7.06) 4p(0.13) | [core] 4s(2.00) 3d(6.00) |
| **Cr@DV** | 1.99 | 0.58 | [core] 4s(0.24) 3d(4.94) 4p(0.27) 4d(0.01) | [core] 4s(1.00) 3d(5.00) |
| **Mn@DV** | 1.99 | 0.83 | [core] 4s(0.26) 3d(5.71) 4p(0.23) | [core] 4s(1.00) 3d(5.00) |
| **Fe@DV** | 1.94 | 0.71 | [core] 4s(0.27) 3d(6.82) 4p(0.22) | [core] 4s(2.00) 3d(6.00) |
| **Fe@Mn@DV** | 1.98 | 0.38 (Mn) | [core] 4s(0.38) 3d(5.55) 4p(0.79) 4d(0.03) 4f(0.02) (Mn) | [core] 4s(1.00) 3d(5.00) (Mn) |
| | | 1.00 (Fe) | [core] 4s(0.06) 3d(7.07) 4p(0.10) 4d(0.01) 4f(0.01) (Fe) | [core] 4s(2.00) 3d(6.00) (Fe) |

**Table S2** The binding energy $E_{bind}$ (eV), TM–C distance (Å), partial charge (e) and electronic configuration of TM@SV, TM@DV and $TM_2$@$TM_1$@DV in the oxidation state III.

| System (III) | TM–C | Charge | Electronic configuration of TM@S(D)V | Electronic configuration of TM |
|---|---|---|---|---|
| **Cr@SV** | 1.90 | 0.71 | [core] 4s(0.18) 3d(4.99) 4p(0.17) 4d(0.01) | [core] 4s(1.00) 3d(5.00) |
| **Mn@SV** | 1.83 | 0.71 | [core] 4s(0.20) 3d(6.01) 4p(0.13) | [core] 4s(1.00) 3d(5.00) |
| **Fe@SV** | 1.82 | 0.84 | [core] 4s(0.21) 3d(6.84) 4p(0.14) | [core] 4s(2.00) 3d(6.00) |
| **Cr@DV** | 2.01 | 0.69 | [core] 4s(0.22) 3d(4.85) 4p(0.27) 4d(0.01) | [core] 4s(1.00) 3d(5.00) |
| **Mn@DV** | 1.98 | 0.94 | [core] 4s(0.26) 3d(5.60) 4p(0.23) | [core] 4s(1.00) 3d(5.00) |
| **Fe@DV** | 1.93 | 0.94 | [core] 4s(0.27) 3d(6.55) 4p(0.27) | [core] 4s(2.00) 3d(6.00) |
| **Fe@Mn@DV** | 1.98 | 0.15 (Mn) | [core] 4s(0.41) 3d(5.74) 4p(0.72) 4d(0.01) (Mn) | [core] 4s(1.00) 3d(5.00) (Mn) |
| | | 0.78 (Fe) | [core]4s(0.65) 3d(6.57) 4p(0.03) | [core] 4s(2.00) 3d(6.00) (Fe) |

**Table S3** The bandgaps (eV) of spin-up (spin-down) channel of TM@SV and TM@DV.

| | Cr@SV | Mn@SV | Fe@SV | Cr@DV | Mn@DV | Fe@DV |
|---|---|---|---|---|---|---|
| **Present** | 0.2 (0.3) | 0.2 (0.3) | 0.0 (0.2) | 0.3 (0.3) | 0.0 (0.0) | 0.0 (0.0) |
| **Literature** [17, 19, 22, 23, 26] | 0.2 (0.2), 0.0 (0.0) | 0.0 (0.0) | 0.0 (0.2) | 0.5 (0.7) | - | - |

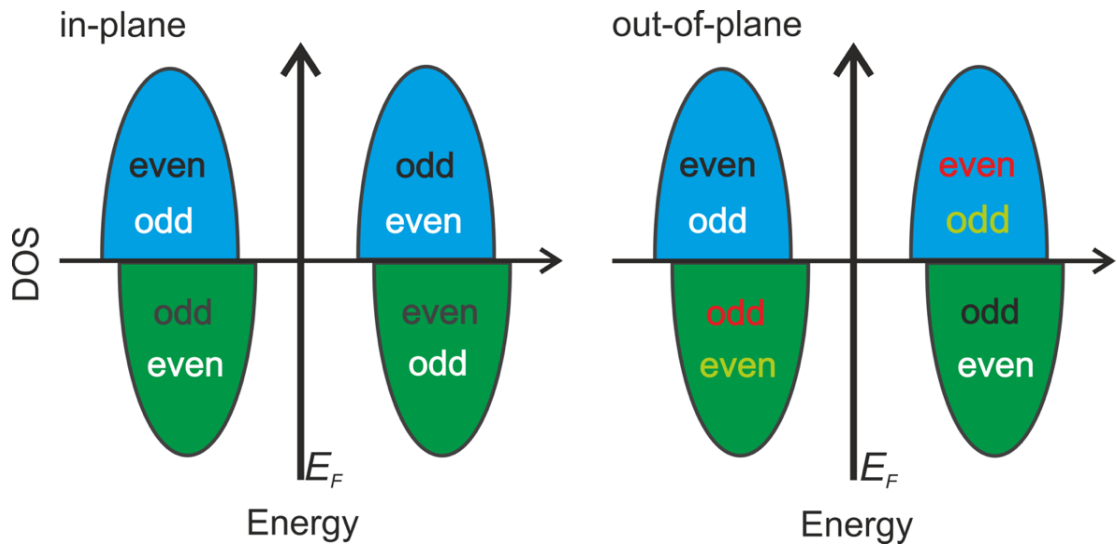

**Figure S4** Scheme showing the coupling of eigenstates with energies above and below Fermi energy through spin-orbit interaction, which favors either perpendicular or in-plane magnetization, depending on the spins and symmetries of the interacting states. For the five $d$ states (per spin), in-plane magnetization is favored if the Fermi level ($E_F$) is located between even $m_l = 0, \pm2$ and odd $m_l = \pm1$ states of the same spin, or between even or odd states of opposite spin. A perpendicular magnetization is favored if $E_F$ located between even (or odd) and odd (or even) states with opposite spins.

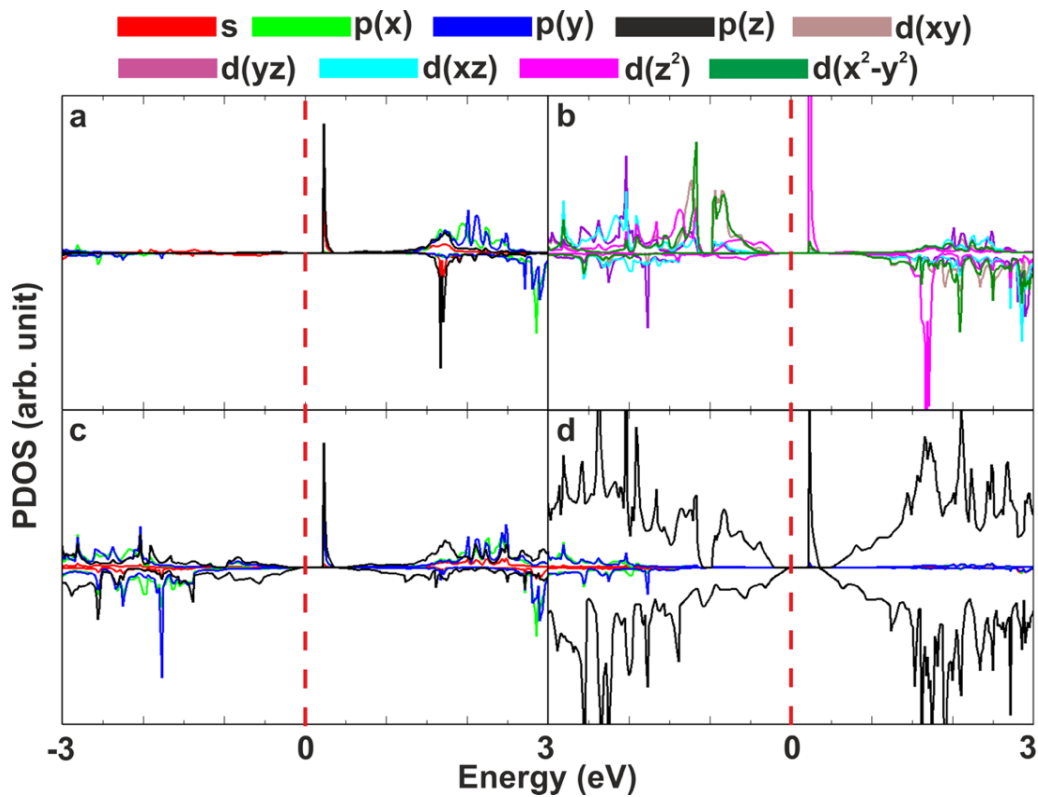

**Figure S5** Atom/orbital decomposed DOS of Cr@SV. **a, b)** transition metal, **c)** nearest carbon atoms, **d)** other carbon atoms.

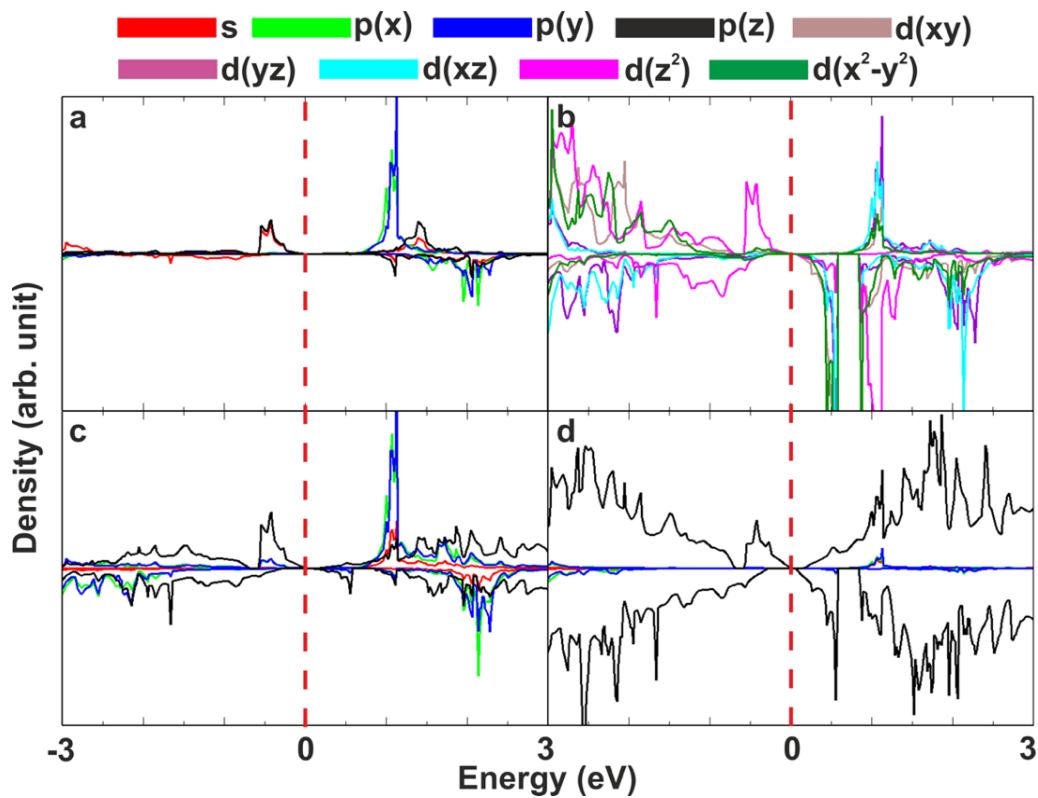

**Figure S6** Atom/orbital decomposed DOS of Mn@SV. Cf. **Figure S5**.

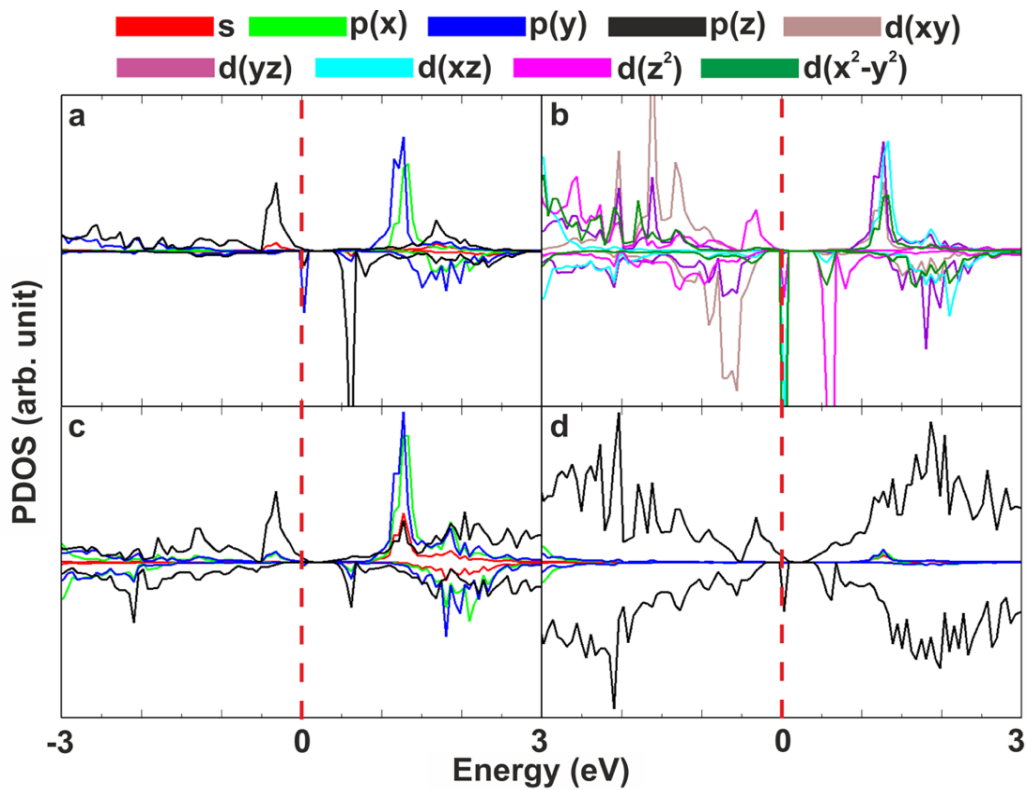

**Figure S7** Atom/orbital decomposed DOS of Fe@SV. Cf. **Figure S5**.

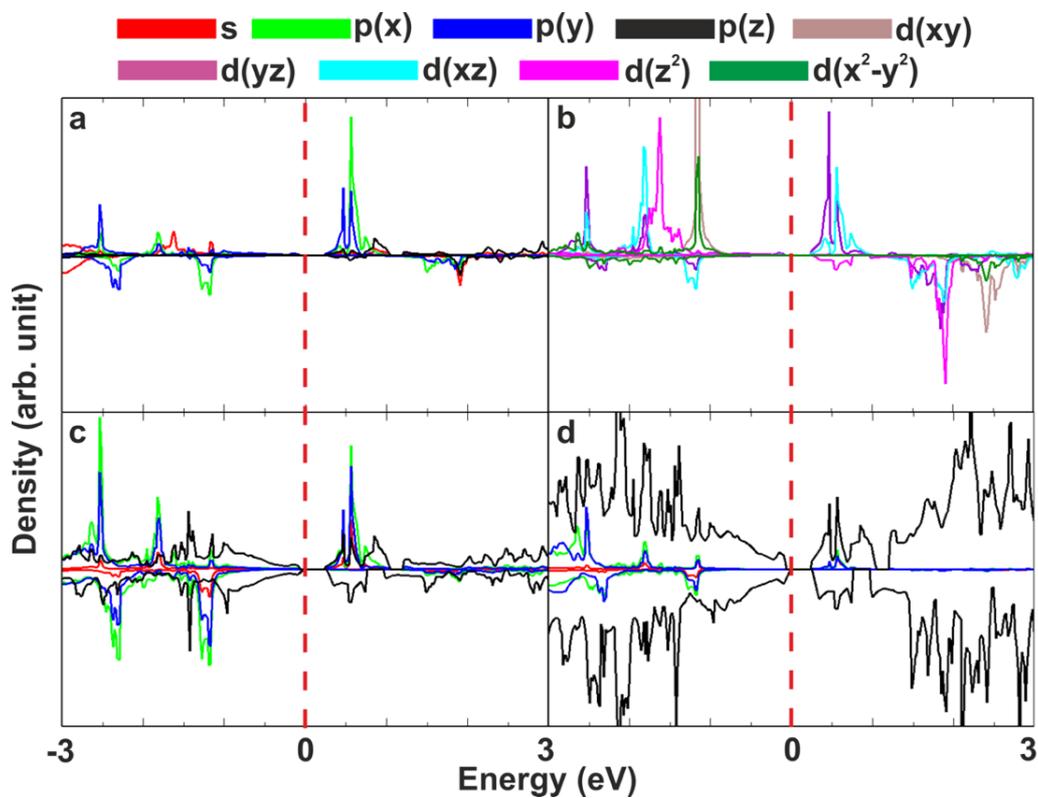

**Figure S8** Atom/orbital decomposed of DOS Cr@DV. Cf. **Figure S5**.

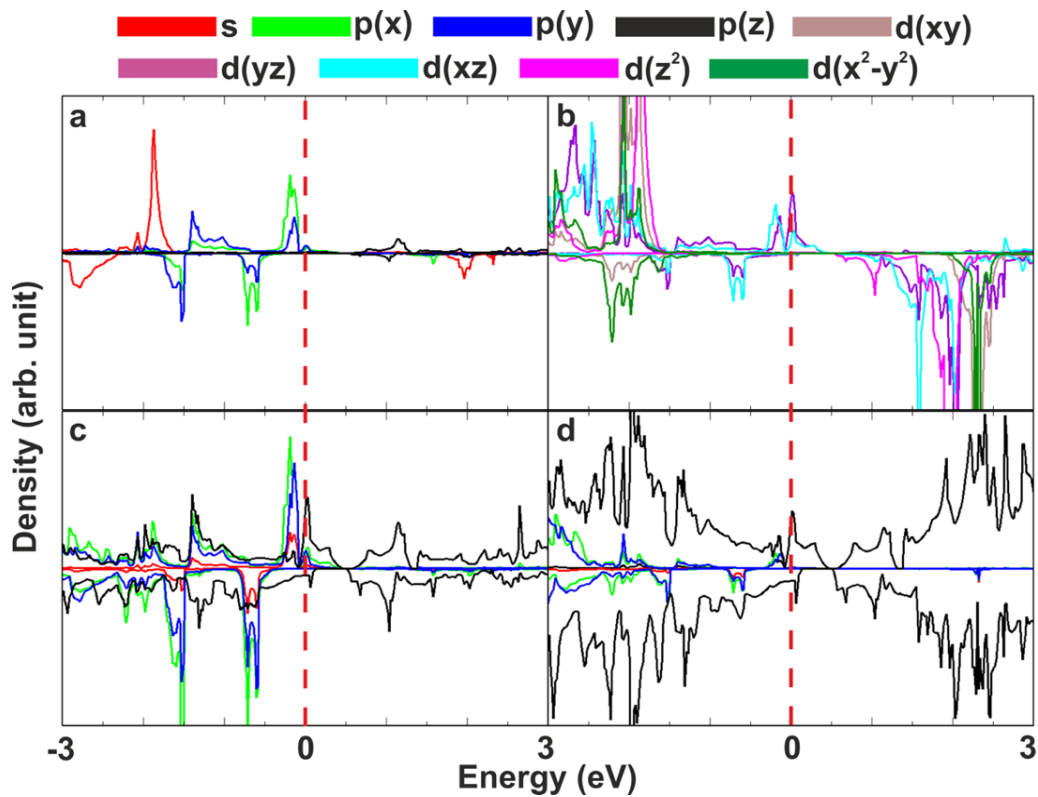

**Figure S9** Atom/orbital decomposed DOS of Mn@DV. Cf. **Figure S5**.

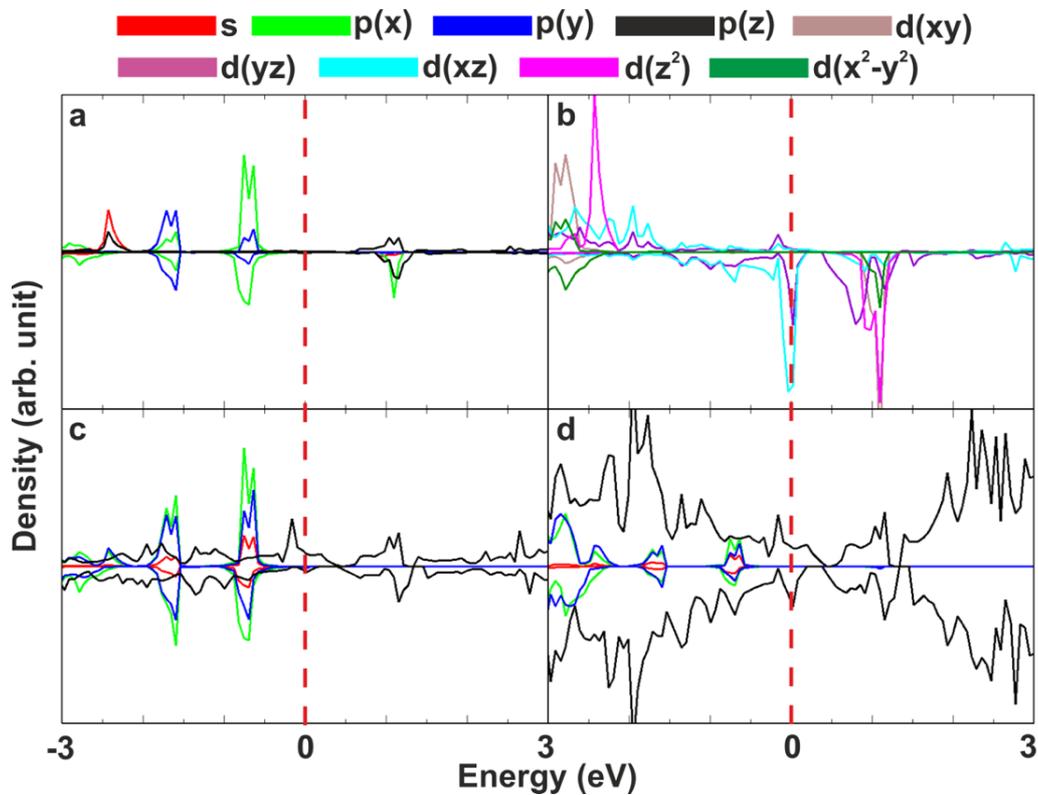

**Figure S10** Atom/orbital decomposed DOS of Fe@DV. Cf. **Figure S5**.

**Table S4** Total spin moments $\mu_{s\_tot}$, local spin moments $\mu_{s\_loc}$ and orbital moments $\mu_l$ (all in $\mu_B$) oriented along the "$x$", and "$z$" magnetization axes and their anisotropies ($\Delta$) for TM@SV and TM@DV shown in **Figure 1** and **Figure S2**.

| system | $\mu_{s\_tot\,(x)}$ | $\mu_{s\_loc\,(x)}$ | $\mu_{l\,(x)}$ | $\mu_{s\_tot\,(z)}$ | $\mu_{s\_loc\,(z)}$ | $\mu_{l\,(z)}$ | $\mu_{s\_tot\,(\Delta)}$ | $\mu_{s\_loc(\Delta)}$ | $\mu_{l\,(\Delta)}$ |
|---|---|---|---|---|---|---|---|---|---|
| **Cr@SV** | 1.00 | 1.04 | 0.01 | 1.00 | 1.04 | 0.01 | 0.00 | 0.00 | 0.00 |
| **Mn@SV** | 1.50 | 1.33 | 0.00 | 1.50 | 1.33 | 0.01 | 0.00 | 0.00 | -0.01 |
| **Fe@SV** | 0.94 | 0.93 | 0.02 | 0.94 | 0.93 | 0.12 | 0.00 | 0.00 | -0.10 |
| **Cr@DV** | 1.00 | 1.00 | 0.03 | 1.00 | 1.00 | 0.01 | 0.00 | 0.00 | 0.02 |
| **Mn@DV** | 1.65 | 1.57 | 0.00 | 1.65 | 1.57 | 0.00 | 0.00 | 0.00 | 0.00 |
| **Fe@DV** | 1.61 | 1.52 | 0.04 | 1.61 | 1.52 | 0.05 | 0.00 | 0.00 | -0.01 |

**Table S5** Distance between Mn atoms and Mn and Si atoms (Å), total magnetic moment $\mu_{tot}$ ($\mu_B$), magnetic moment on Mn $\mu_{Mn1}$ ($\mu_B$), and magnetic moment on the second adatom (Mn or Si) $\mu_{Mn2}/\mu_{Si}$ ($\mu_B$) of $Mn_1Mn_2$@SV and/or $Mn_1Si$@SV.

| system | distance | $\mu_{tot}$ | $\mu_{Mn1}$ | $\mu_{Mn2}/\mu_{Si}$ |
|---|---|---|---|---|
| **$Mn_1Si$@SV** (Figure S11-a) | 4.00 | 3.00 | 2.50 | 0.00 |
| **$Mn_1Mn_2$@SV** (Figure S12-20) | 3.87 | 6.00 | 2.92 | 2.92 |
| **$Mn_1Si$@SV** (Figure S11-b) | 6.64 | 3.00 | 2.49 | 0.01 |
| **$Mn_1Mn_2$@SV** (Figure S12-3) | 6.54 | 6.00 | 2.90 | 2.90 |

**Table S6** Distance between Mn atoms and Mn and Si atoms (Å), total magnetic moment $\mu_{tot}$ ($\mu_B$), magnetic moment on Mn $\mu_{Mn1}$ ($\mu_B$), and magnetic moment on the second adatom (Mn or Si) $\mu_{Mn2}/\mu_{Si}$ ($\mu_B$) for Mn$_1$Mn$_2$@DV and/or Mn$_1$Si@DV.

| system | distance | $\mu_{tot}$ | $\mu_{Mn1}$ | $\mu_{Mn2}/\mu_{Si}$ |
|---|---|---|---|---|
| **Mn$_1$Si@DV** (Figure S11-c) | 5.77 | 3.00 | 3.05 | 0.01 |
| **Mn$_1$Mn$_2$@DV** (Figure S13-2) | 5.77 | 8.00 | 3.68 | 3.70 |
| **Mn$_2$Si@SV** (Figure S11-d) | 9.68 | 3.45 | 3.42 | -0.02 |
| **Mn$_1$Mn$_2$@DV** (Figure S13-6) | 9.70 | 7.40 | 3.75 | 3.63 |

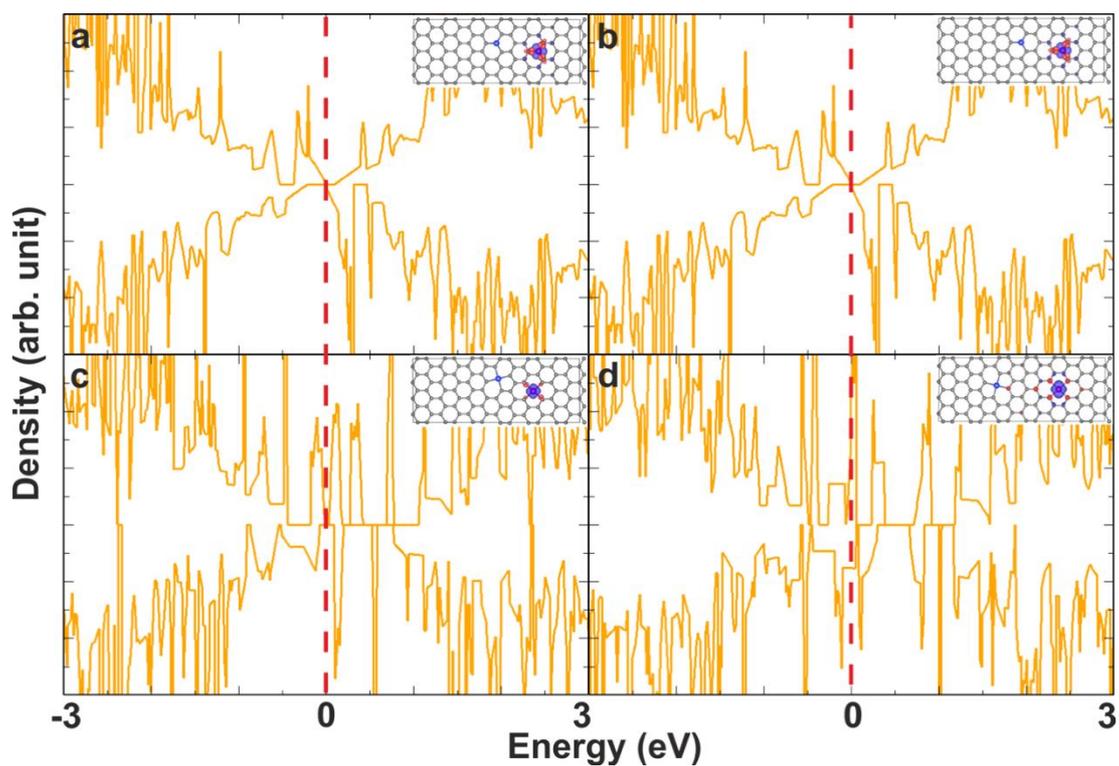

**Figure S11** Density of state plots of **a,b)** Mn$_1$Si@SV and **c,d)** Mn$_1$Si@DV.

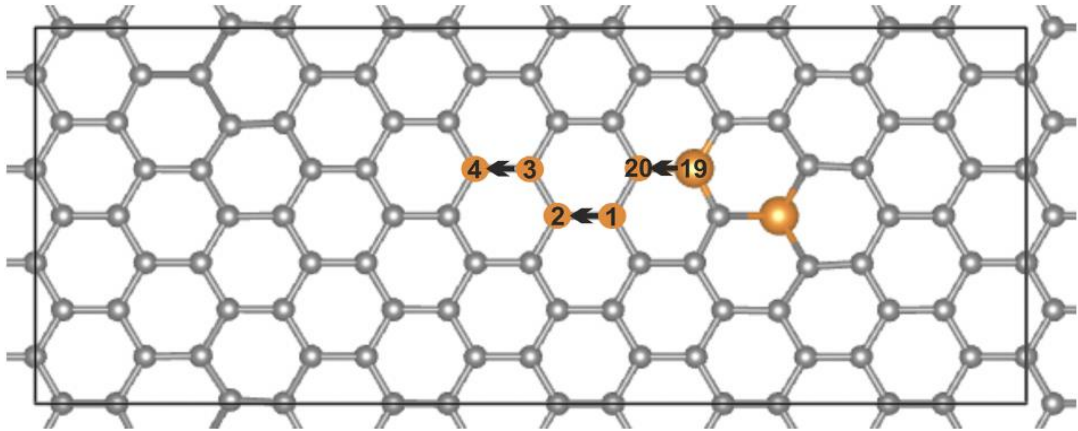

**Figure S12** Scheme depicting the manipulation of a TM-atom in SV-graphene (TM$_1$TM$_2$@SV) to monitor changes of properties depending on the distance between TM-atoms.

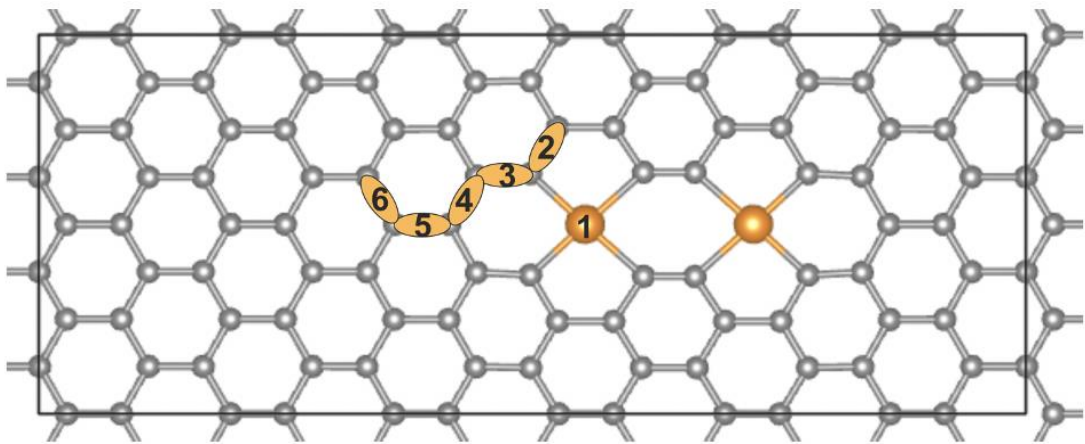

**Figure S13** Scheme depicting the manipulation of a TM-atom in DV-graphene (TM$_1$TM$_2$@DV) to monitor changes of properties depending on the distance between the TM-atoms.

**Table S7** The binding energy $E_{bind}$ (eV), TM–TM distance (Å), total magnetic moment $\mu_{tot}$ ($\mu_B$), magnetic moment of TM $\mu_{TM}$ ($\mu_B$), and MAE (meV) of $TM_1TM_2$@SV. Structures correspond to **Figure S12**.

| structure | 19 | 20 | 1 | 2 | 3 | 4 |
|---|---|---|---|---|---|---|
| | | | CrCr@SV | | | |
| $E_{bind}$ | -11.11 | -10.03 | -11.60 | -10.82 | -11.21 | -11.11 (AFM), -11.01 |
| **Cr–Cr** | 2.73 | 4.05 | 5.05 | 5.90 | 6.59 | 7.96 |
| $\mu_{tot}$ | 4.02 | 4.37 | 4.00 | 4.72 | 4.00 | 0.00 (AFM), 4.34 |
| $\mu_{TM}$ | 2.63 | 2.64 | 2.60 | 2.69 | 2.60 | ±2.64 (AFM), 2.62 |
| **MAE** | 0.10 | -0.18 | -0.14 | -0.54 | -0.20 | 0.36 (AFM), -0.24 |
| | | | MnMn@SV | | | |
| $E_{bind}$ | -8.50 | -11.83, (AFM) -11.76 | -10.56, (AFM) -10.49 | -8.47 | -9.81 | -10.06 |
| **Mn–Mn** | 2.45 | 3.87 | 4.44 | 5.96 | 6.54 | 7.96 |
| $\mu_{tot}$ | 4.00 | 0.00 (AFM), 6.00 | 0.00 (AFM), 6.00 | 5.76 | 6.00 | 5.88 |
| $\mu_{TM}$ | 2.57 | ±2.92 (AFM), 2.92 | ±2.85 (AFM), 2.90 | 2.56 | 2.90 | 2.89 |
| **MAE** | -0.02 | -0.9 (AFM), -0.35 | -0.50 (AFM), -0.48 | -0.40 | -0.48 | -0.47 |
| | | | FeFe@SV | | | |
| $E_{bind}$ | -12.49 | -11.36 | -10.49 | -7.47 | -5.29 | -9.69 |
| | | (AFM), -11.26 | | | | |
| **Fe–Fe** | 2.73 | 3.33 | 4.34 | 5.96 | 6.65 | 7.99 |
| $\boldsymbol{\mu_{tot}}$ | 4.00 | 0.00 (AFM), 3.92 | 6.00 | 2.52 | 2.61 | 3.64 |
| $\boldsymbol{\mu_{TM}}$ | 2.32 | ±1.75 (AFM), 1.84 | 2.90 | 1.36 | 1.35 | 1.77 |
| **MAE** | -0.59 | 0.53 (AFM), -0.14 | -0.25 | 0.01 | -0.21 | 0.20 |

**Table S8** The binding energy $E_{bind}$ (eV), TM–TM distance (Å), total magnetic moment $\mu_{tot}$ ($\mu_B$), magnetic moment of TM $\mu_{TM}$ ($\mu_B$), and MAE (meV) of $TM_1TM_2@DV$. Structures correspond to **Figure S13**.

| structure | 1 | 2 | 3 | 4 | 5 | 6 |
|---|---|---|---|---|---|---|
| | | | **CrCr@DV** | | | |
| $E_{bind}$ | -11.29 | -11.53 | -11.00 | -11.57 (AFM), -11.54 | -10.96 | -11.26 |
| **Cr–Cr** | 4.38 | 5.47 | 6.57 | 7.59 | 8.58 | 9.71 |
| $\mu_{tot}$ | 6.06 | 6.17 | 8.00 | 0.00 (AFM), 8.00 | 8.00 | 8.00 |
| $\mu_{TM}$ | 3.10 | 2.88 | 3.19 | ±3.22 (AFM), 3.18 | 3.17 | 3.21 |
| **MAE** | 0.95 | 0.60 | 0.94 | 1.6 (AFM), 0.58 | 0.96 | 0.57 |
| | | | **MnMn@DV** | | | |
| $E_{bind}$ | -13.67 | -14.46 | -13.72 | -14.79 | -13.83 | -13.54 |
| **Mn–Mn** | 4.32 | 5.77 | 6.53 | 7.59 | 8.54 | 9.70 |
| $\mu_{tot}$ | 8.00 | 8.00 | 7.28 | 8.00 | 8.06 | 7.40 |
| $\mu_{TM}$ | 3.74 | 3.70 | 3.68 | 3.68 | 3.70 | 3.63 |
| **MAE** | 1.26 | 1.29 | 1.34 | 1.37 | 1.10 | 1.35 |
| | | | **FeFe@DV** | | | |
| $E_{bind}$ | -13.18 | -12.98 | -14.49 | -16.02 (AFM), -15.99 | -15.63 | -16.08 |
| **Fe–Fe** | 4.34 | 5.78 | 6.50 | 7.54 | 8.52 | 9.67 |
| $\mu_{tot}$ | 5.81 | 6.21 | 5.55 | 0.00 (AFM), 6.00 | 6.87 | 6.00 |
| $\mu_{TM}$ | 2.83 | 2.91 | 3.14 | ±3.11 (AFM), 3.00 | 3.22 | 3.10 |
| **MAE** | -0.46 | 0.42 | -0.22 | 1.48 (AFM), 0.84 | 0.15 | 0.88 |

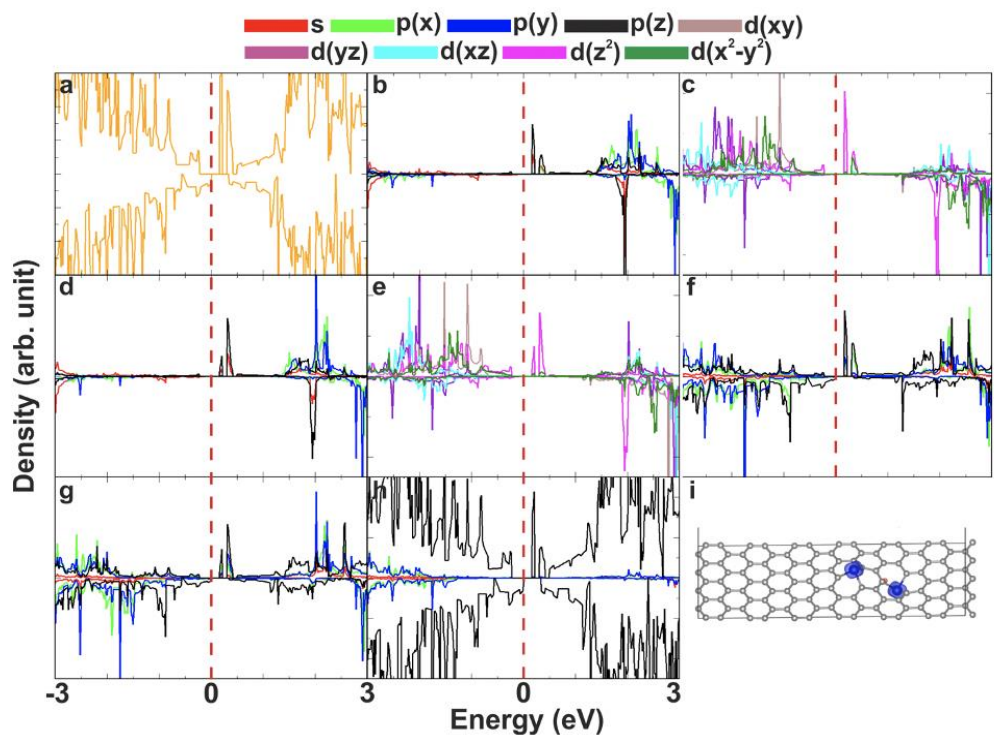

**Figure S14** Atom/orbital decomposed DOS of CrCr@SV (Cf. **Figure S12-1**). **a)** total DOS, **b,c)** $Cr_1$ atom, **d,e)** $Cr_2$ atom, **f)** nearest carbon atoms to $Cr_1$, **g)** nearest carbon atoms to $Cr_2$, **h)** other carbon atoms, **i)** spin-polarized structure.

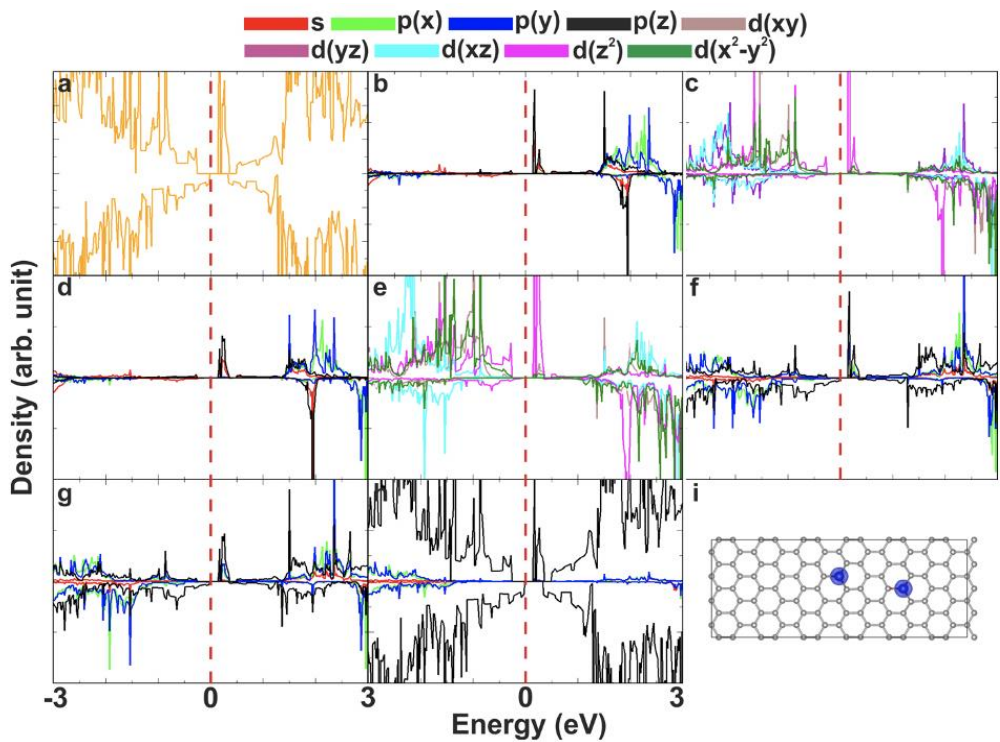

**Figure S15** Atom/orbital decomposed DOS of CrCr@SV (Cf. **Figure S12-3**). **a)** total DOS, **b,c)** $Cr_1$ atom, **d,e)** $Cr_2$ atom, **f)** nearest carbon atoms to $Cr_1$, **g)** nearest carbon atoms to $Cr_2$, **h)** other carbon atoms, **i)** spin-polarized structure.

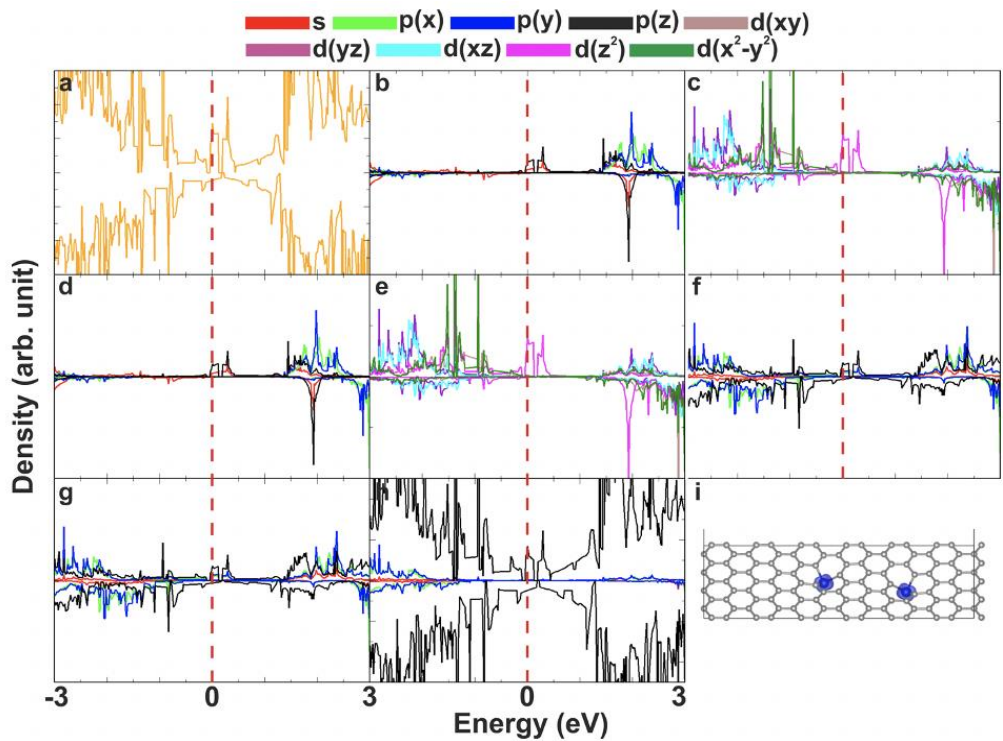

**Figure S16** Atom/orbital decomposed DOS of CrCr@SV (Cf. **Figure S12-4**). **a)** total DOS, **b,c)** $Cr_1$ atom, **d,e)** $Cr_2$ atom, **f)** nearest carbon atoms to $Cr_1$, **g)** nearest carbon atoms to $Cr_2$, **h)** other carbon atoms, **i)** spin-polarized structure.

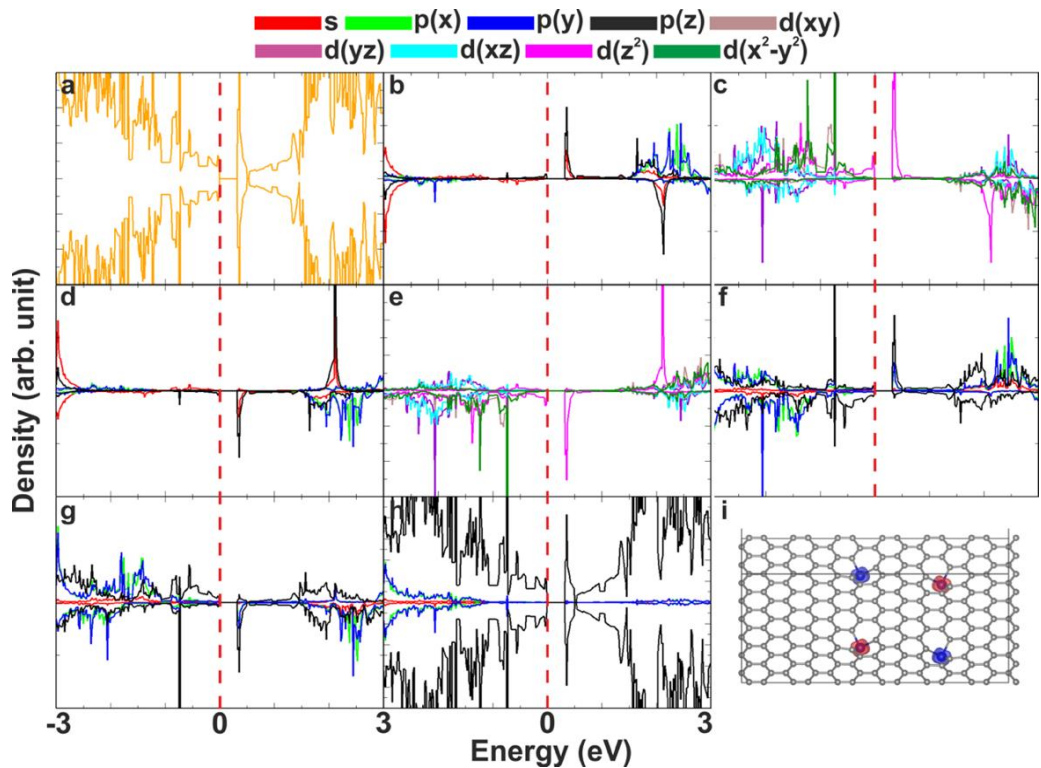

**Figure S17** Atom/orbital decomposed DOS of CrCr@SV, AFM alignment (Cf. **Figure S12-4**). **a)** total DOS, **b,c)** $Cr_1$ atom, **d,e)** $Cr_2$ atom, **f)** nearest carbon atoms to $Cr_1$, **g)** nearest carbon atoms to $Cr_2$, **h)** other carbon atoms, **i)** spin-polarized structure.

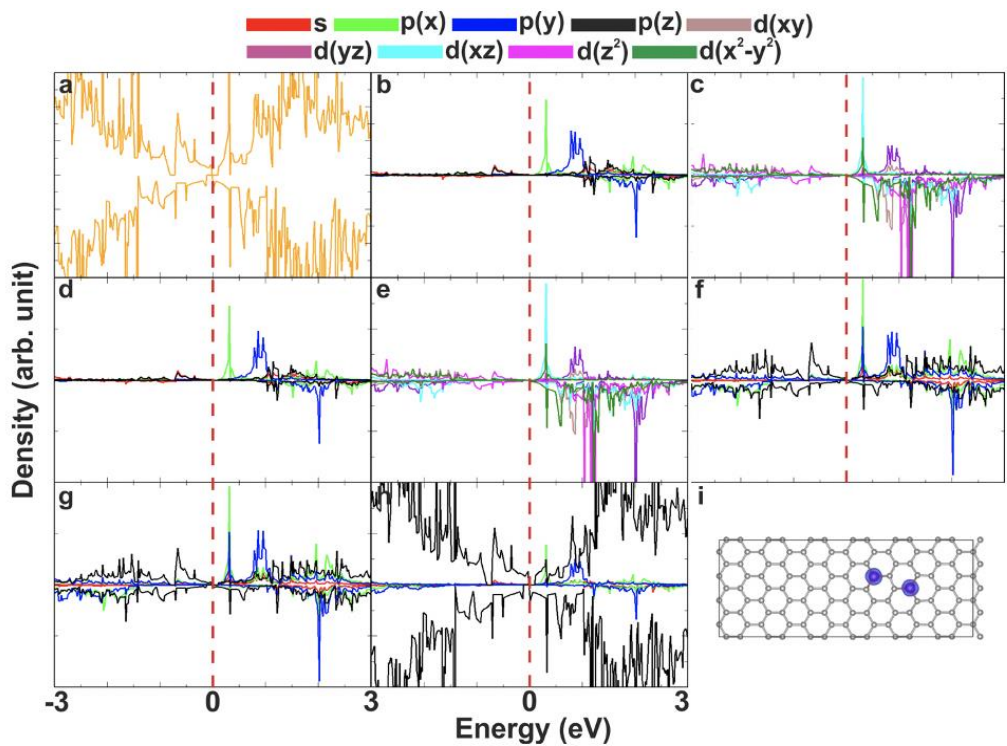

**Figure S18** Atom/orbital decomposed DOS of MnMn@SV (Cf. **Figure S12-20**). **a)** total DOS, **b,c)** $Mn_1$ atom, **d,e)** $Mn_2$ atom, **f)** nearest carbon atoms to $Mn_1$, **g)** nearest carbon atoms to $Mn_2$, **h)** other carbon atoms, **i)** spin-polarized structure.

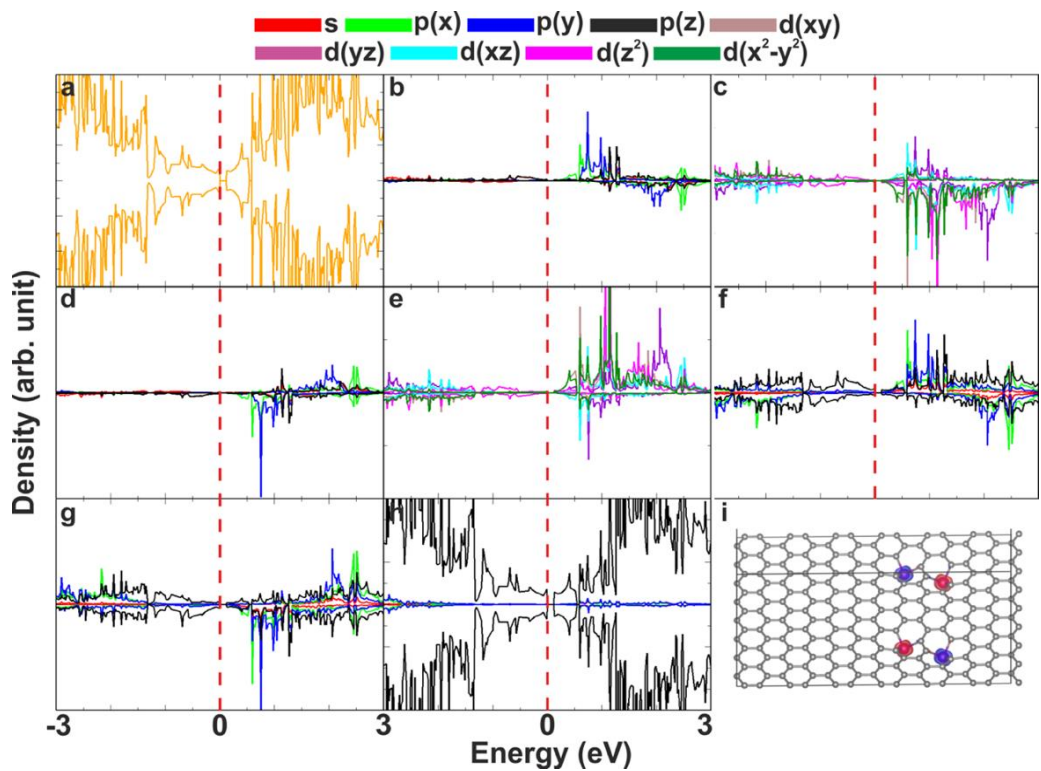

**Figure S19** Atom/orbital decomposed DOS of MnMn@SV, AFM alignment (Cf. **Figure S12-20**). **a)** total DOS, **b,c)** $Mn_1$ atom, **d,e)** $Mn_2$ atom, **f)** nearest carbon atoms to $Mn_1$, **g)** nearest carbon atoms to $Mn_2$, **h)** other carbon atoms, **i)** spin-polarized structure.

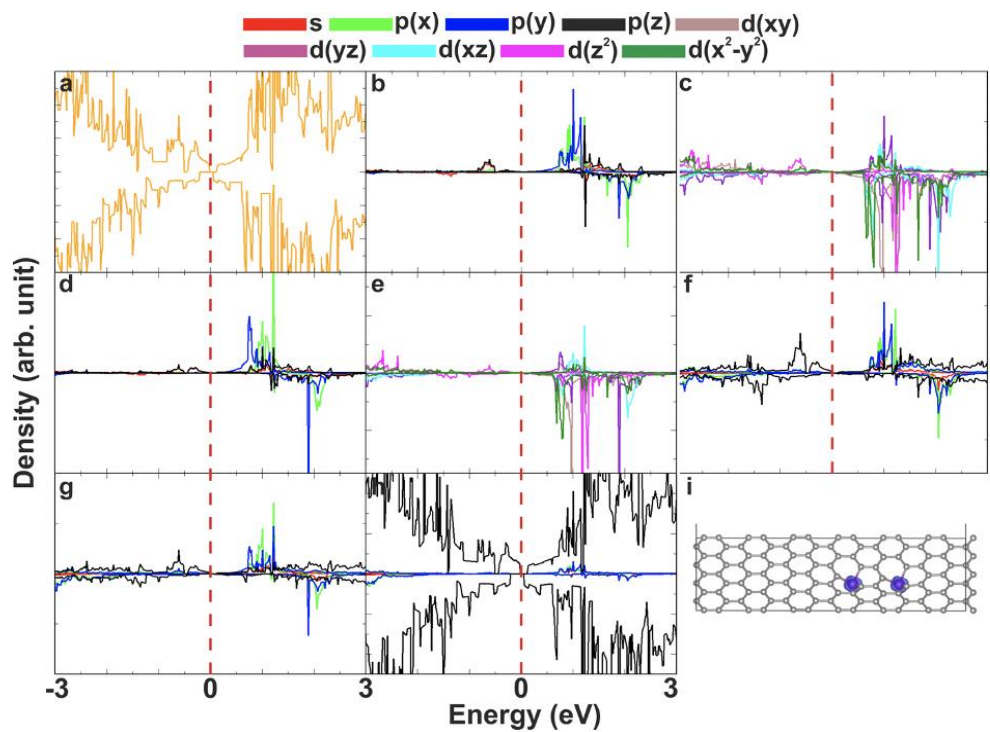

**Figure S20** Atom/orbital decomposed DOS of MnMn@SV (Cf. **Figure S12-1**). **a)** total DOS, **b,c)** $Mn_1$ atom, **d,e)** $Mn_2$ atom, **f)** nearest carbon atoms to $Mn_1$, **g)** nearest carbon atoms to $Mn_2$, **h)** other carbon atoms, **i)** spin-polarized structure.

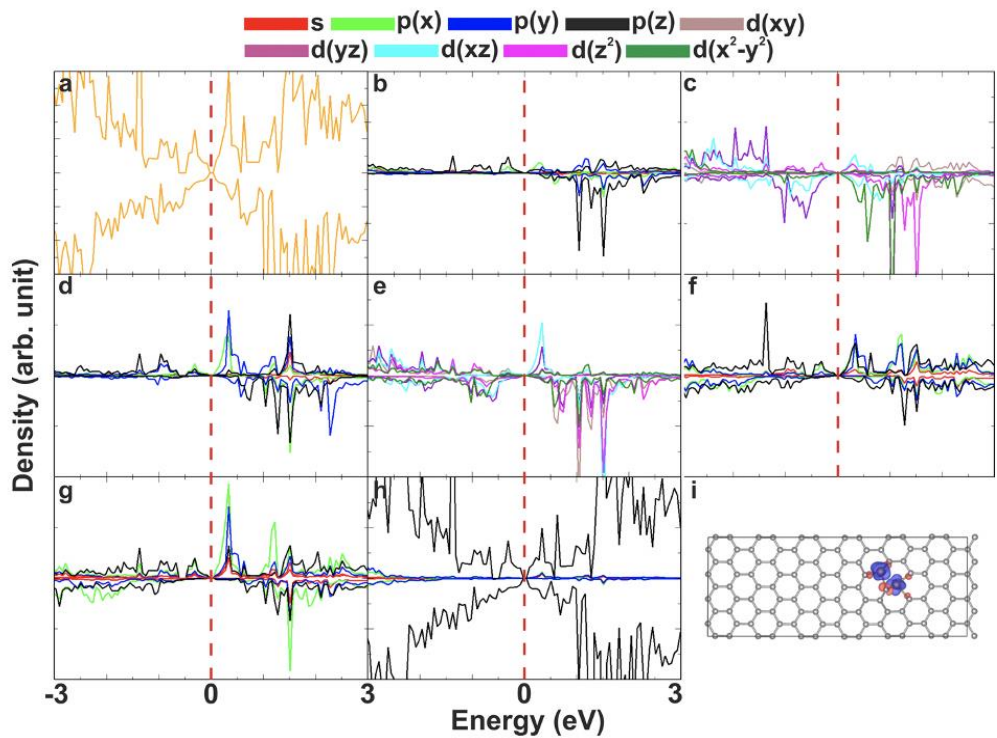

**Figure S21** Atom/orbital decomposed DOS of FeFe@SV (Cf. **Figure S12-19**). **a)** total DOS, **b,c)** $Fe_1$ atom, **d,e)** $Fe_2$ atom, **f)** nearest carbon atoms to $Fe_1$, **g)** nearest carbon atoms to $Fe_2$, **h)** other carbon atoms, **i)** spin-polarized structure.

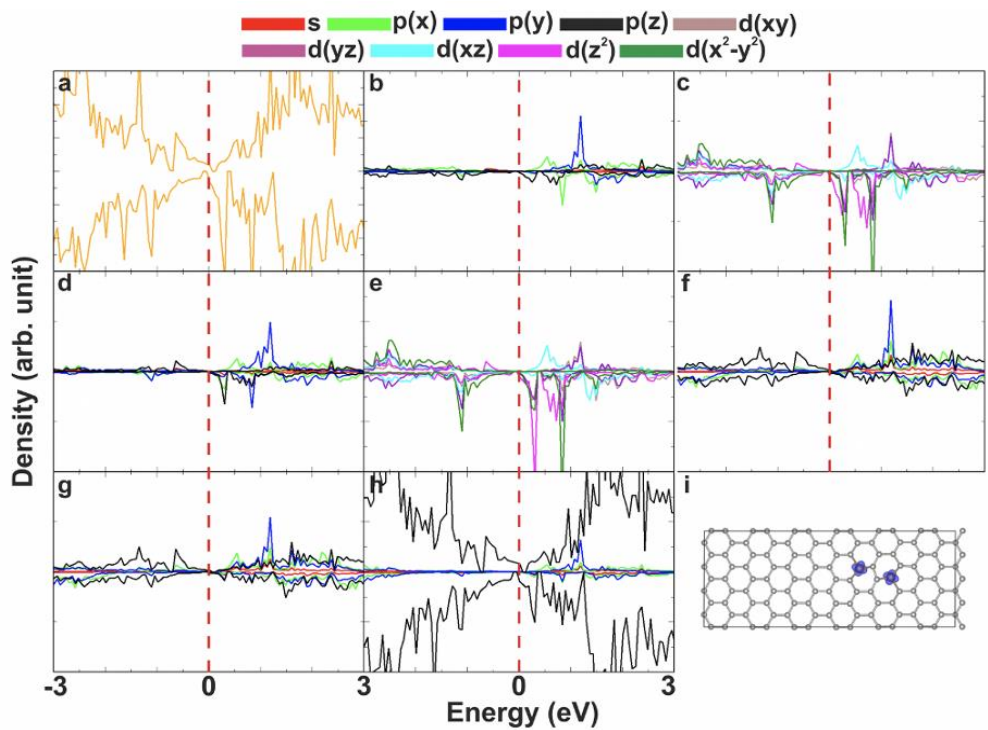

**Figure S22** Atom/orbital decomposed DOS of FeFe@SV (Cf. **Figure S12-20**). **a)** total DOS, **b,c)** Fe$_1$ atom, **d,e)** Fe$_2$ atom, **f)** nearest carbon atoms to Fe$_1$, **g)** nearest carbon atoms to Fe$_2$, **h)** other carbon atoms, **i)** spin-polarized structure.

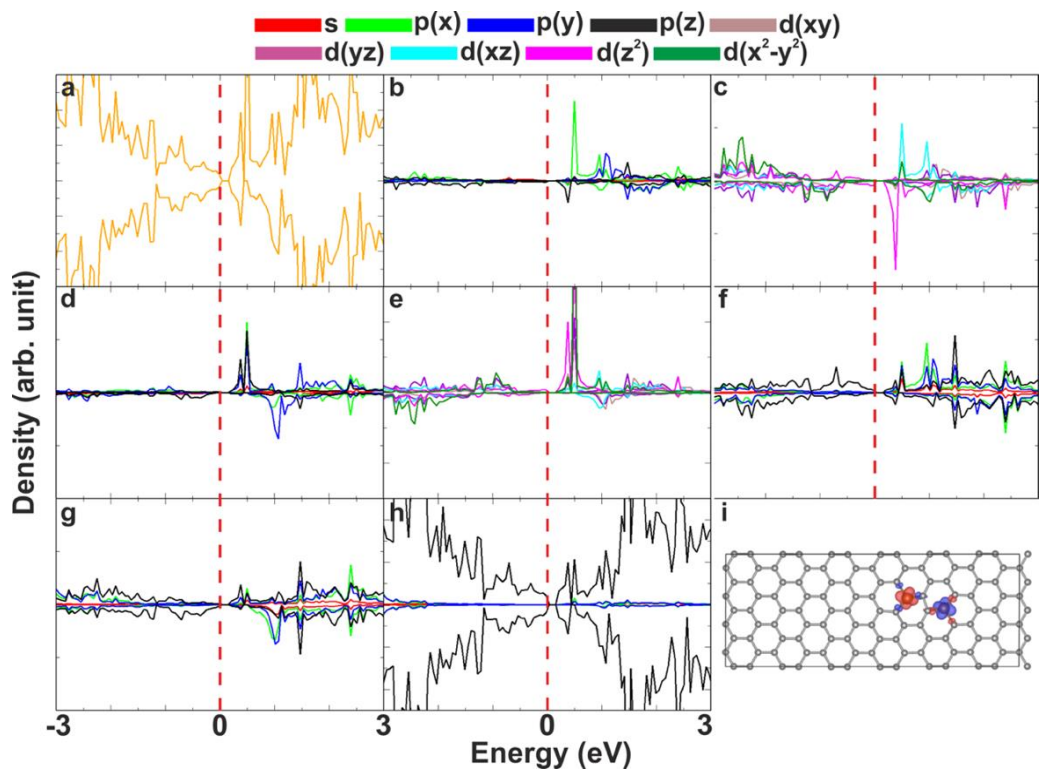

**Figure S23** Atom/orbital decomposed DOS of FeFe@SV, AFM alignment (Cf. **Figure S12-20**). **a)** total DOS, **b,c)** Fe$_1$ atom, **d,e)** Fe$_2$ atom, **f)** nearest carbon atoms to Fe$_1$, **g)** nearest carbon atoms to Fe$_2$, **h)** other carbon atoms, **i)** spin-polarized structure.

**Table S9** The comparison of spin-up (spin-down) bandgaps (eV) of TM@SV, TM@DV and TM$_1$TM$_2$@SV, TM$_1$TM$_2$@DV.

| system | bandgap |
|---|---|
| **Cr@SV** | 0.2 (0.3) |
| **CrCr@SV-structure 1** | 0.5 (0.3) |
| **CrCr@SV-structure 3** | 0.5 (0.3) |
| **CrCr@SV-structure 4** | 0.0 (0.0) |
| **CrCr@SV-structure 4 (AFM)** | 0.3 (0.3) |
| **Mn@SV** | 0.2 (0.3) |
| **MnMn@SV-structure 20** | 0.0 (0.0) |
| **MnMn@SV-structure 20 (AFM)** | 0.1 (0.1) |
| **MnMn@SV-structure 1** | 0.1 (0.2) |
| **Fe@SV** | 0.0 (0.2) |
| **FeFe@SV-structure 19** | 0.1 (0.1) |
| **FeFe@SV-structure 20** | 0.0 (0.0) |
| **FeFe@SV-structure 20 (AFM)** | 0.2 (0.2) |
| **Cr@DV** | 0.3 (0.3) |
| **CrCr@DV-structure 4** | 0.1 (0.0) |
| **CrCr@DV-structure 4 (AFM)** | 0.0 (0.2) |
| **CrCr@DV-structure 2** | 0.0 (0.0) |
| **Mn@DV** | 0.0 (0.0) |
| **MnMn@DV-structure 4** | 0.3 (0.1) |
| **MnMn@DV-structure 2** | 0.3 (0.0) |
| **Fe@DV** | 0.0 (0.0) |
| **FeFe@DV-structure 6** | 0.2 (0.0) |
| **FeFe@DV-structure 4** | 0.3 (0.0) |
| **FeFe@DV-structure 4 (AFM)** | 0.1 (0.1) |

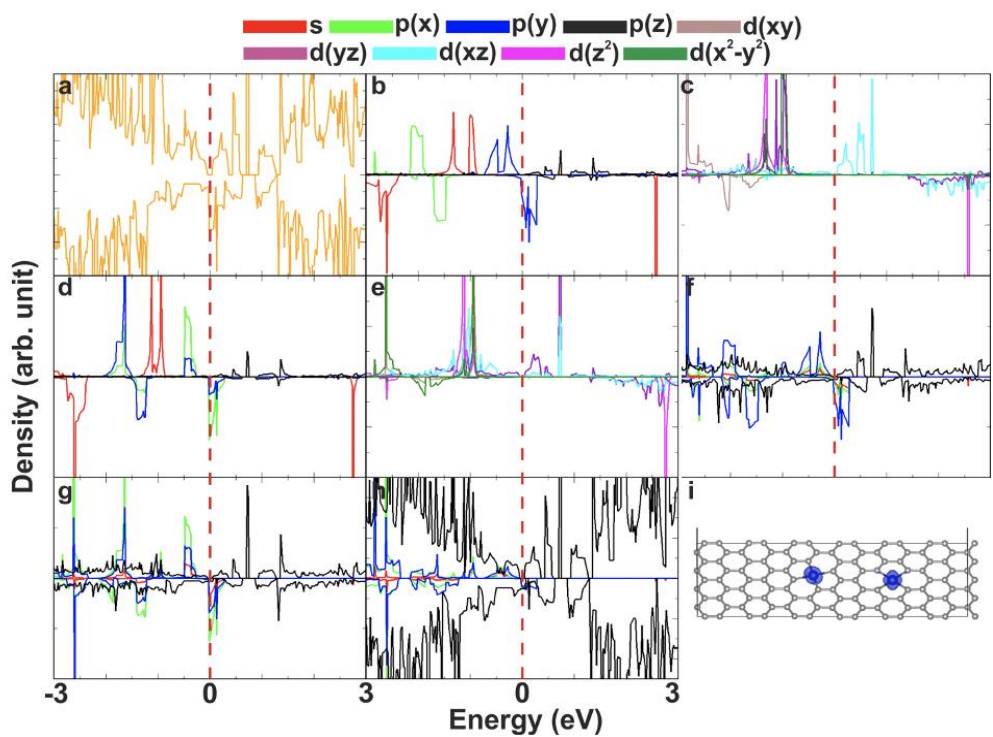

**Figure S24** Atom/orbital decomposed DOS of CrCr@DV (**Figure S13-4**). **a)** total DOS, **b,c)** Cr$_1$ atom, **d,e)** Cr$_2$ atom, **f)** nearest carbon atoms to Cr$_1$, **g)** nearest carbon atoms to Cr$_2$, **h)** other carbon atoms, **i)** spin-polarized structure.

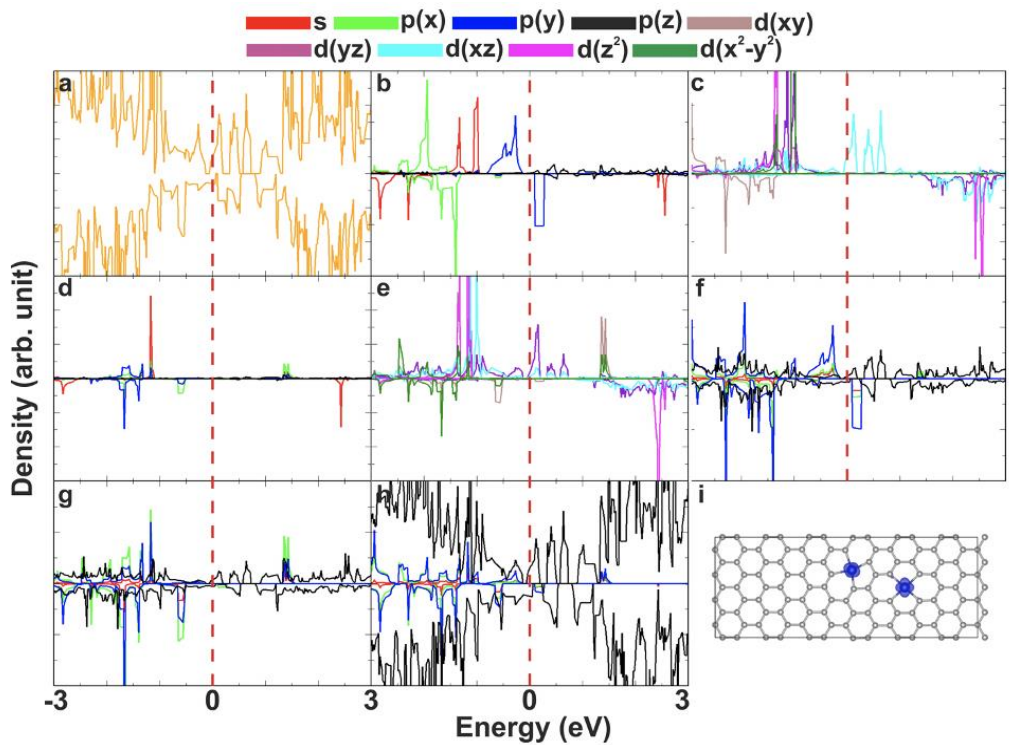

**Figure S25** Atom/orbital decomposed DOS of CrCr@DV (**Figure S13-2**). **a)** total DOS, **b,c)** Cr$_1$ atom, **d,e)** Cr$_2$ atom, **f)** nearest carbon atoms to Cr$_1$, **g)** nearest carbon atoms to Cr$_2$, **h)** other carbon atoms, **i)** spin-polarized structure.

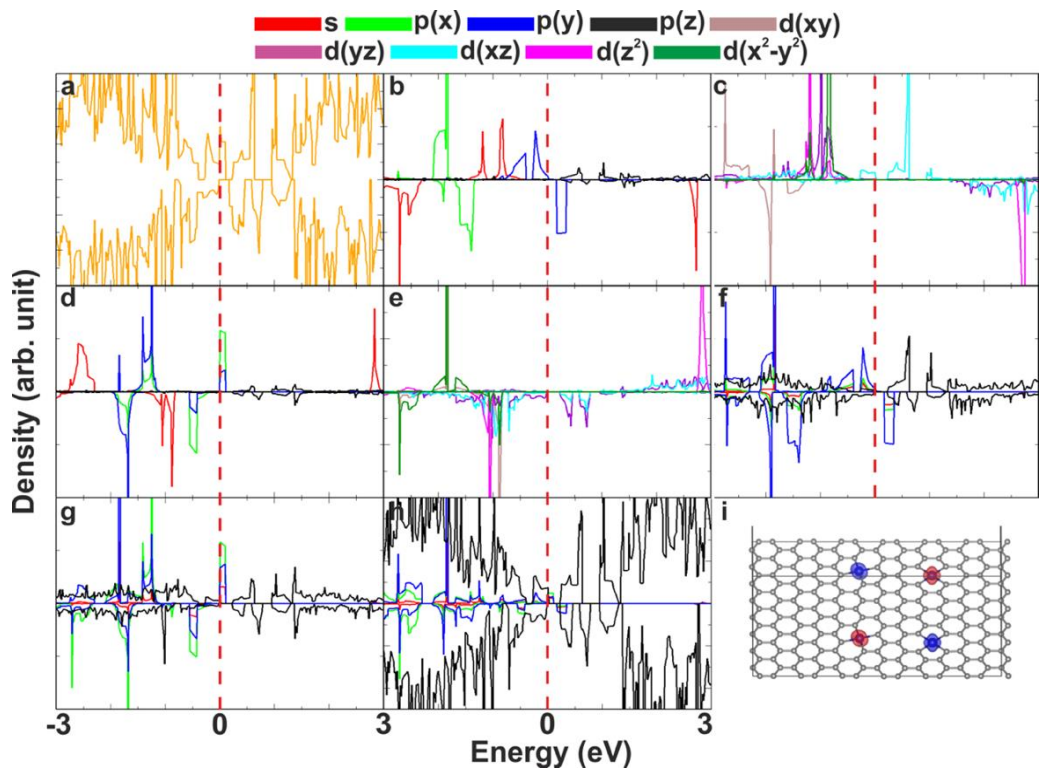

**Figure S26** Atom/orbital decomposed DOS of CrCr@DV, AFM alignment (**Figure S13-4**). **a)** total DOS, **b,c)** $Cr_1$ atom, **d,e)** $Cr_2$ atom, **f)** nearest carbon atoms to $Cr_1$, **g)** nearest carbon atoms to $Cr_2$, **h)** other carbon atoms, **i)** spin-polarized structure.

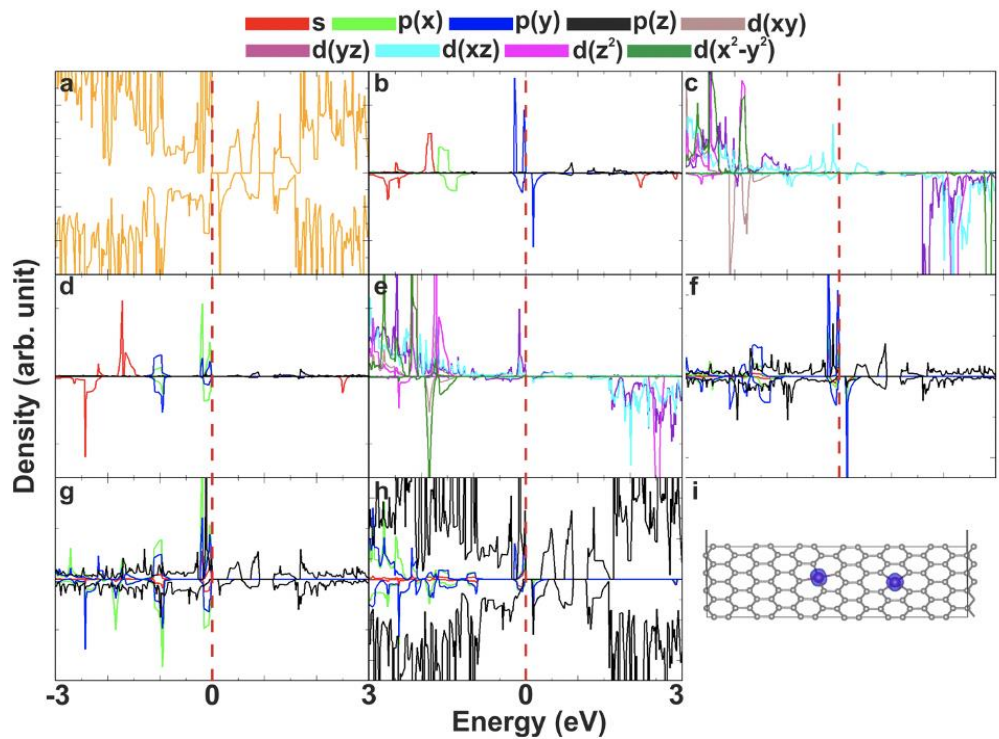

**Figure S27** Atom/orbital decomposed DOS of MnMn@DV (**Figure S13-4**). **a)** total DOS, **b,c)** $Mn_1$ atom, **d,e)** $Mn_2$ atom, **f)** nearest carbon atoms to $Mn_1$, **g)** nearest carbon atoms to $Mn_2$, **h)** other carbon atoms, **i)** spin-polarized structure.

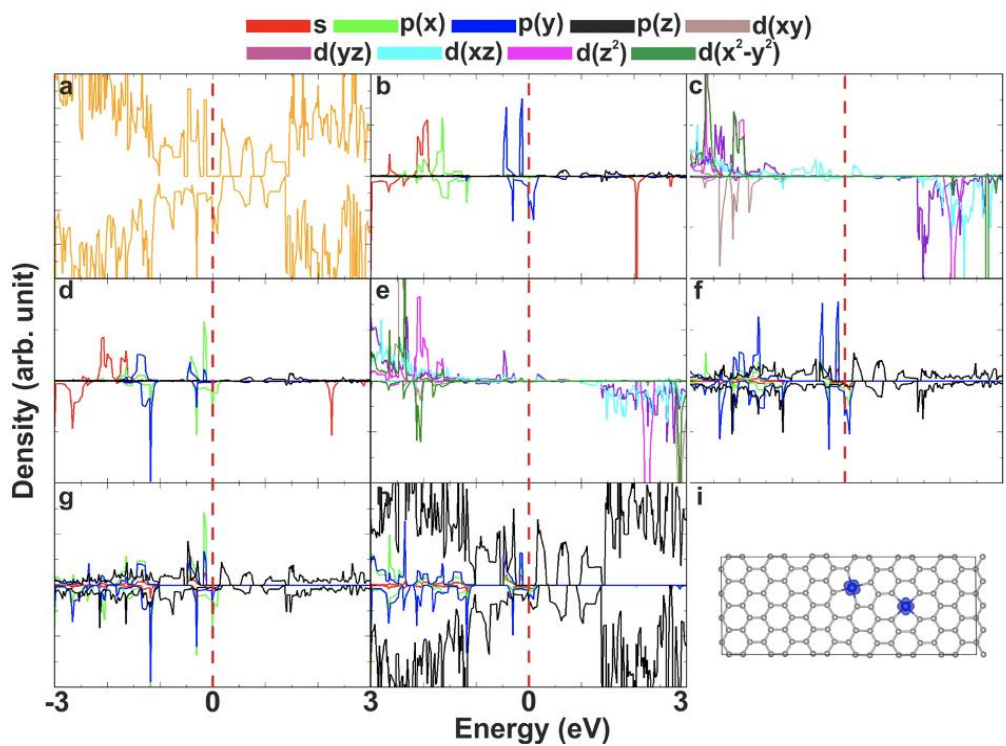

**Figure S28** Atom/orbital decomposed DOS of MnMn@DV (**Figure S13-2**). **a)** total DOS, **b,c)** $Mn_1$ atom, **d,e)** $Mn_2$ atom, **f)** nearest carbon atoms to $Mn_1$, **g)** nearest carbon atoms to $Mn_2$, **h)** other carbon atoms, **i)** spin-polarized structure.

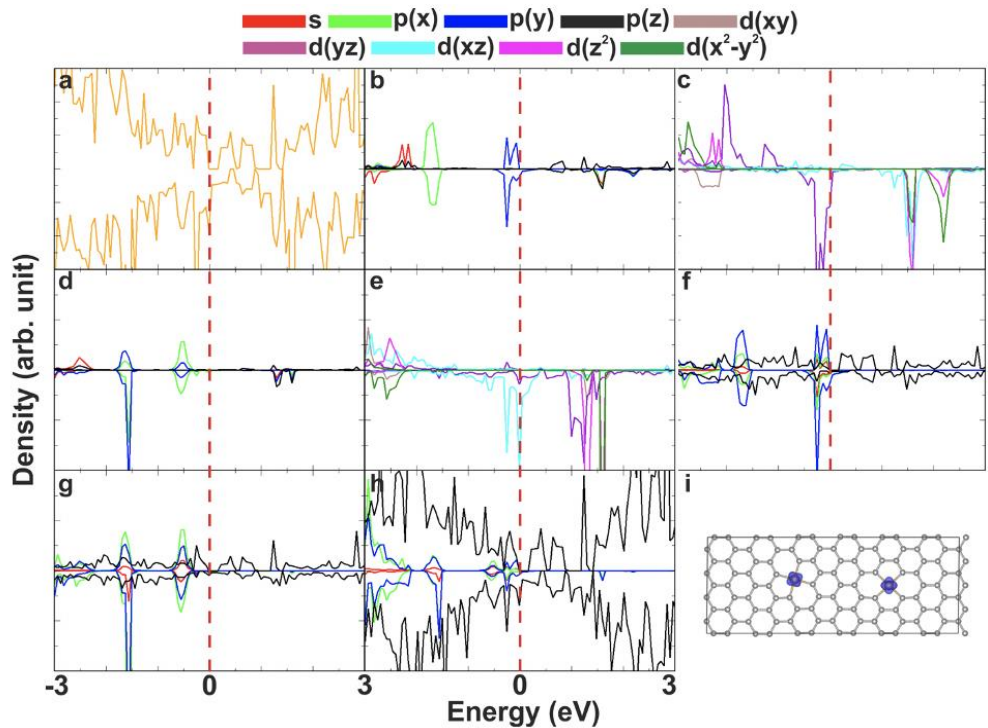

**Figure S29** Atom/orbital decomposed DOS of FeFe@DV (**Figure S13-6**). **a)** total DOS, **b,c)** $Fe_1$ atom, **d,e)** $Fe_2$ atom, **f)** nearest carbon atoms to $Fe_1$, **g)** nearest carbon atoms to $Fe_2$, **h)** other carbon atoms, **i)** spin-polarized structure.

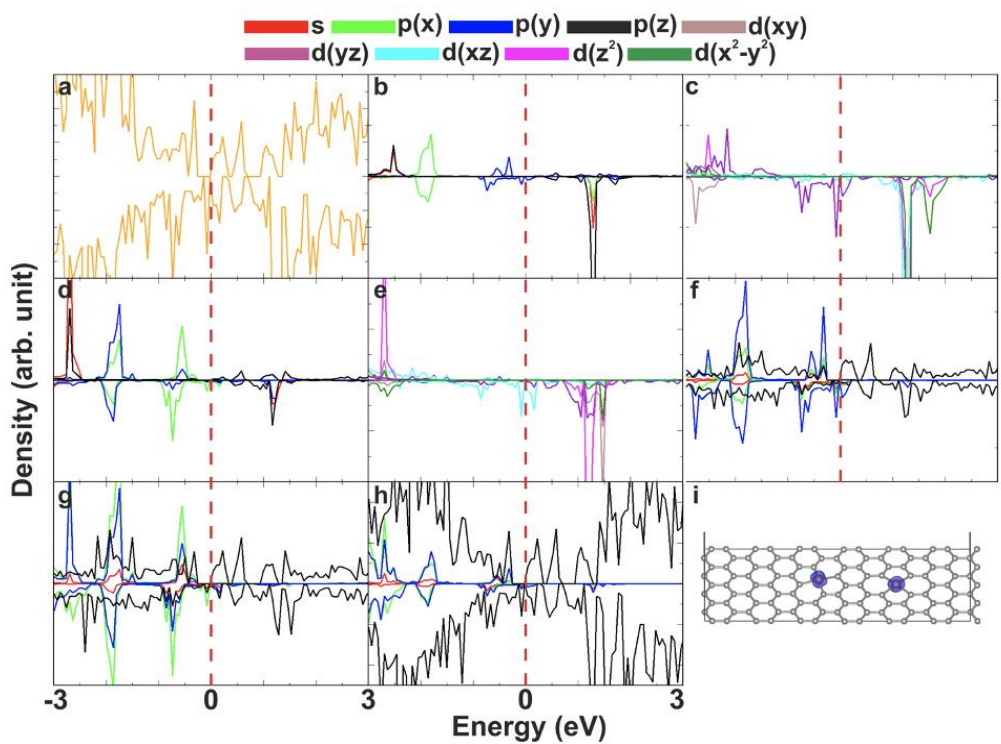

**Figure S30** Atom/orbital decomposed DOS of FeFe@DV (**Figure S13-4**). **a)** total DOS, **b,c)** $Fe_1$ atom, **d,e)** $Fe_2$ atom, **f)** nearest carbon atoms to $Fe_1$, **g)** nearest carbon atoms to $Fe_2$, **h)** other carbon atoms, **i)** spin-polarized structure.

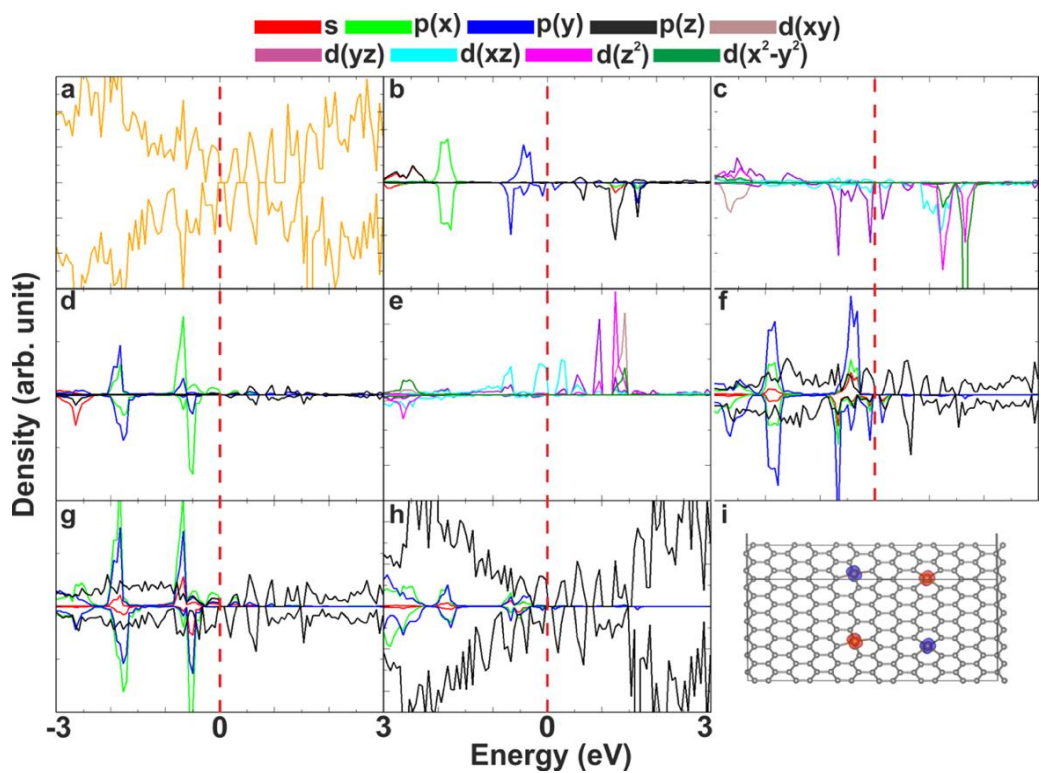

**Figure S31** Atom/orbital decomposed DOS of FeFe@DV, AFM alignment (**Figure S13-4**). **a)** total DOS, **b,c)** $Fe_1$ atom, **d,e)** $Fe_2$ atom, **f)** nearest carbon atoms to $Fe_1$, **g)** nearest carbon atoms to $Fe_2$, **h)** other carbon atoms, **i)** spin-polarized structure.

**Table S10** Total spin moments $\mu_{s\_tot}$, local spin moments $\mu_{s\_loc}$ and orbital moments $\mu_l$ (all in $\mu_B$) oriented along the "$x$", and "$z$" magnetization axes and their anisotropy ($\Delta$) for $TM_1TM_2@SV$ and $TM_1TM_2@DV$ shown in **Figures S32-S38**.

| system | $\mu_{s\_tot\,(x)}$ | $\mu_{s\_loc\,(x)}$ | $\mu_{l\,(x)}$ | $\mu_{s\_tot\,(z)}$ | $\mu_{s\_loc\,(z)}$ | $\mu_{l\,(z)}$ | $\mu_{s\_tot\,(\Delta)}$ | $\mu_{s\_loc(\Delta)}$ | $\mu_{l\,(\Delta)}$ |
|---|---|---|---|---|---|---|---|---|---|
| **CrFe@SV** | 1.49 | 1.64 | 0.06 | 1.51 | 1.66 | 0.13 | -0.03 | -0.03 | -0.07 |
| **CrMn@SV** | 2.50 | 2.36 | 0.01 | 2.50 | 2.36 | 0.00 | 0.00 | 0.00 | 0.01 |
| **FeMn@SV** | 1.83 | 1.80 | 0.07 | 1.83 | 1.80 | 0.04 | 0.00 | 0.00 | 0.03 |
| **CrFe@DV (AFM)** | 0.05 | 0.05 | 0.00 | 0.03 | 0.03 | 0.00 | 0.02 | 0.02 | 0.00 |
| **CrFe@DV** | 3.00 | 2.76 | 0.02 | 3.00 | 2.76 | 0.02 | 0.00 | 0.00 | 0.00 |
| **CrMn@DV** | 2.62 | 2.55 | 0.03 | 2.62 | 2.55 | 0.02 | 0.00 | 0.00 | 0.01 |
| **FeMn@DV** | 1.85 | 1.75 | 0.04 | 2.04 | 1.89 | 0.05 | -0.20 | -0.14 | -0.02 |

**Table S11** The spin-up (spin-down) bandgaps of $TM_1TM_2@SV$, $TM_1TM_2@DV$.

| system | bandgap |
|---|---|
| **CrFe@SV** | 0.1 (0.3) |
| **CrMn@SV** | 0.2 (0.3) |
| **FeMn@SV** | 0.0 (0.0) |
| **CrFe@DV** | 0.3 (0.2) |
| **CrFe@DV (AFM)** | 0.3 (0.2) |
| **CrMn@DV** | 0.0 (0.5) |
| **FeMn@DV** | 0.0 (0.0) |

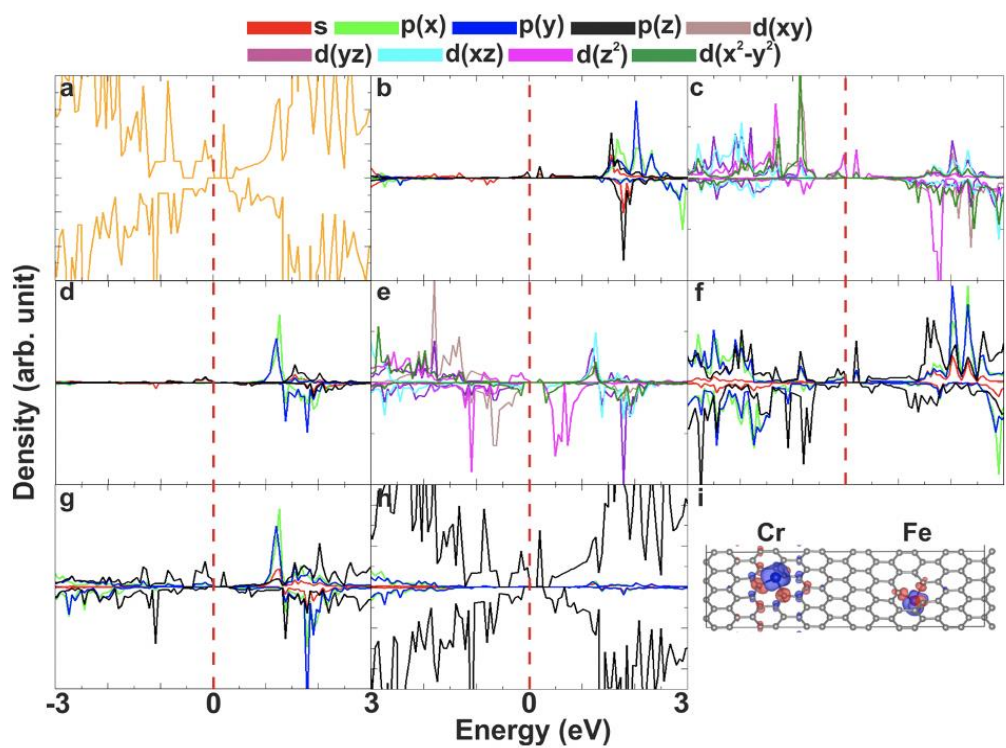

**Figure S32** Atom/orbital decomposed DOS of CrFe@SV. **a)** total DOS, **b**,**c)** Cr atom, **d**,**e)** Fe atom, **f)** nearest carbon atoms to Cr, **g)** nearest carbon atoms to Fe, **h)** other carbon atoms, **i)** spin-polarized structure.

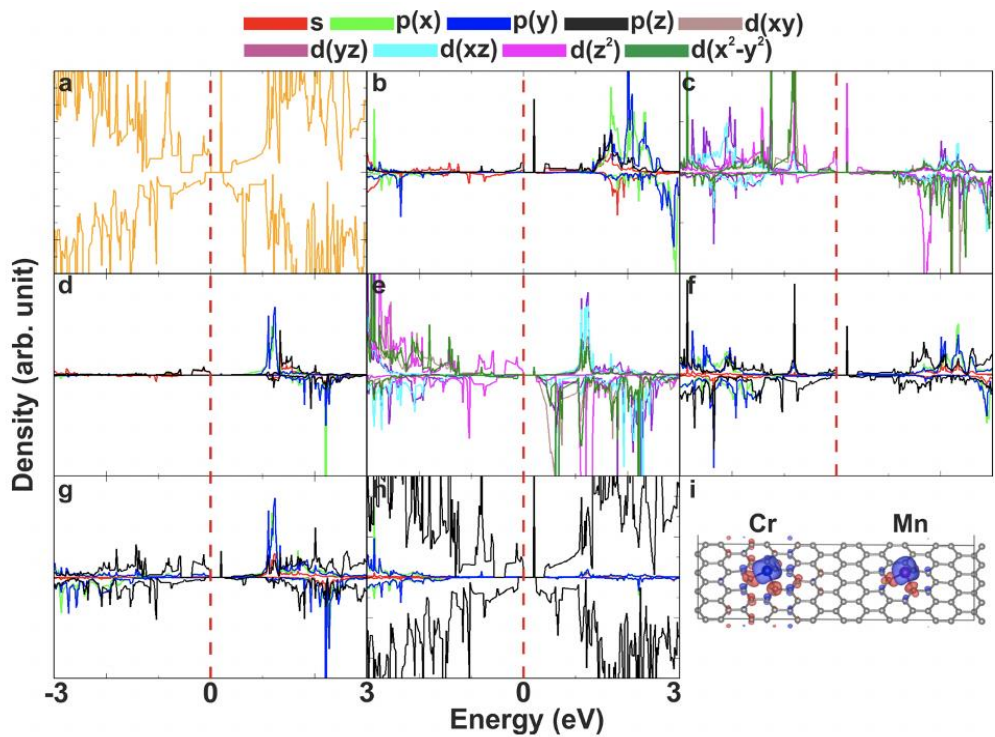

**Figure S33** Atom/orbital decomposed DOS of CrMn@SV. **a)** total DOS, **b**,**c)** Cr atom, **d**,**e)** Mn atom, **f)** nearest carbon atoms to Cr, **g)** nearest carbon atoms to Mn, **h)** other carbon atoms, **i)** spin-polarized structure.

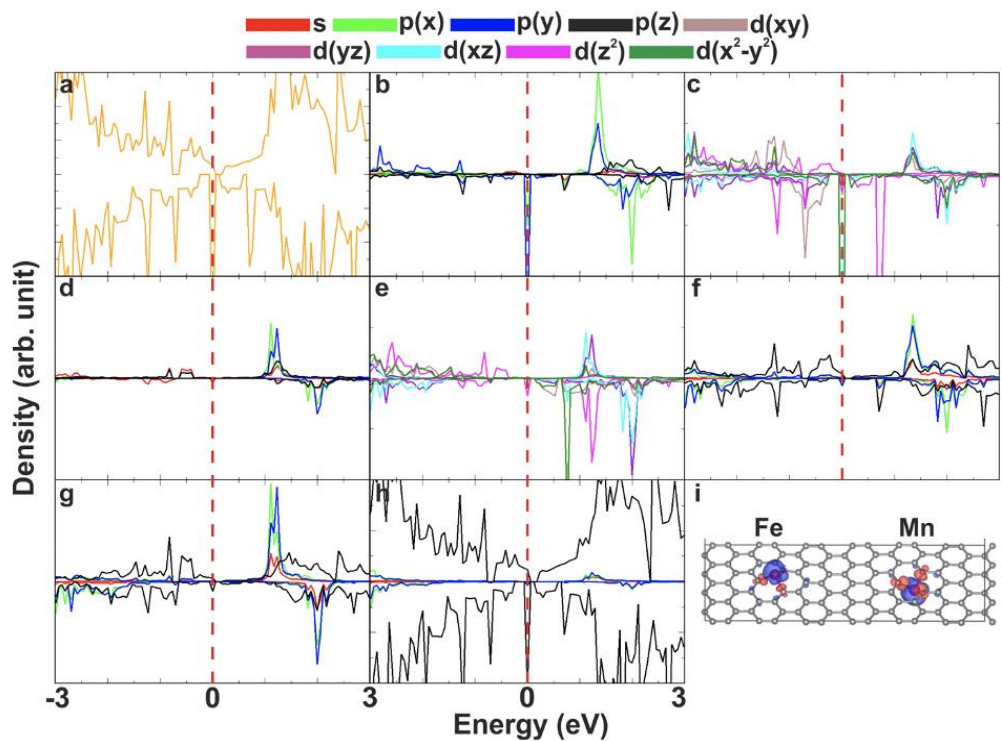

**Figure S34** Atom/orbital decomposed DOS of FeMn@SV. **a)** total DOS, **b,c)** Fe atom, **d,e)** Mn atom, **f)** nearest carbon atoms to Fe, **g)** nearest carbon atoms to Mn, **h)** other carbon atoms, **i)** spin-polarized structure.

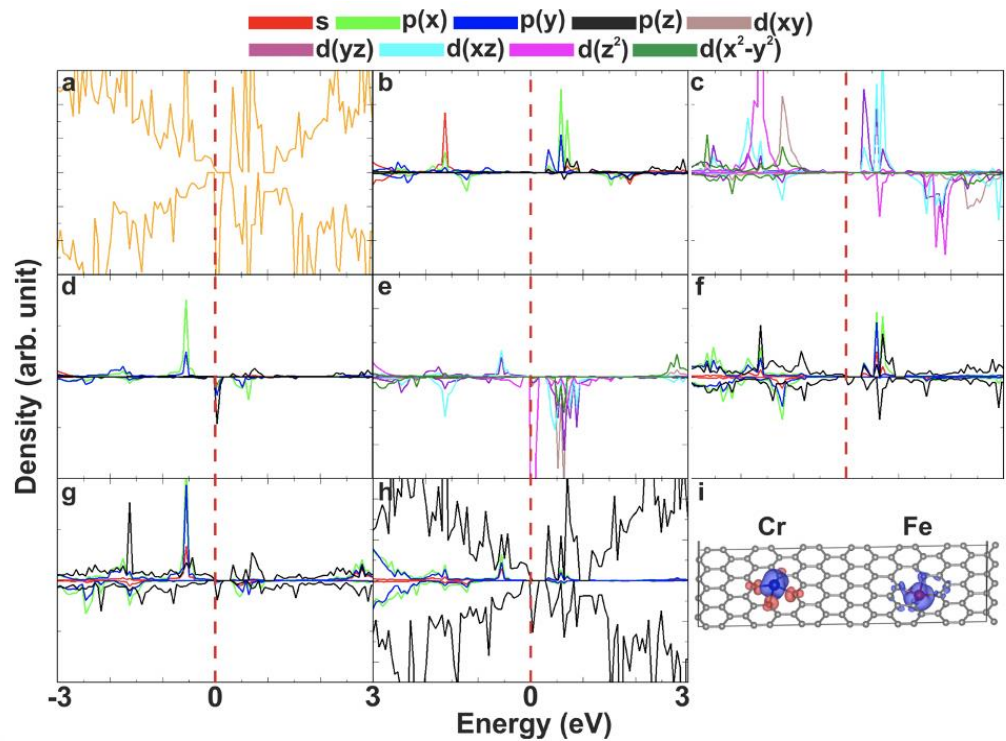

**Figure S35** Atom/orbital decomposed DOS of CrFe@DV. **a)** total DOS, **b,c)** Cr atom, **d,e)** Fe atom, **f)** nearest carbon atoms to Cr, **g)** nearest carbon atoms to Fe, **h)** other carbon atoms, **i)** spin-polarized structure.

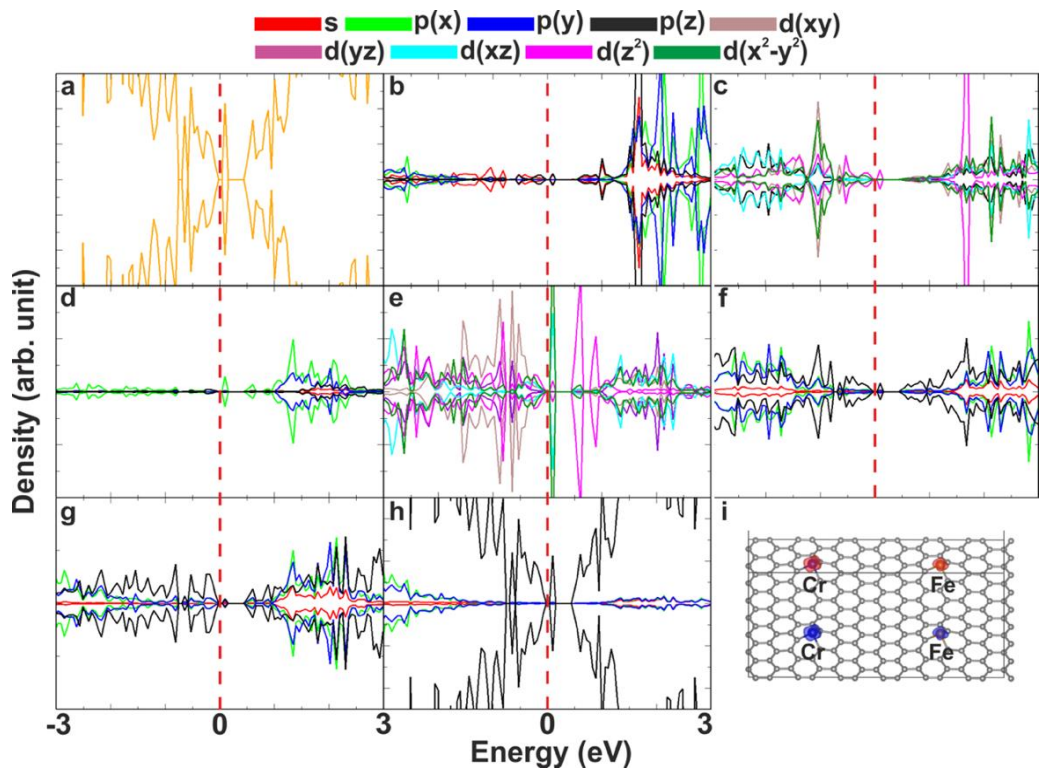

**Figure S36** Atom/orbital decomposed DOS of CrFe@DV, AFM alignment. **a)** total DOS, **b,c)** Cr atom, **d,e)** Fe atom, **f)** nearest carbon atoms to Cr, **g)** nearest carbon atoms to Fe, **h)** other carbon atoms, **i)** spin-polarized structure.

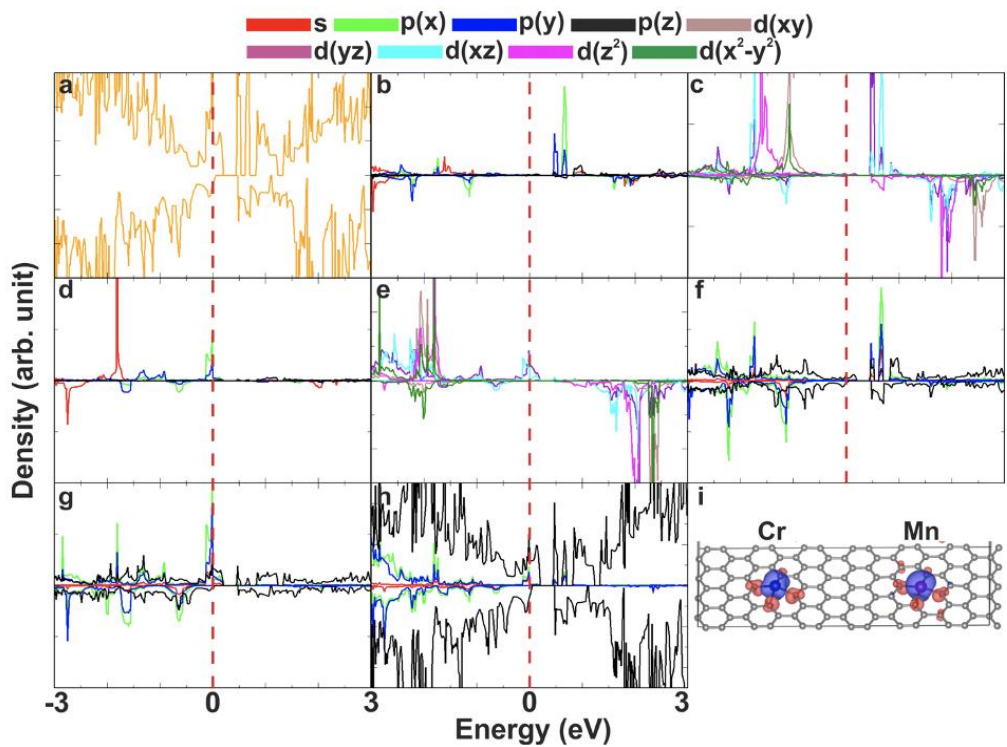

**Figure S37** Atom/orbital decomposed DOS of CrMn@DV. **a)** total DOS, **b,c)** Cr atom, **d,e)** Mn atom, **f)** nearest carbon atoms to Cr, **g)** nearest carbon atoms to Mn, **h)** other carbon atoms, **i)** spin-polarized structure.

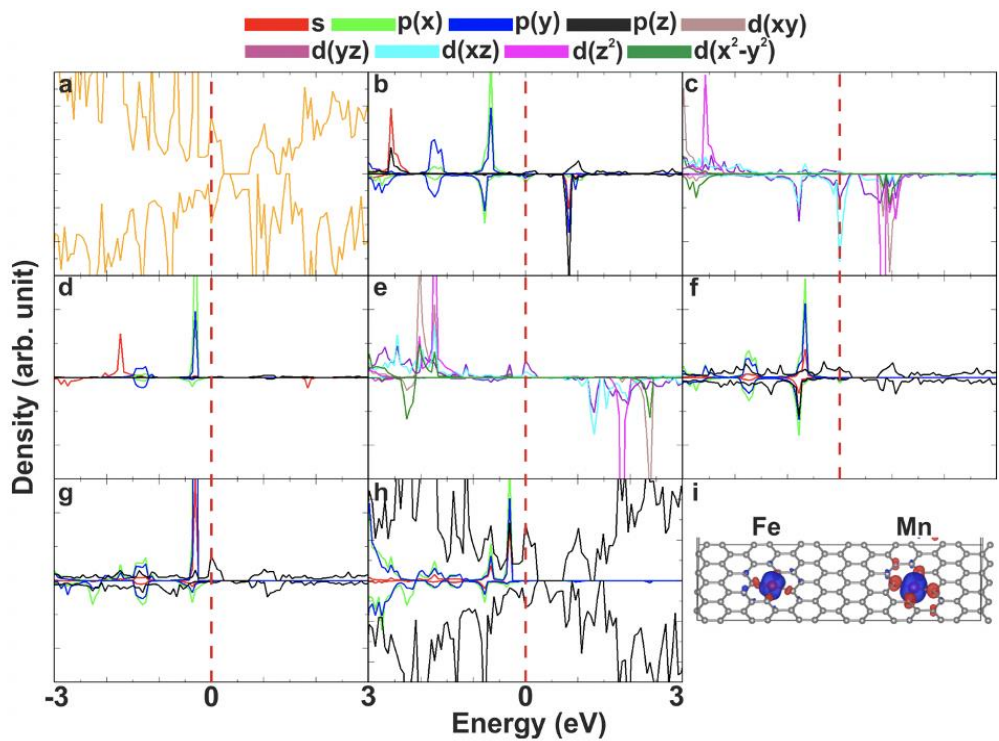

**Figure S38** Atom/orbital decomposed DOS of FeMn@DV. **a)** total DOS, **b,c)** Fe atom, **d,e)** Mn atom, **f)** nearest carbon atoms to Fe, **g)** nearest carbon atoms to Mn, **h)** other carbon atoms, **i)** spin-polarized structure.

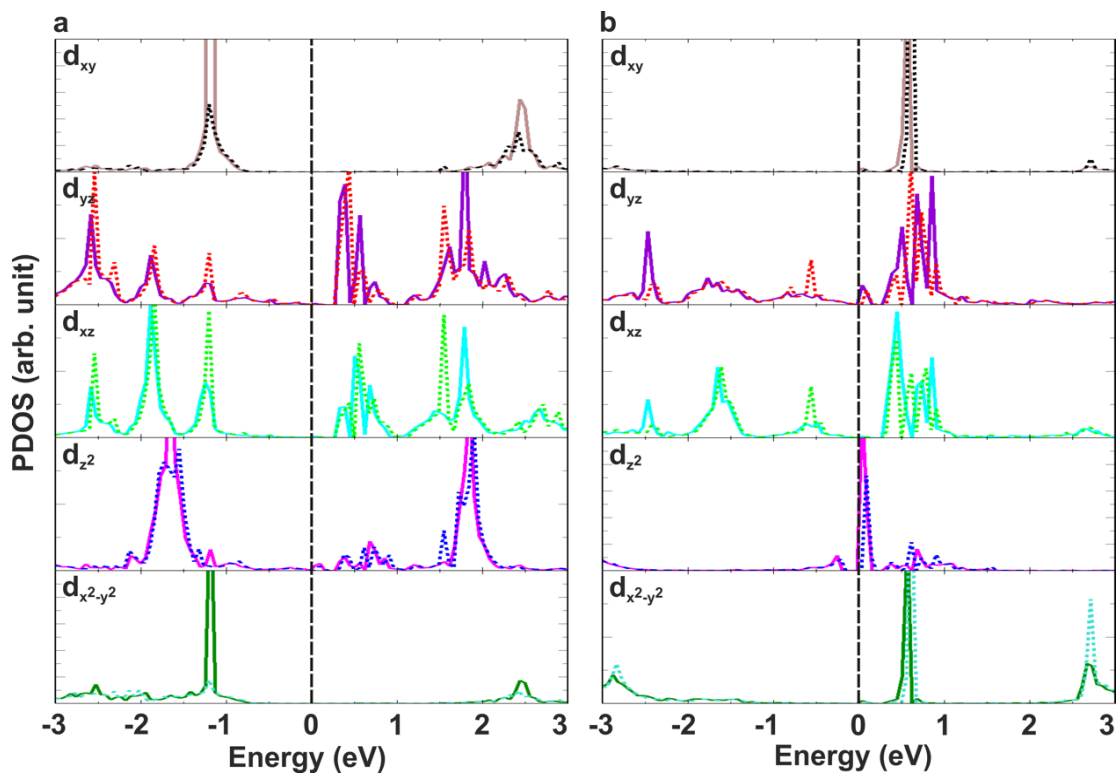

**Figure S39** Relativistic partial atom/orbital-resolved densities of states (PDOS) for CrFe@DV for in-plane (solid lines) and perpendicular magnetization (dashed lines). **a)** Cr atom and **b)** Fe atom.

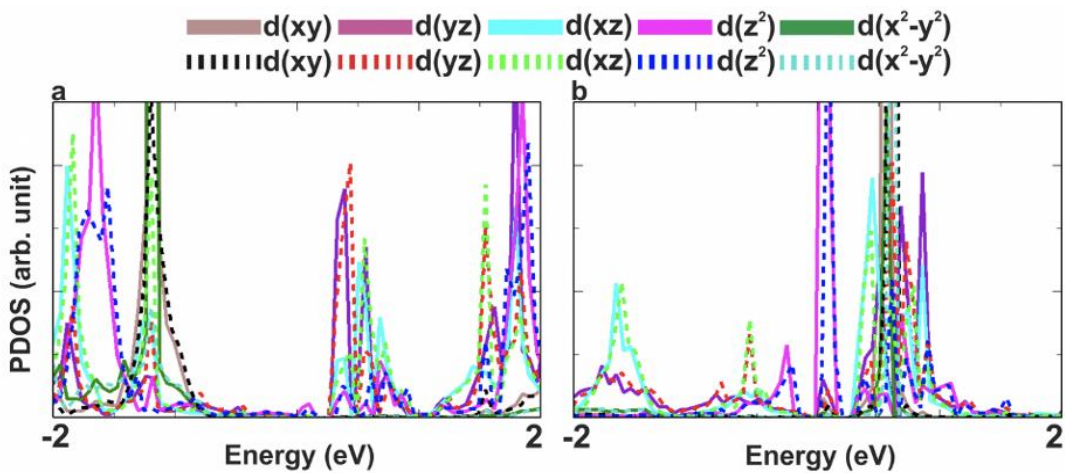

**Figure S40** The zoomed partial atom/orbital contribution of CrFe@DV. **a)** Cr atom and **b)** Fe atom. Full lines denote in-plane magnetization, dashed lines denote perpendicular magnetization.

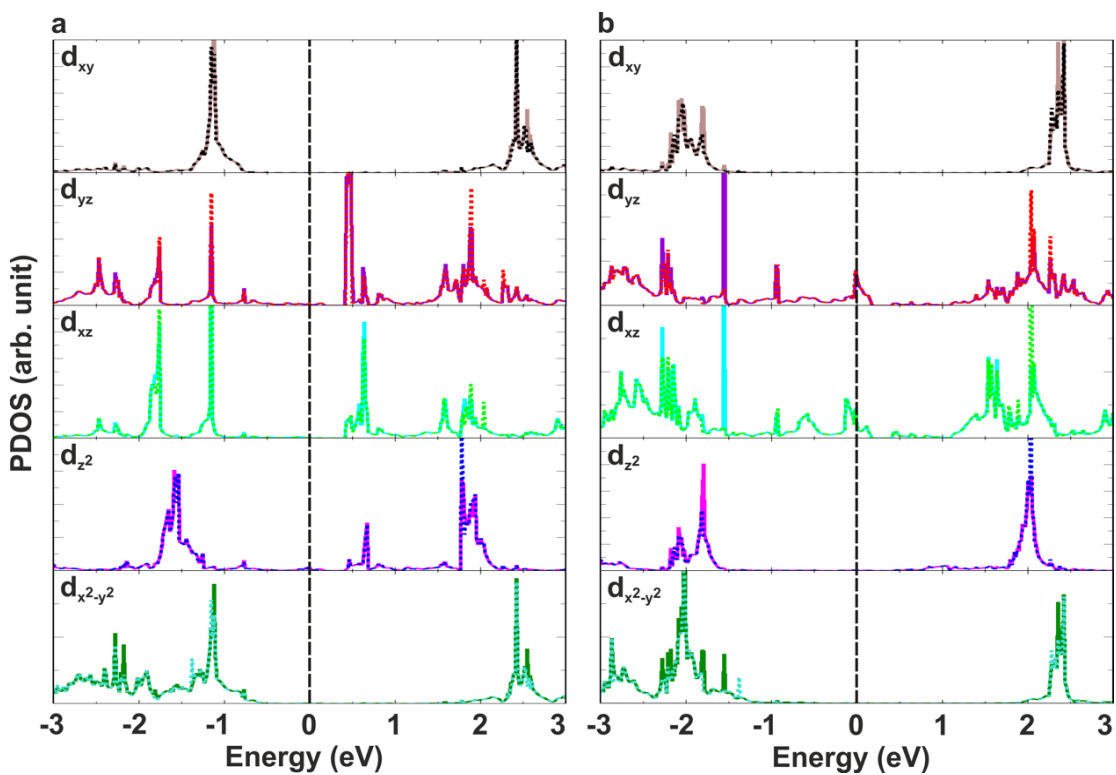

**Figure S41** Relativistic partial atom/orbital-resolved densities of states for CrMn@DV for in-plane (solid lines) and perpendicular magnetization (dashed lines). **a)** Cr atom and **b)** Mn atom.

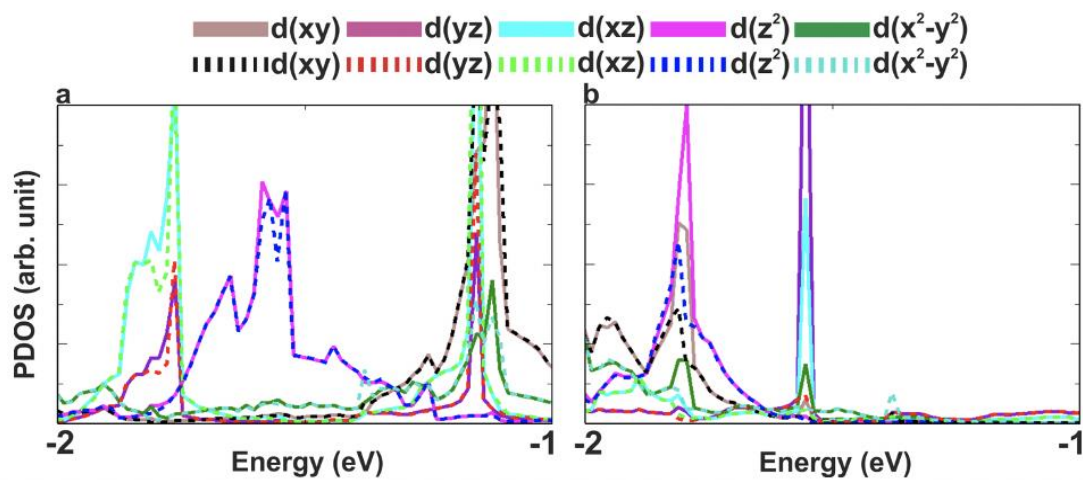

**Figure S42** The zoomed partial atom/orbital contribution of CrMn@DV. **a)** Cr atom and **b)** Mn atom. Full lines denote in-plane magnetization, dashed lines denote perpendicular magnetization.

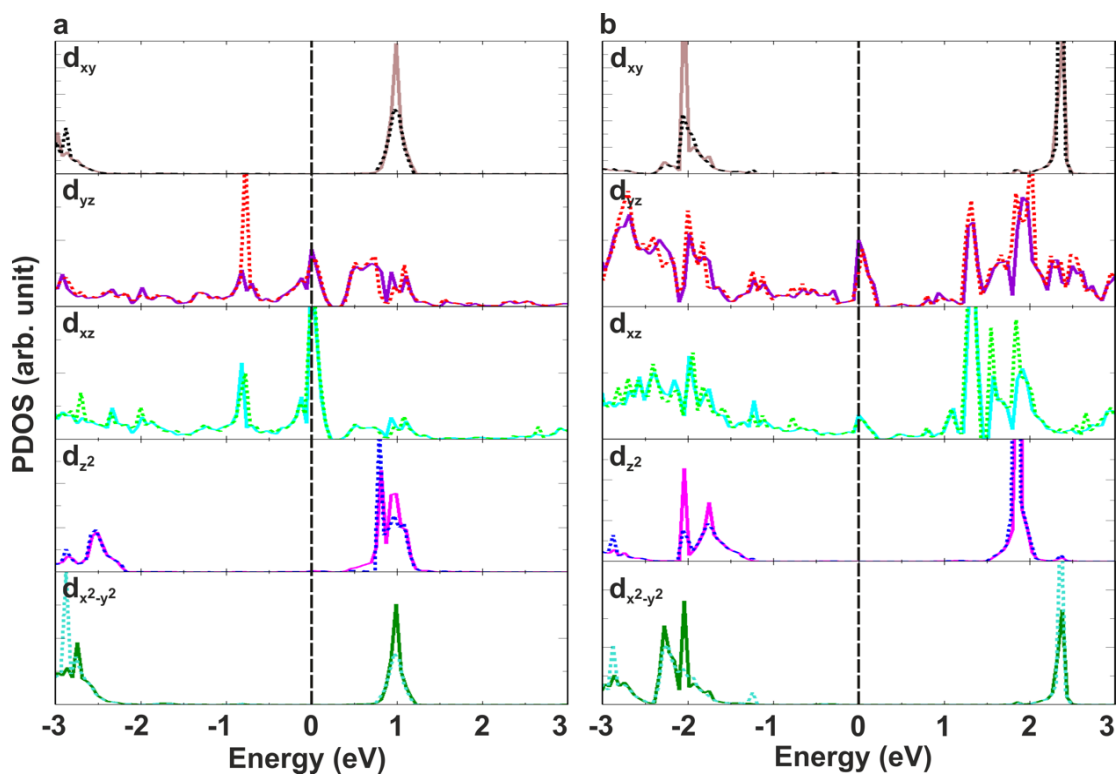

**Figure S43** Relativistic partial atom/orbital-resolved densities of states for FeMn@DV for in-plane (solid lines) and perpendicular magnetization (dashed lines). **a)** Fe atom and **b)** Mn atom.

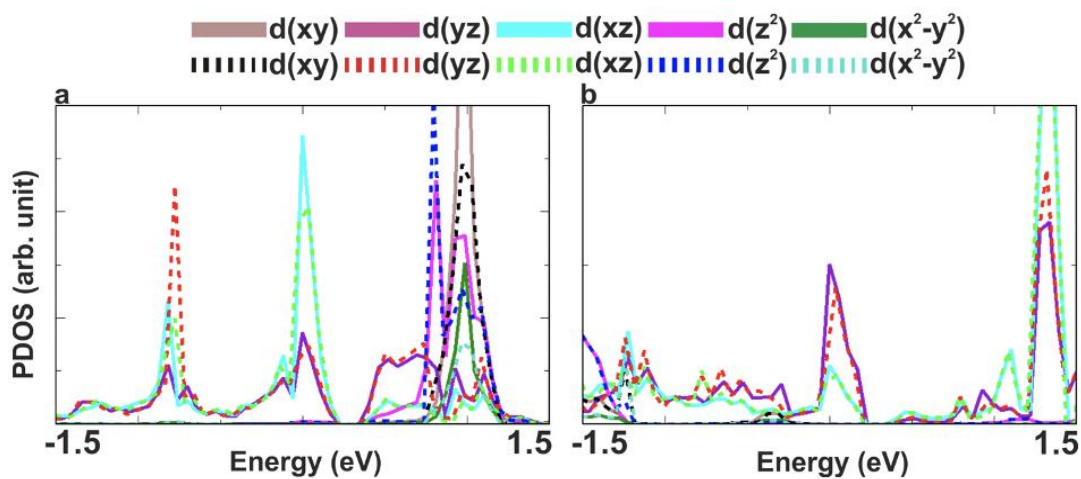

**Figure S44** The zoomed partial atom/orbital contribution of FeMn@DV. **a)** Fe atom and **b)** Mn atom. Full lines denote in-plane magnetization, dashed lines denote perpendicular magnetization.

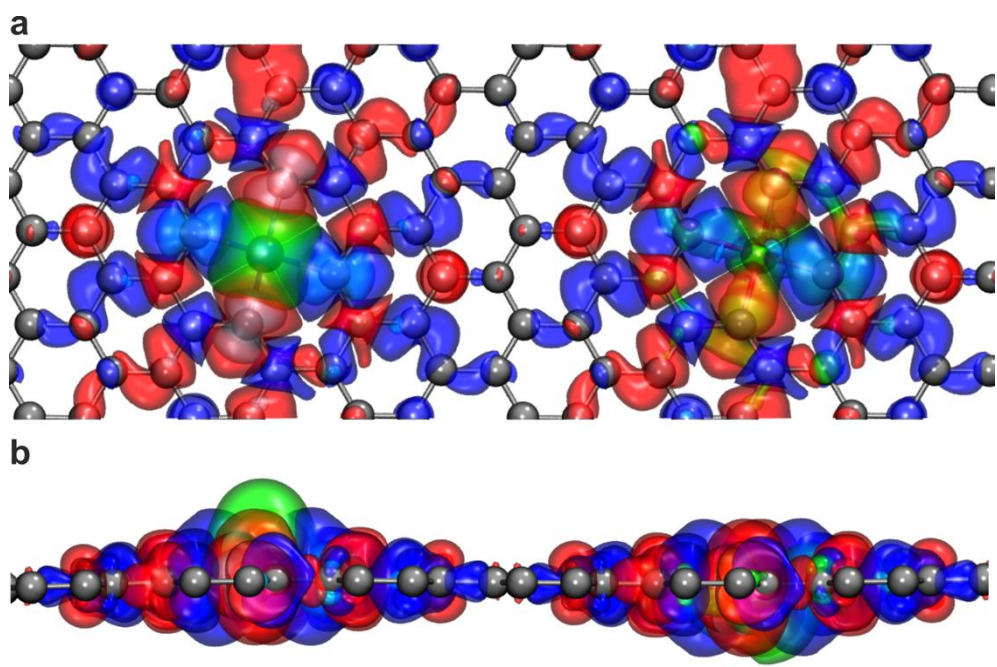

**Figure S45 a)** Top and **b)** side view of spin densities for the system of Cr and Fe atoms bound to separate DV defects plotted at ±0.01 e⁻Å⁻³ isovalues for Cr and Fe (displayed in green/cyan for spin densities corresponding to positive/negative magnetic moments) and ±0.001 e⁻Å⁻³ (shown in blue/red) for DV-graphene.

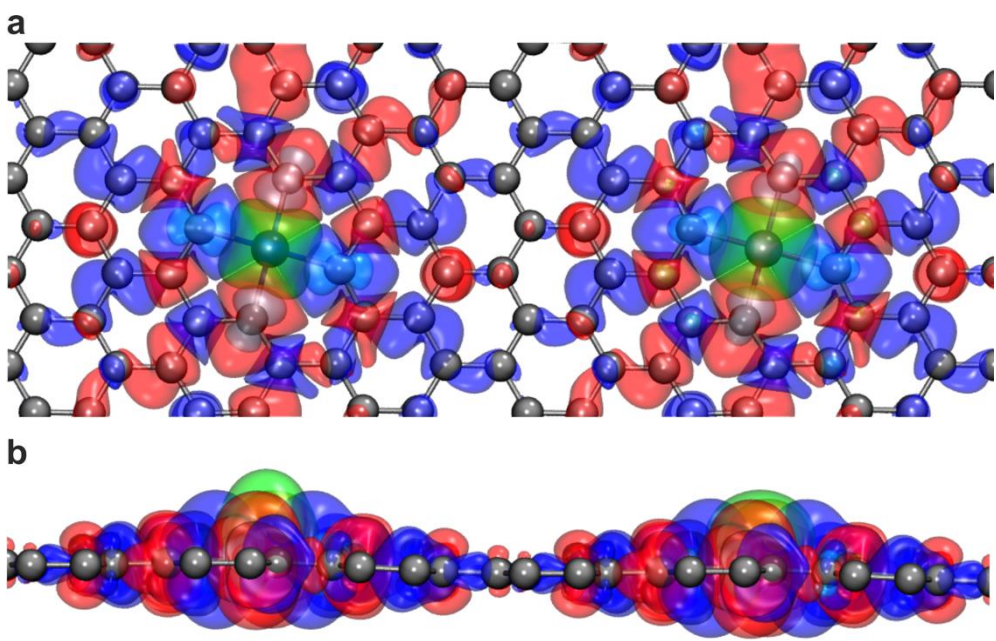

**Figure S46 a)** Top and **b)** side view of spin densities for the system of Cr and Mn atoms bound to separate DV defects plotted at ±0.01 e⁻Å⁻³ isovalues for Cr and Mn (displayed in green/cyan for spin densities corresponding to positive/negative magnetic moments) and ±0.001 e⁻Å⁻³ (shown in blue/red) for DV-graphene.

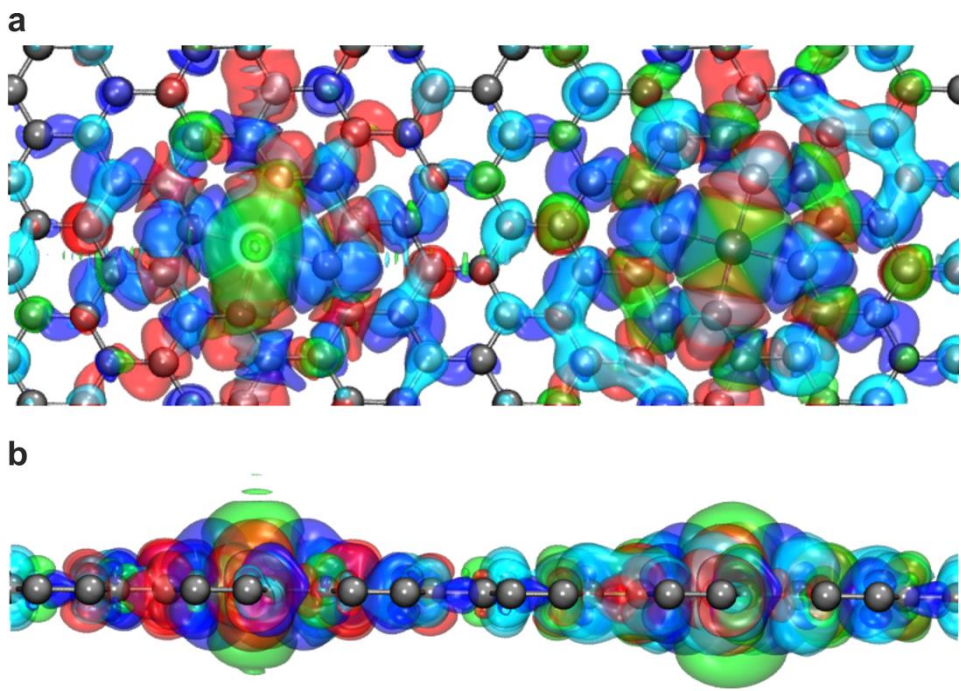

**Figure S47 a)** Top and **b)** side view of spin densities for the system of Fe and Mn atoms bound to separate DV defects plotted at ±0.01 e⁻Å⁻³ isovalues for Fe and Mn (displayed in green/cyan for spin densities corresponding to positive/negative magnetic moments) and ±0.001 e⁻Å⁻³ (shown in blue/red) for DV-graphene.

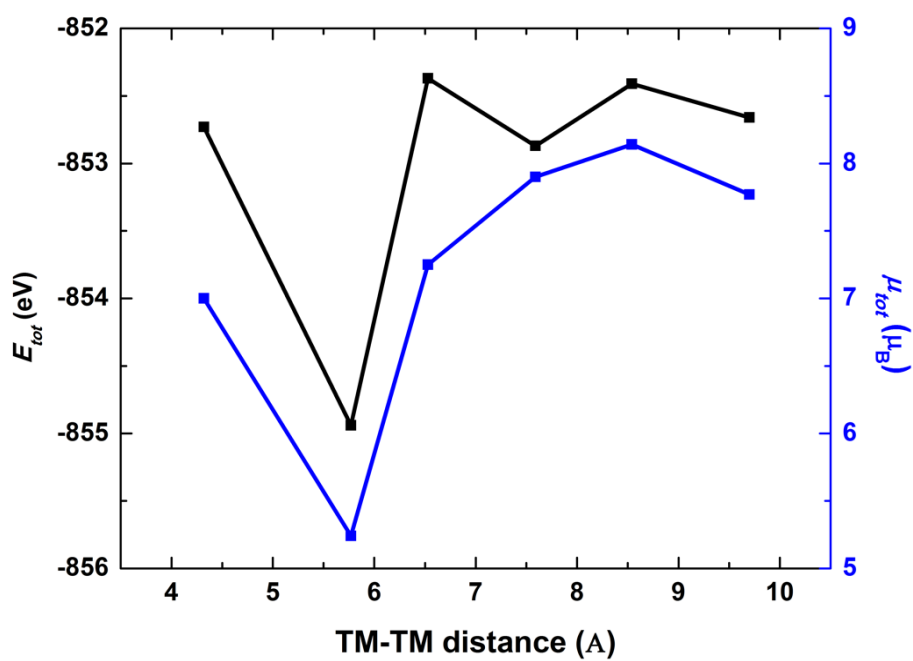

**Figure S48** Variation in the calculated total energies (in black) and magnetic moments (in blue) with the distance between TM atoms in the CrFe@DV system.

**Table S12** Barrier for diffusion (in eV) of the TM atom ($TM_1$ in **a** and $TM_2$ in **b**) from the vacancy center. The label corresponds to the position of the adsorbed TM atom in **Figure S49**.

| system | a | b | c | d | e | f |
|---|---|---|---|---|---|---|
| **CrFe@DV (a)** | 4.12 | 4.60 | 4.70 | 4.55 | 3.97 | 5.01 |
| **CrFe@DV (b)** | 5.22 | 4.84 | 4.80 | 4.51 | 4.39 | 7.51 |
| **CrMn@DV (a)** | 4.50 | 6.25 | 3.28 | 5.77 | 3.36 | 3.48 |
| **CrMn@DV (b)** | 2.85 | 2.95 | 3.06 | 3.03 | 3.11 | 3.19 |
| **FeMn@DV (a)** | 4.23 | 5.34 | 5.13 | 4.75 | 5.19 | 4.63 |
| **FeMn@DV (b)** | 4.26 | 5.38 | 5.02 | 4.83 | 5.29 | 4.78 |

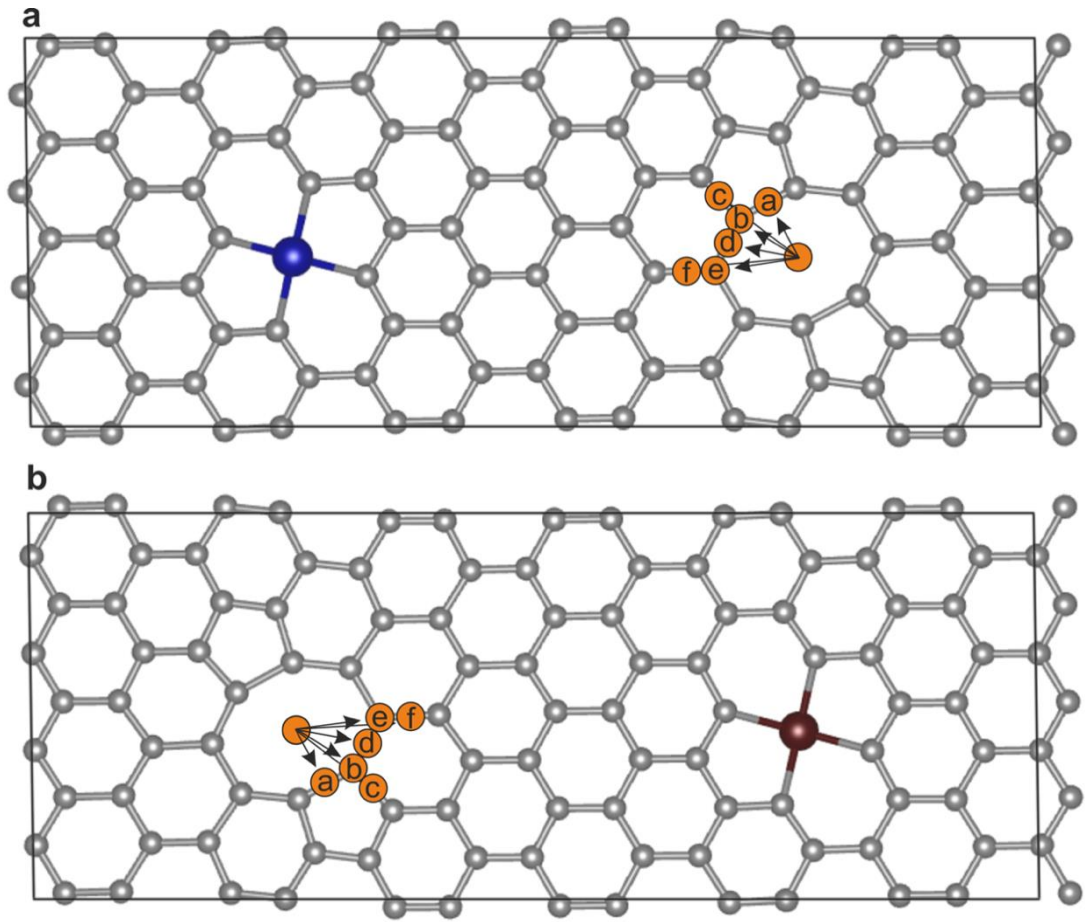

**Figure S49** Scheme showing a diffusion of the TM atom, $TM_1$ in **a)** and $TM_2$ in **b)**, from vacancy center. The label corresponds to the diffusion energy of the TM atom gathered in **Table S12**.

**Table S13** The binding energy $E_{bind}$ (eV), $TM_1$–$TM_2$ distance (Å), total magnetic moment $\mu_{tot}$ ($\mu_B$), magnetic moments of $TM_1$ and $TM_2$ $\mu_{TM_1}$, $\mu_{TM_2}$ ($\mu_B$), and $MAE_{(TE)}$ (in meV per computational cell) of TM atoms doped into both defect types (**Figure S50**).

| system | $E_{bind}$ | $TM_1$–$TM_2$ | $\mu_{tot}$ | $\mu_{TM_1}$, $\mu_{TM_2}$ | $MAE_{(TE)}$ |
|---|---|---|---|---|---|
| **$Cr_1Cr_2$@SVDV (a)** | -11.85 | 7.84 | 4.00 | 2.58, 2.61 | 0.09 |
| **$Cr_1Cr_2$@SVDV (b)** | -12.04 | 9.41 | 4.00 | 2.61. 2.48 | 8.12 |
| **CrFe@SVDV (c)** | -12.73 | 7.84 | 4.00 | 2.63, 1.67 | 1.60 |
| **CrFe@SVDV (d)** | -12.93 | 9.49 | 4.00 | 2.62, 1.63 | 1.25 |

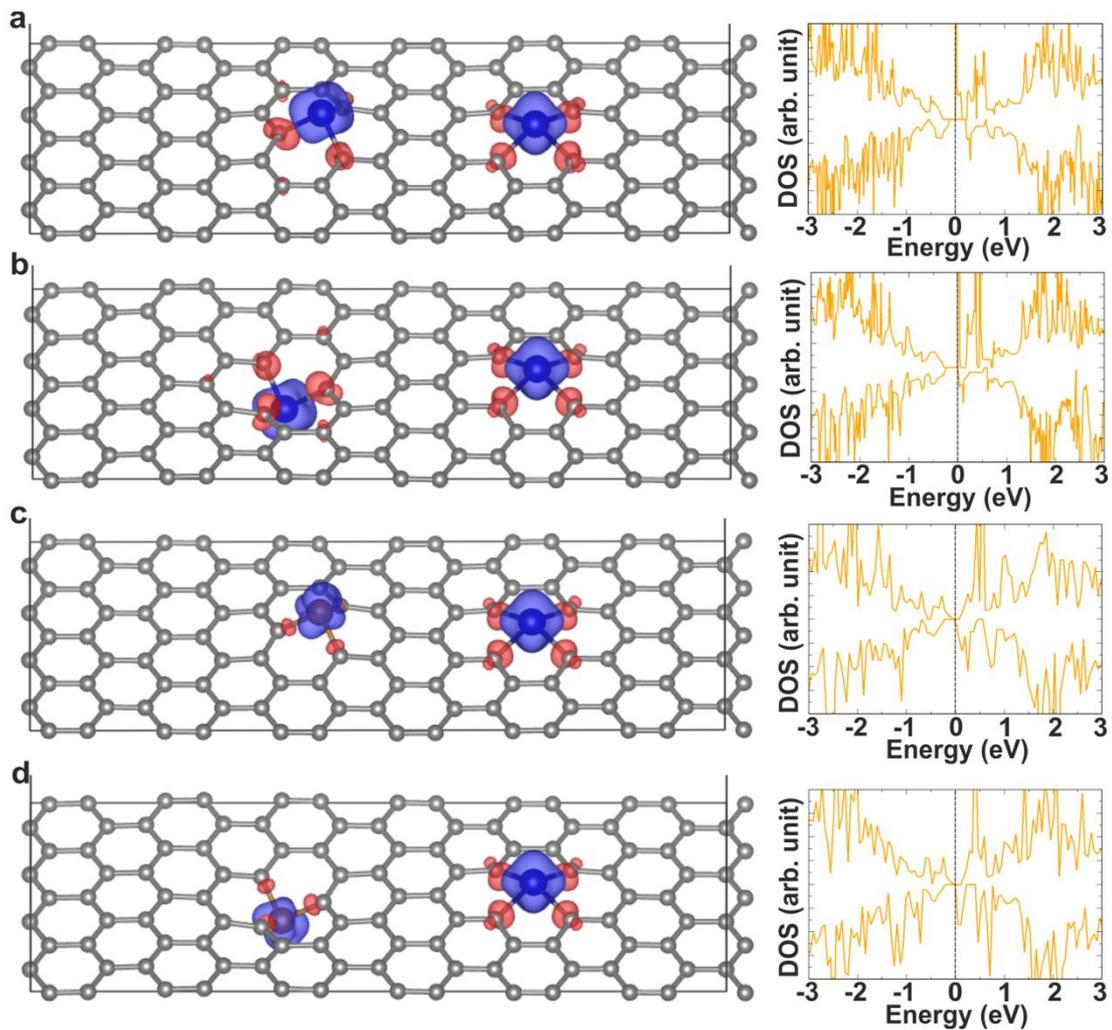

**Figure S50** Structure and density of states (DOS) for a system of two TM atoms, one embedded into SV and the other into DV defect (cf. **Table S13**). Spin densities corresponding to positive/negative magnetic moments are displayed in blue/red at an isosurface value of $\pm 0.01$ e$^-$Å$^{-3}$.

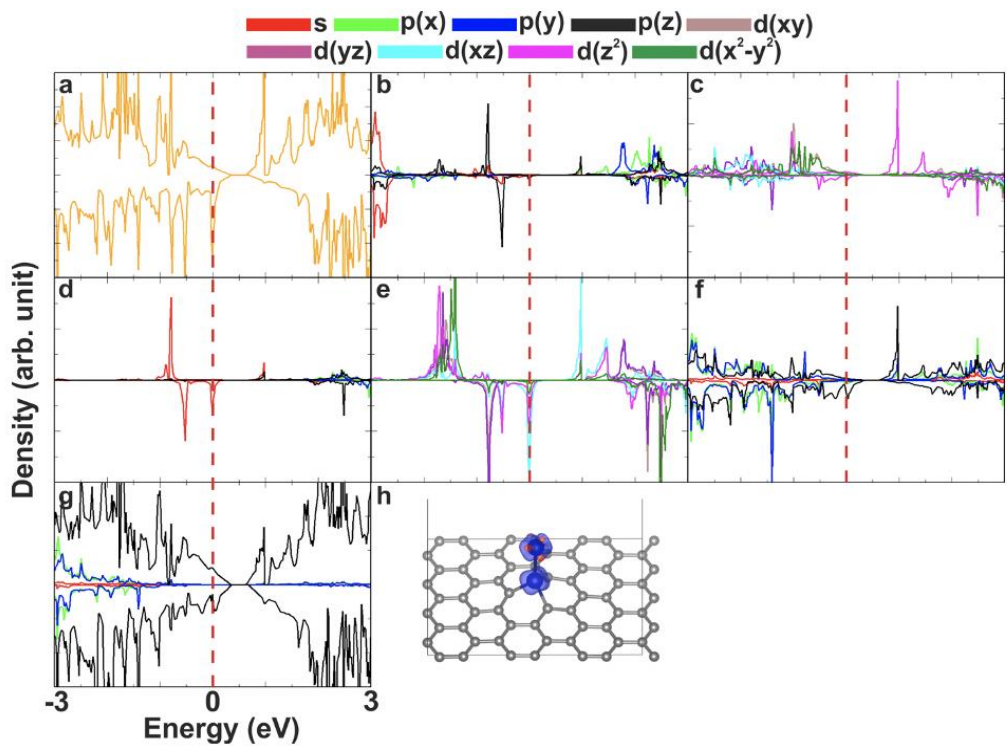

**Figure S51** Atom/orbital decomposed DOS of Cr@Cr@SV. **a)** total DOS, **b**,**c)** Cr atom, **d**,**e)** Cr atom, **f)** nearest carbon atoms, **g)** other carbon atoms, **h)** spin-polarized structure.

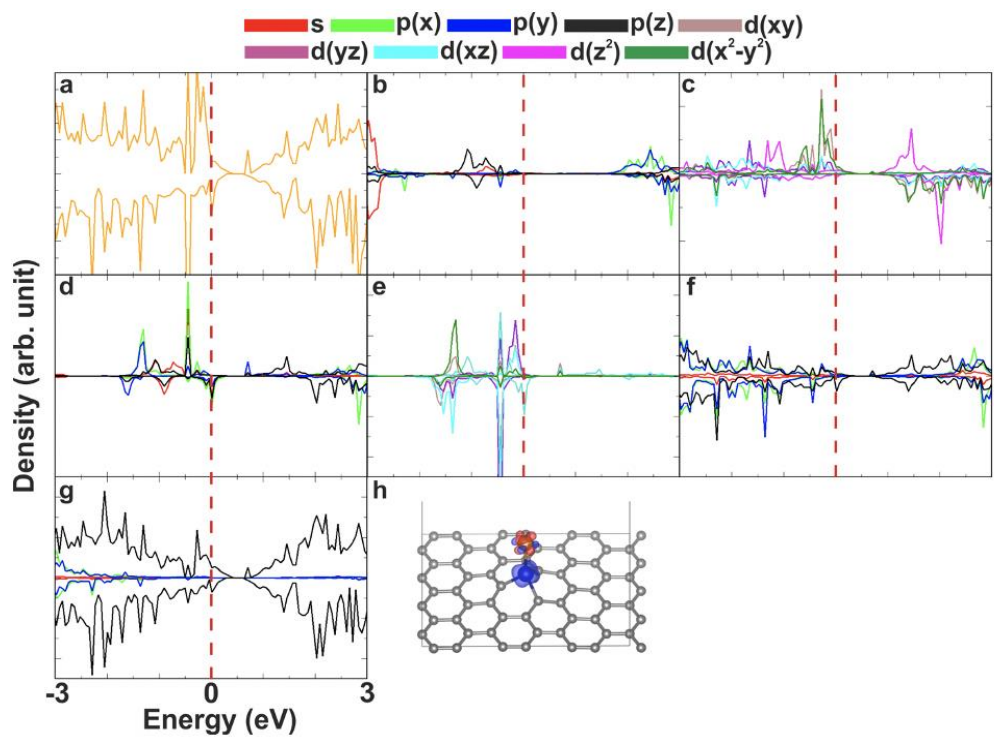

**Figure S52** Atom/orbital decomposed DOS of Fe@Cr@SV. **a)** total DOS, **b**,**c)** Cr atom, **d**,**e)** Fe atom, **f)** nearest carbon atoms, **g)** other carbon atoms, **h)** spin-polarized structure.

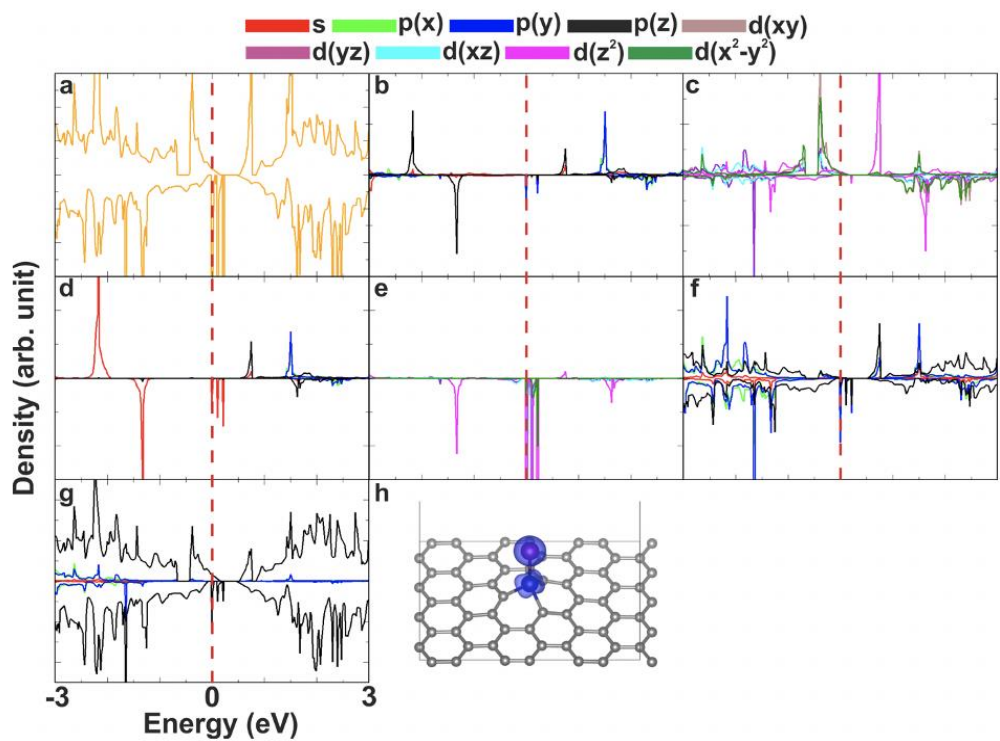

**Figure S53** Atom/orbital decomposed DOS of Mn@Cr@SV. **a)** total DOS, **b,c)** Cr atom, **d**,**e**) Mn atom, **f)** nearest carbon atoms, **g)** other carbon atoms, **h**) spin-polarized structure.

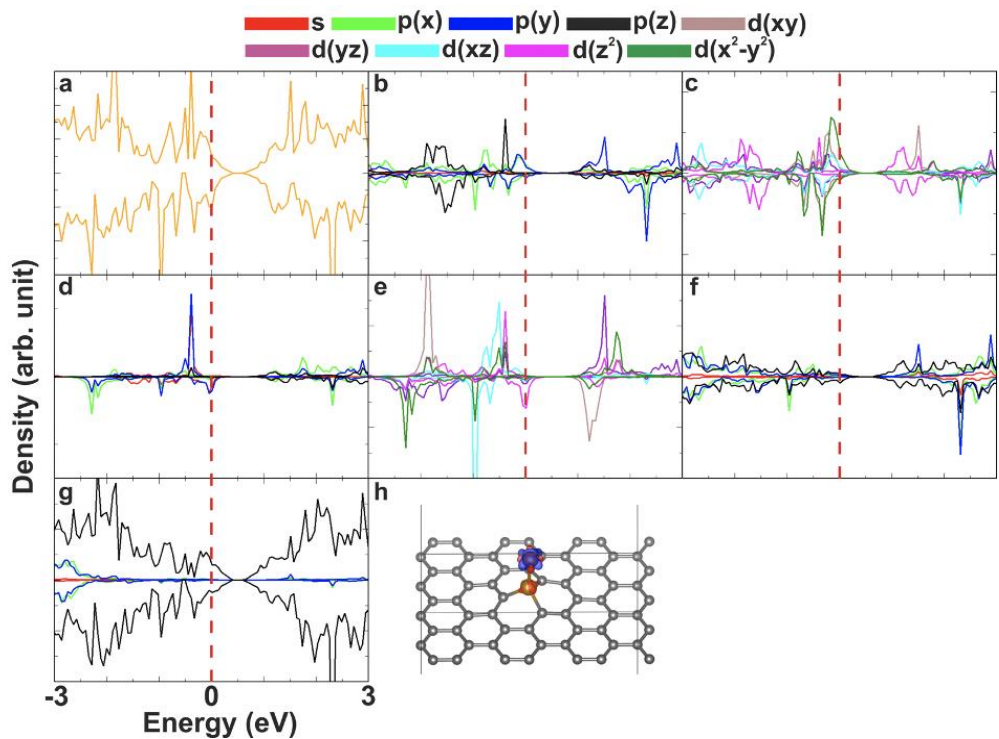

**Figure S54** Atom/orbital decomposed DOS of Fe@Fe@SV. **a)** total DOS, **b,c)** Fe atom, **d**,**e**) Fe atom, **f)** nearest carbon atoms, **g)** other carbon atoms, **h**) spin-polarized structure.

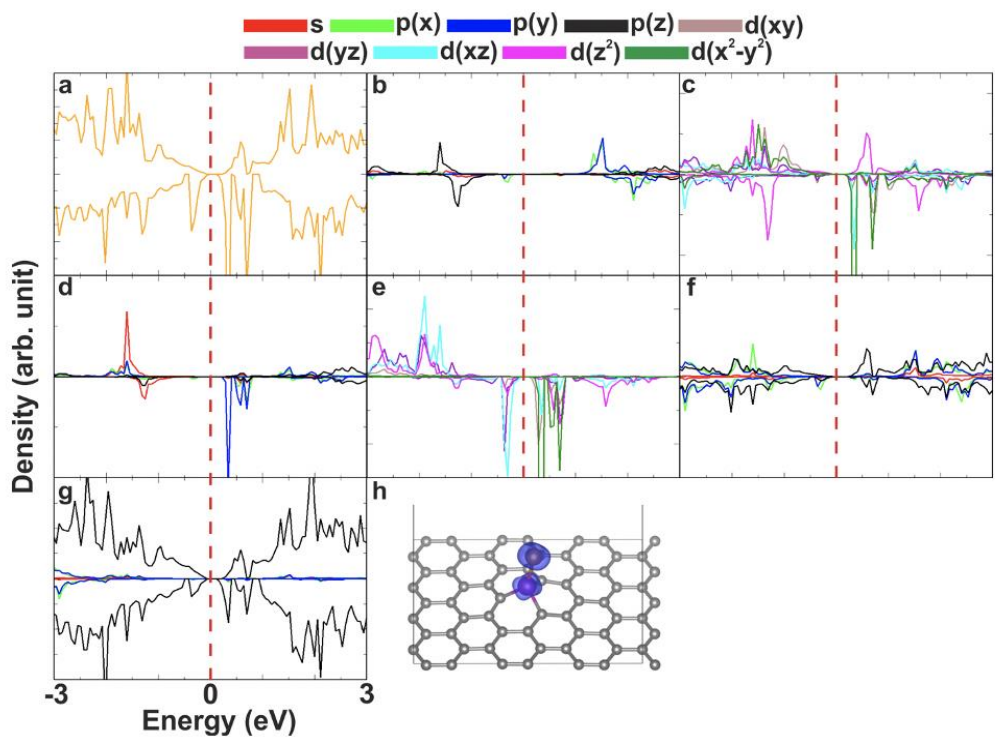

**Figure S55** Atom/orbital decomposed DOS of Fe@Mn@SV. **a)** total DOS, **b**,**c)** Mn atom, **d**,e**)** Fe atom, **f)** nearest carbon atoms, **g)** other carbon atoms, **h)** spin-polarized structure.

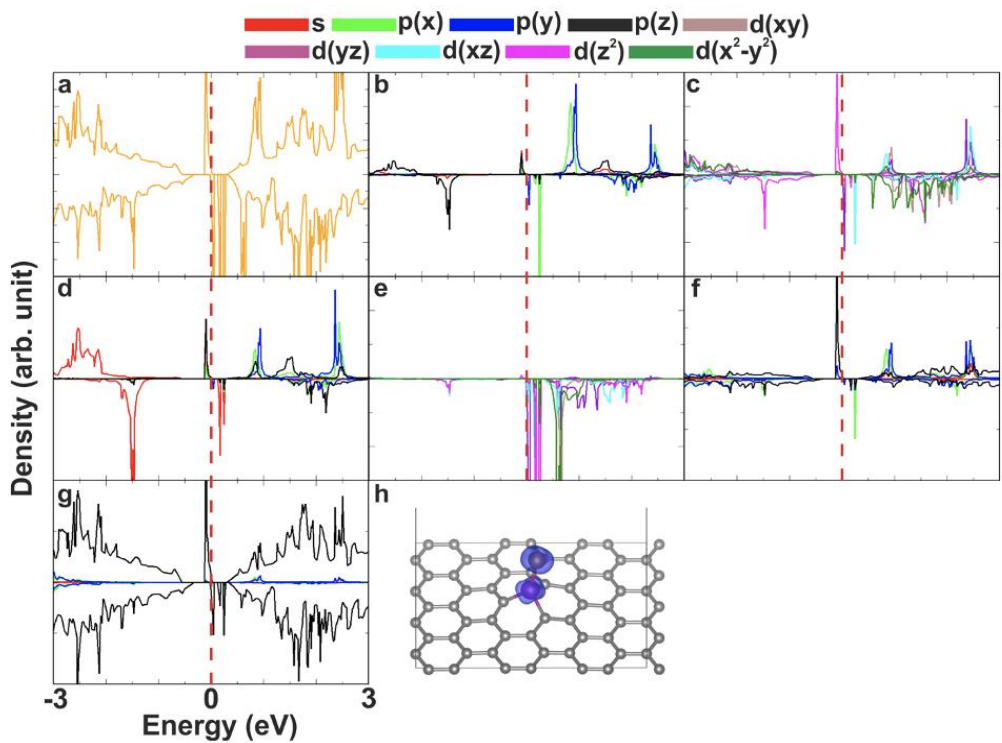

**Figure S56** Atom/orbital decomposed DOS of Mn@Mn@SV. **a)** total DOS, **b**,**c)** Mn atom, **d**,e**)** Mn atom, **f)** nearest carbon atoms, **g)** other carbon atoms, **h)** spin-polarized structure.

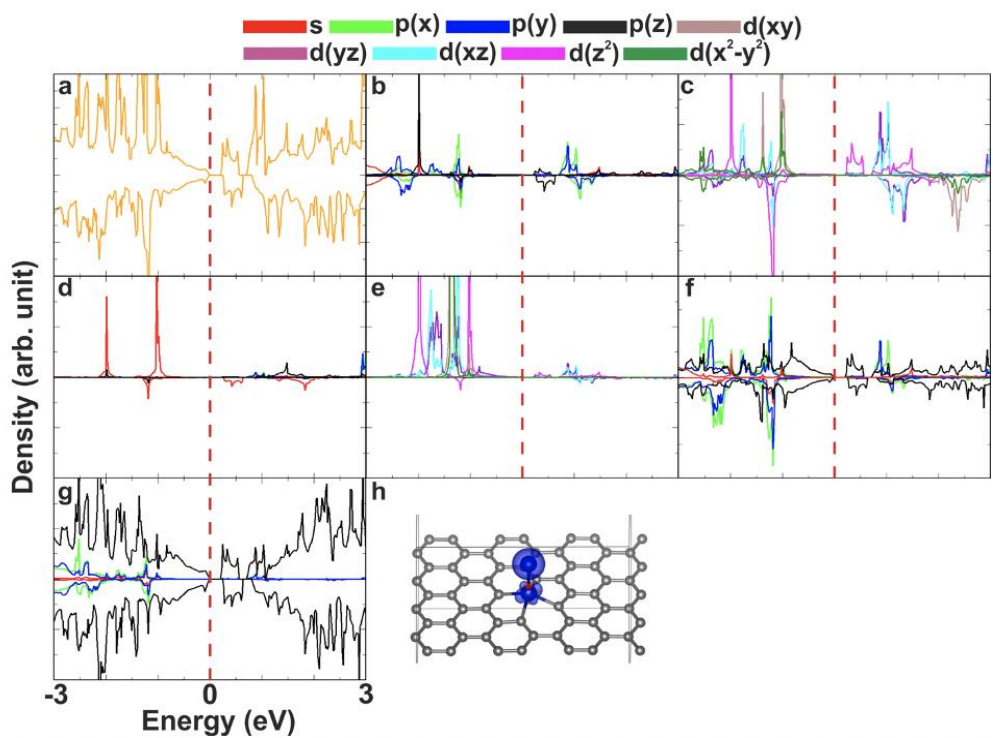

**Figure S57** Atom/orbital decomposed DOS of Cr@Cr@DV. **a)** total DOS, **b,c)** Cr atom, **d,e)** Cr atom, **f)** nearest carbon atoms, **g)** other carbon atoms, **h)** spin-polarized structure.

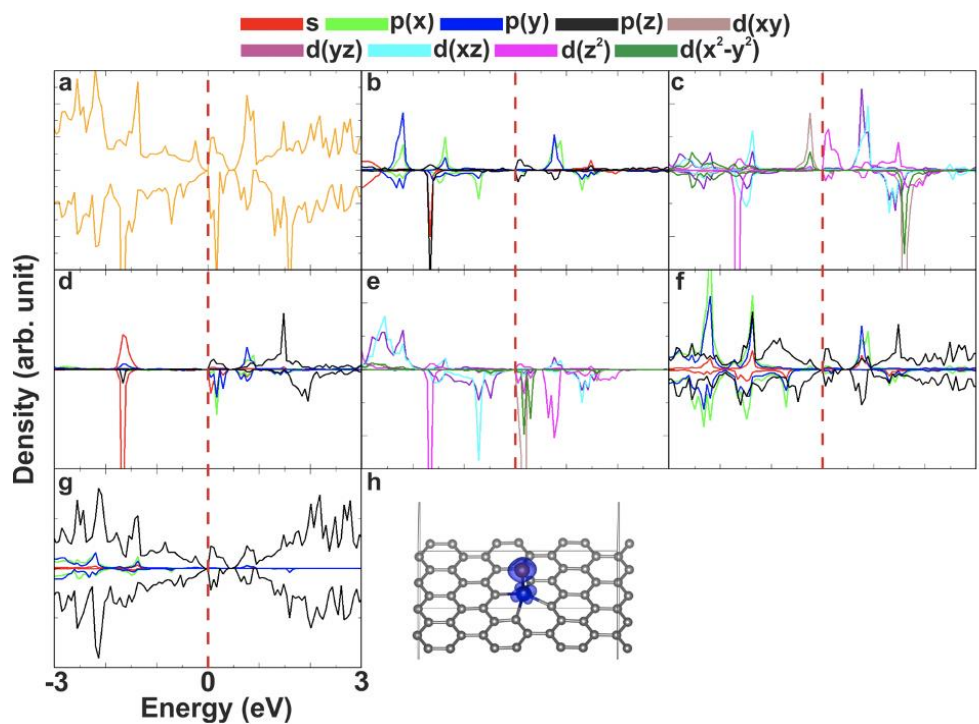

**Figure S58** Atom/orbital decomposed DOS of Fe@Cr@DV. **a)** total DOS, **b,c)** Cr atom, **d,e)** Fe atom, **f)** nearest carbon atoms, **g)** other carbon atoms, **h)** spin-polarized structure.

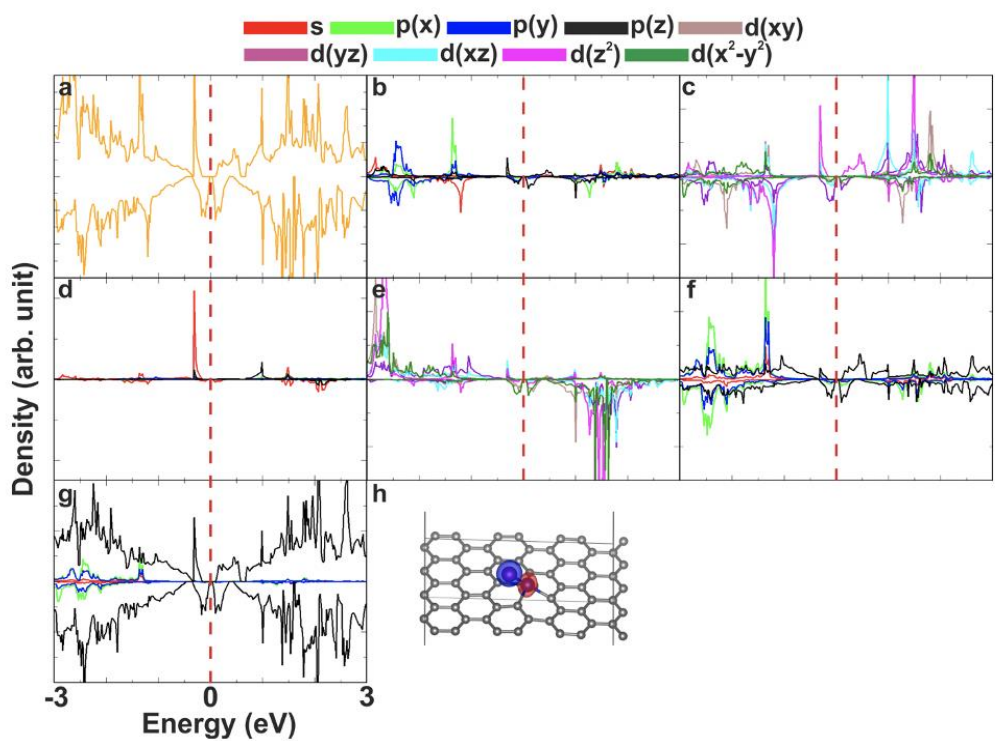

**Figure S59** Atom/orbital decomposed DOS of Mn@Cr@DV. **a)** total DOS, **b,c)** Cr atom, **d,e)** Mn atom, **f)** nearest carbon atoms, **g)** other carbon atoms, **h)** spin-polarized structure.

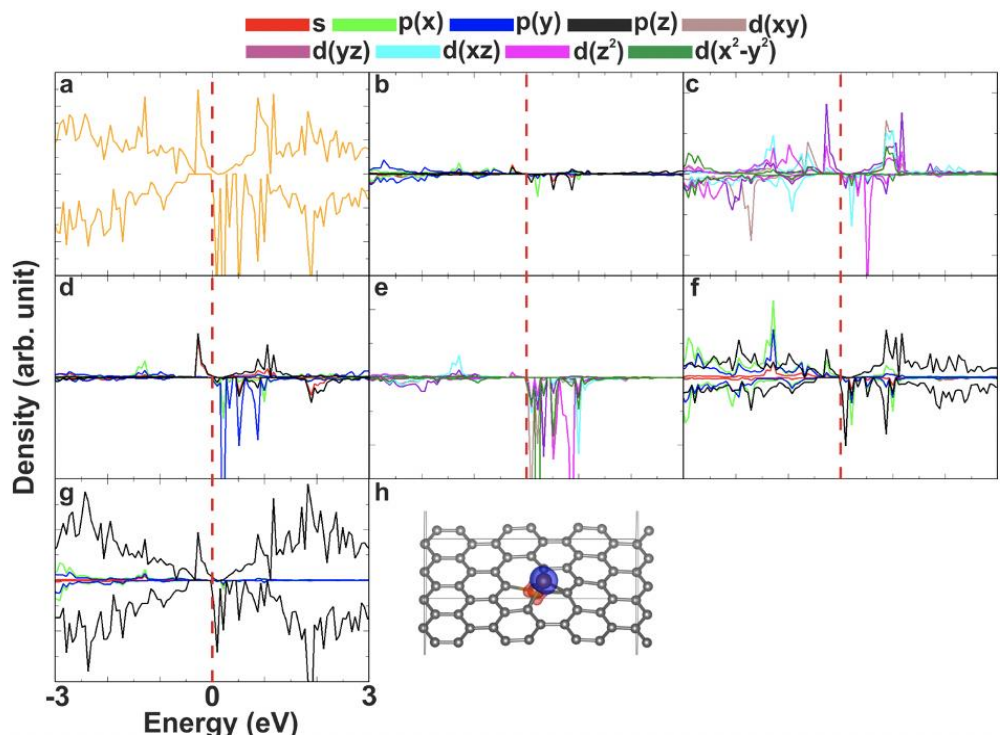

**Figure S60** Atom/orbital decomposed DOS of Fe@Fe@DV. **a)** total DOS, **b,c)** Fe atom, **d,e)** Fe atom, **f)** nearest carbon atoms, **g)** other carbon atoms, **h)** spin-polarized structure.

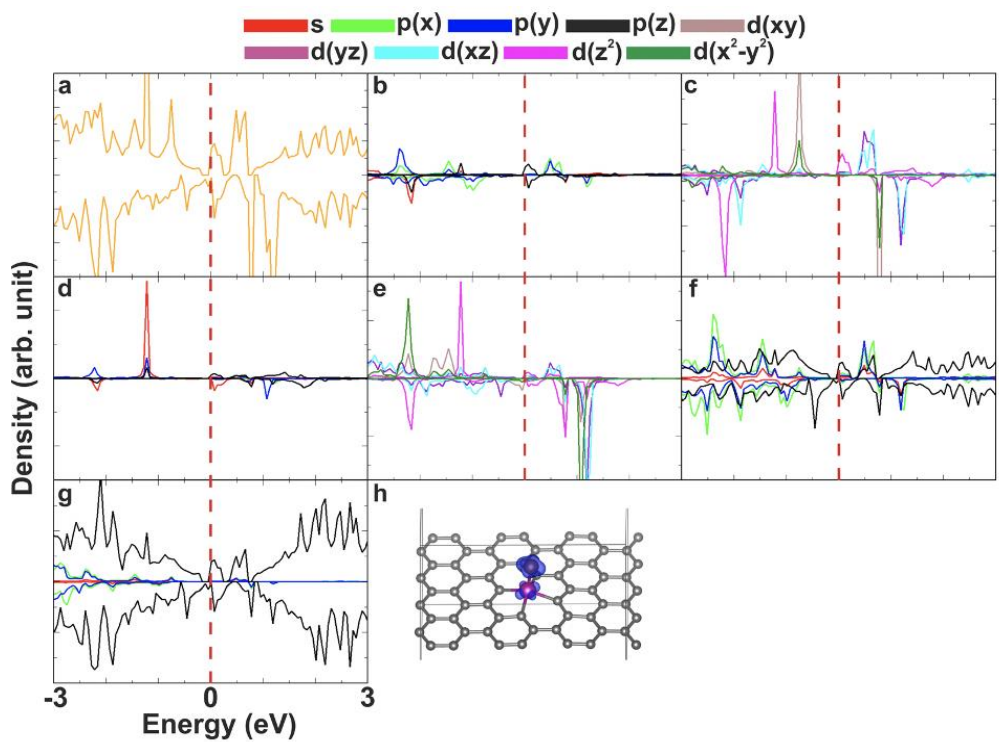

**Figure S61** Atom/orbital decomposed DOS of Fe@Mn@DV. **a)** total DOS, **b,c)** Mn atom, **d,e)** Fe atom, **f)** nearest carbon atoms, **g)** other carbon atoms, **h)** spin-polarized structure.

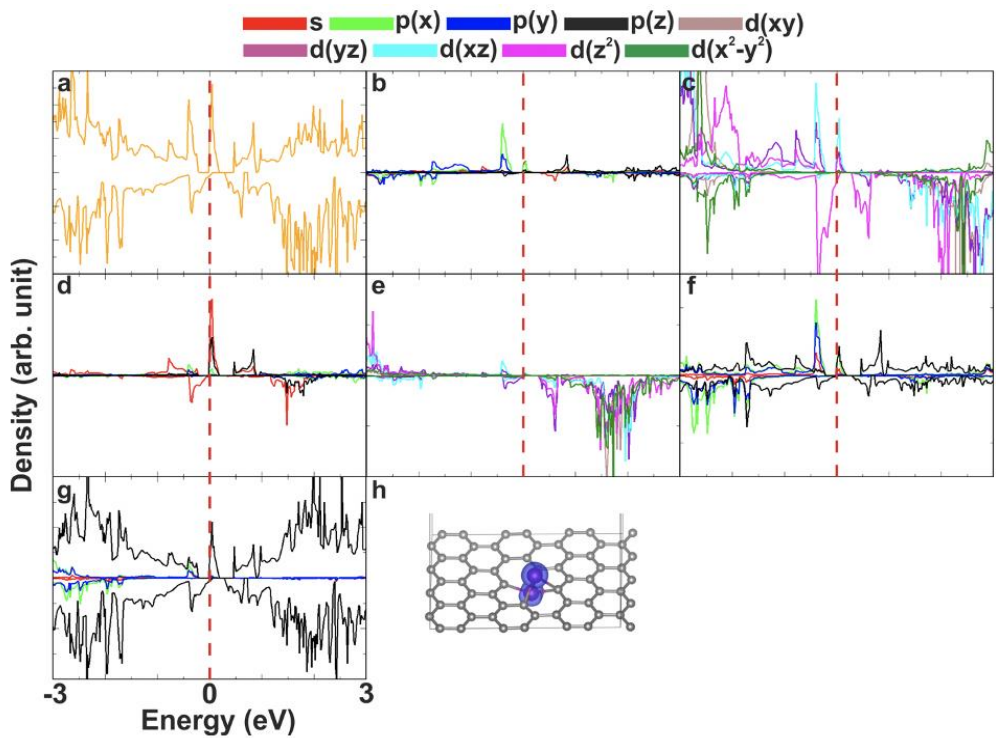

**Figure S62** Atom/orbital decomposed DOS of Mn@Mn@DV. **a)** total DOS, **b,c)** Mn atom, **d,e)** Mn atom, **f)** nearest carbon atoms, **g)** other carbon atoms, **h)** spin-polarized structure.

**Table S14** The binding energy $E_{bind}$ (eV) of a TM atom to graphene while the other TM atom binds to graphene on the other side, TM$_1$–TM$_2$ distance (Å), total magnetic moment $\mu_{tot}$ ($\mu_B$), magnetic moments $\mu_{TM_1}$, $\mu_{TM_2}$ of TM$_1$ and TM$_2$ ($\mu_B$), and MAE$_{(TE)}$ (per computational cell, in meV) (**Figure S63**).

| System | $E_{bind}$ | TM$_1$–TM$_2$ | $\mu_{tot}$ | $\mu_{TM_1}$, $\mu_{TM_2}$ | MAE$_{(TE)}$ |
|---|---|---|---|---|---|
| **FeMn@DV** | −6.66 | 2.18 | 2.51 | 2.10, 0.69 | −0.71 |
| **FeCr@SV** | −7.43 | 2.58 | 6.00 | 4.05, 1.61 | −1.45 |
| **FeFe@SV** | −8.02 | 2.46 | 4.00 | 1.21, 2.85 | 0.26 |
| **MnMn@SV** | −6.78 | 2.61 | 6.00 | 3.16, 3,18 | 0.84 |

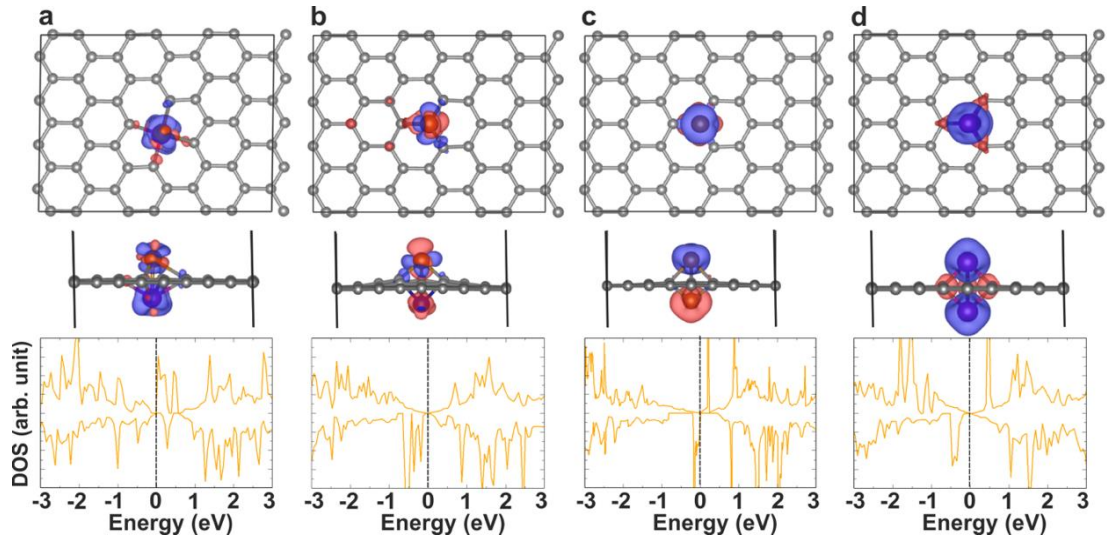

**Figure S63** Structures, spin densities and densities of states (DOS) of graphene with TM atoms binding to graphene, one above and one below the lattice. **a)** FeMn@DV, **b)** FeCr@SV, **c)** FeFe@SV, and **d)** MnMn@SV.

**Table S15** Total spin moments $\mu_{s\_tot}$, local spin moments $\mu_{s\_loc}$ and orbital moments $\mu_l$ (all in $\mu_B$) oriented along the "$x$", and "$z$" magnetization axes and their anisotropy ($\Delta$) for TM$_2$@TM$_1$@SV and dimers TM$_2$@TM$_1$@DV depicted in **Figures S51-S62**.

| system | $\mu_{s\_tot\,(x)}$ | $\mu_{s\_loc\,(x)}$ | $\mu_{l\,(x)}$ | $\mu_{s\_tot\,(z)}$ | $\mu_{s\_loc\,(z)}$ | $\mu_{l\,(z)}$ | $\mu_{s\_tot\,(\Delta)}$ | $\mu_{s\_loc(\Delta)}$ | $\mu_{l\,(\Delta)}$ |
|---|---|---|---|---|---|---|---|---|---|
| **Cr@Cr@SV** | 1.72 | 1.65 | 0.05 | 1.72 | 1.65 | 0.04 | 0.00 | 0.00 | 0.02 |
| **Fe@Cr@SV** | 1.68 | 1.74 | 0.02 | 1.65 | 1.70 | 0.04 | 0.03 | 0.05 | -0.02 |
| **Mn@Cr@SV** | 3.45 | 3.20 | 0.04 | 3.45 | 3.20 | 0.01 | 0.00 | 0.00 | 0.03 |
| **Fe@Fe@SV** | 1.09 | 1.24 | 0.14 | 1.03 | 1.09 | 0.11 | 0.06 | 0.15 | 0.03 |
| **Fe@Mn@SV** | 2.50 | 2.54 | 0.15 | 2.50 | 2.54 | 0.03 | 0.00 | 0.00 | 0.12 |
| **Mn@Mn@SV** | 1.81 | 1.56 | 0.01 | 1.96 | 1.71 | 0.02 | -0.16 | -0.15 | -0.01 |
| **Cr@Cr@DV** | 3.00 | 2.61 | 0.02 | 3.00 | 2.61 | 0.00 | 0.00 | 0.00 | 0.02 |
| **Fe@Cr@DV** | 2.00 | 1.89 | 0.13 | 2.00 | 1.89 | 0.02 | 0.00 | 0.00 | 0.11 |
| **Mn@Cr@DV** | 1.50 | 1.20 | 0.02 | 1.50 | 1.20 | 0.03 | 0.00 | 0.00 | -0.01 |
| **Fe@Fe@DV** | 1.05 | 0.97 | 0.07 | 1.07 | 0.99 | 0.09 | -0.02 | -0.02 | -0.02 |
| **Fe@Mn@DV** | 0.50 | 0.38 | 0.77 | 0.50 | 0.37 | 0.83 | 0.00 | 0.01 | -0.06 |
| **Mn@Mn@DV** | 4.00 | 3.62 | 0.01 | 4.00 | 3.62 | 0.01 | 0.00 | 0.00 | 0.00 |

**Table S16** The spin-up (spin-down) bandgaps (eV) of $TM_2@TM_1@SV$, $TM_2@TM_1@DV$.

| system | bandgap |
| --- | --- |
| **Cr@Cr@SV** | 0.0 (0.0) |
| **Fe@Cr@SV** | 0.0 (0.0) |
| **Mn@Cr@SV** | 0.5 (0.1) |
| **Fe@Fe@SV** | 0.0 (0.0) |
| **Fe@Mn@SV** | 0.3 (0.2) |
| **Mn@Mn@SV** | 0.3 (0.3) |
| **Cr@Cr@DV** | 0.2 (0.2) |
| **Fe@Cr@DV** | 0.0 (0.0) |
| **Mn@Cr@DV** | 0.3 (0.1) |
| **Fe@Fe@DV** | 0.0 (0.4) |
| **Fe@Mn@DV** | 0.0 (0.0) |
| **Mn@Mn@DV** | 0.3 (0.3) |

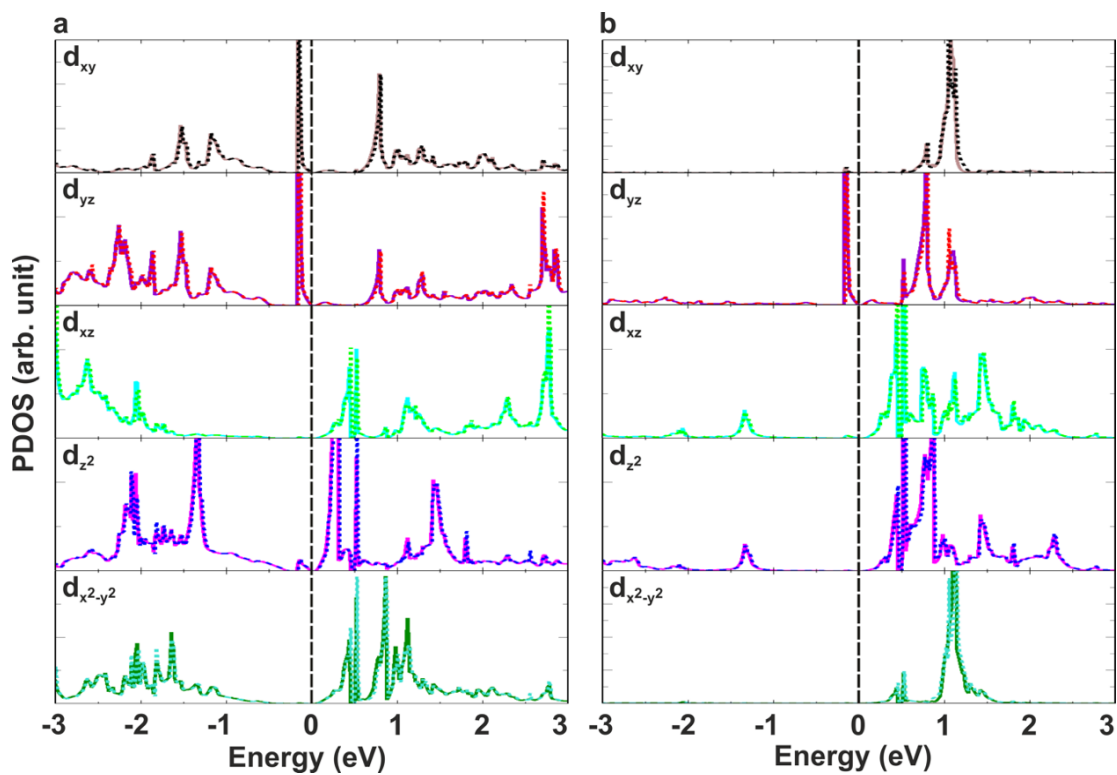

**Figure S64** Relativistic partial atom/orbital-resolved densities of states for Mn@Mn@SV for in-plane (solid lines) and perpendicular magnetization (dashed lines). **a)** Mn$_1$ atom and **b)** Mn$_2$ atom.

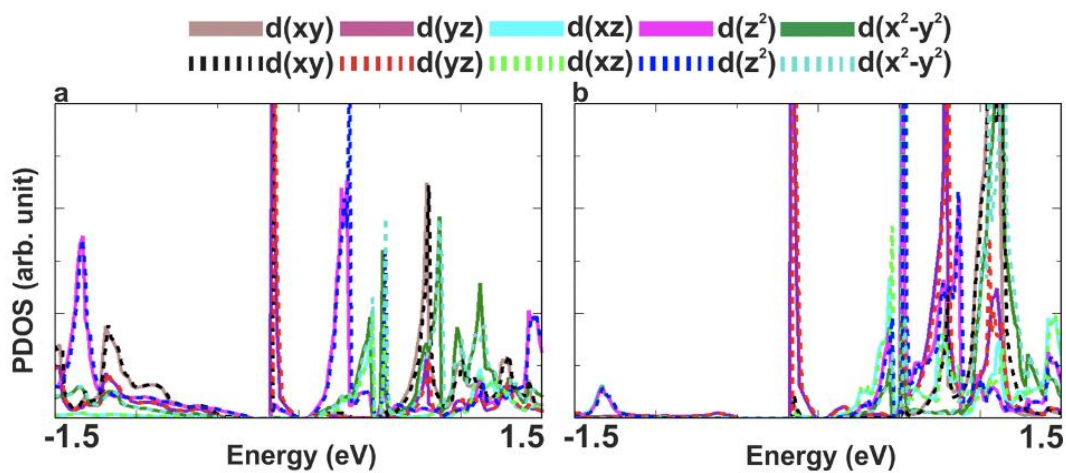

**Figure S65** The zoomed partial atom/orbital contribution of Mn@Mn@SV. **a)** Mn atom and **b)** Mn atom. Full lines denote in-plane magnetization, dashed lines denote perpendicular magnetization.

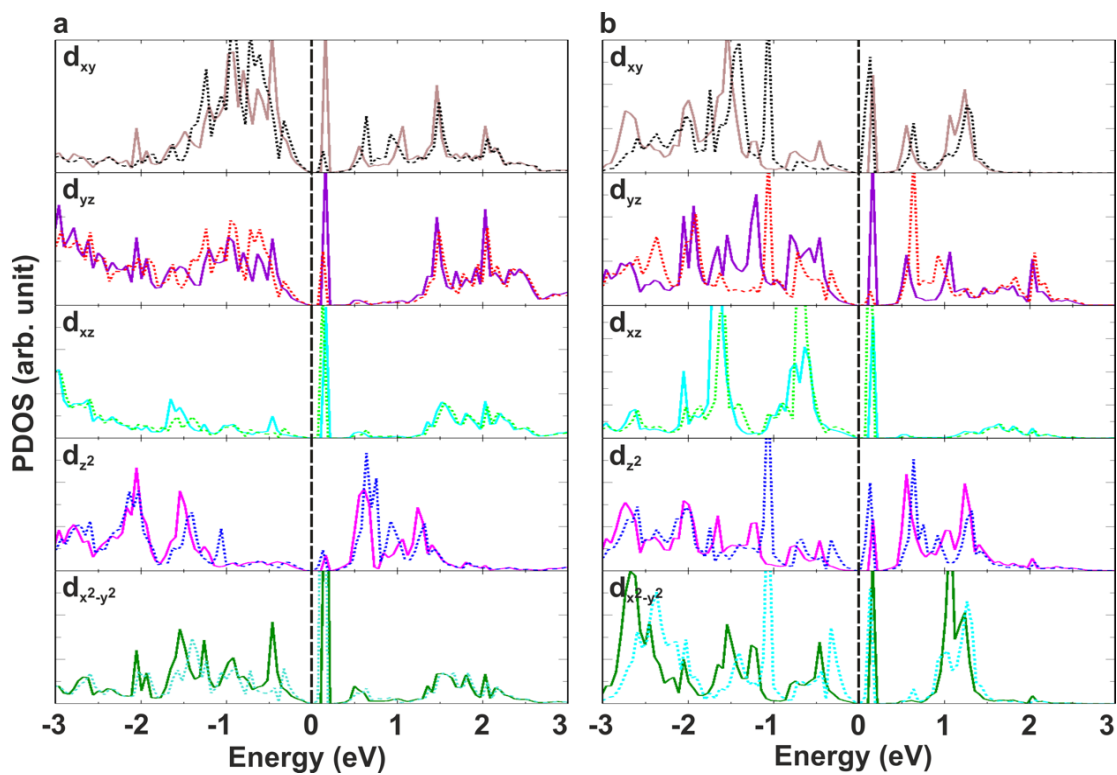

**Figure S66** Relativistic partial atom/orbital-resolved densities of states for Fe@Fe@SV for in-plane (solid lines) and perpendicular magnetization (dashed lines). **a)** Fe atom and **b)** Fe atom.

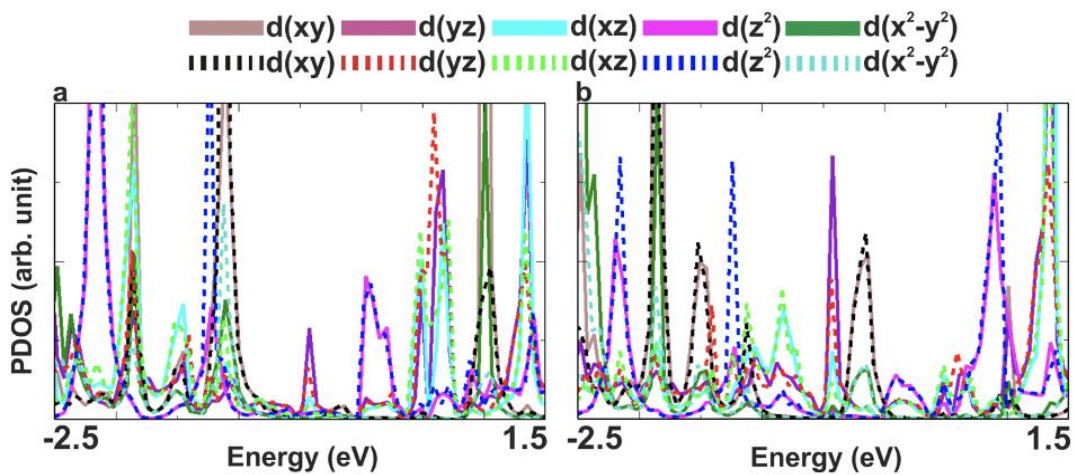

**Figure S67** The zoomed partial atom/orbital contribution of Fe@Mn@DV. **a)** Mn atom and **b)** Fe atom. Full lines denote in-plane magnetization, dashed lines denote perpendicular magnetization.

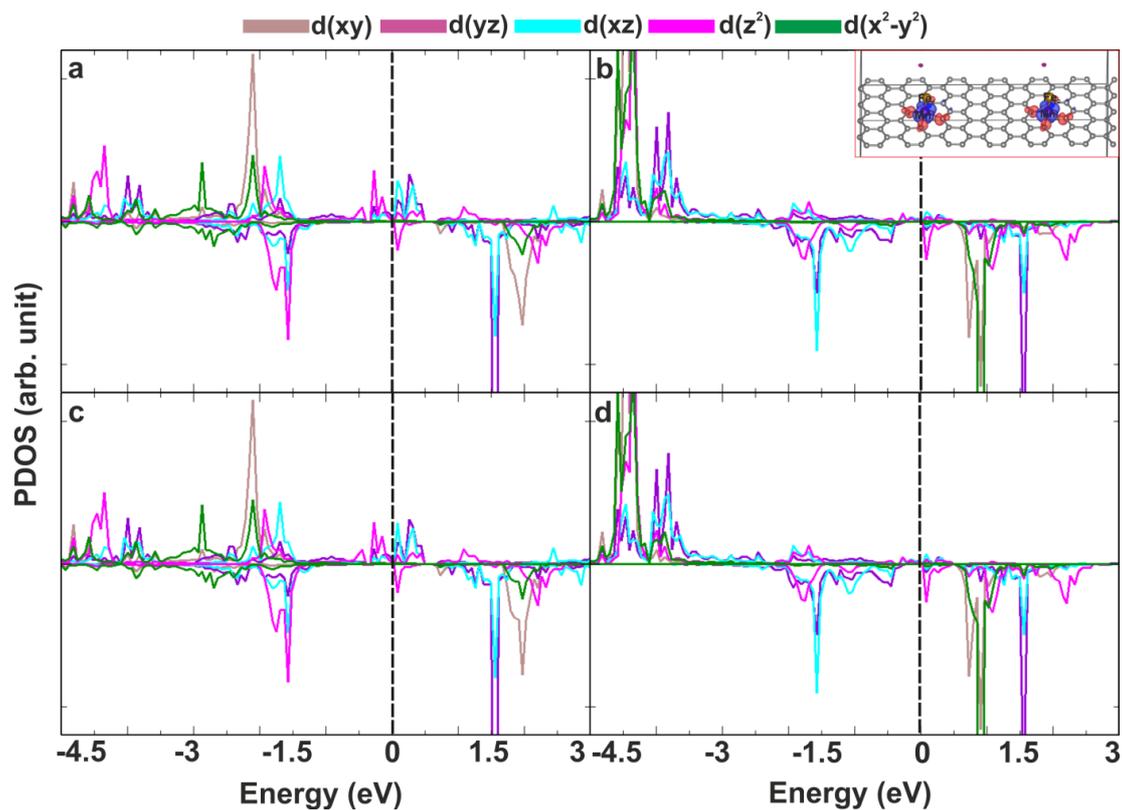

**Figure S68** Atom/orbital decomposed DOS for two Fe—Mn dimers bound to separate DV defects. **a)** Mn$_1$ atom **b)** Fe$_1$ atom, **c)** Mn$_2$ atom, **d)** Fe$_2$ atom. The inset shows the structural model and spin densities.

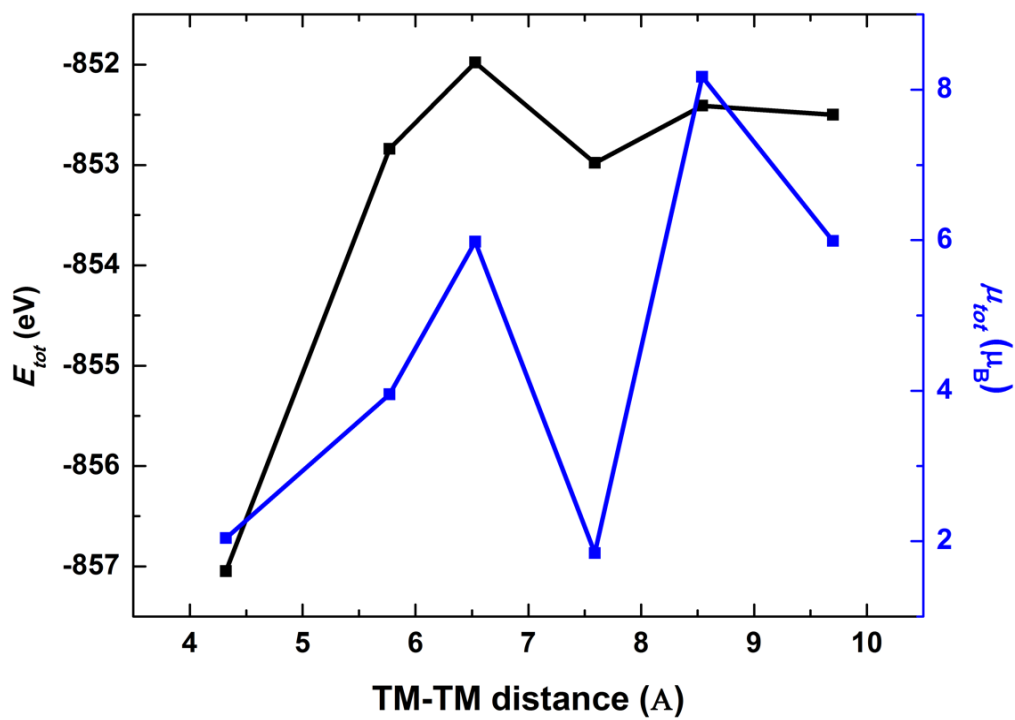

**Figure S69** Oscillations in the calculated total energies (in black) and magnetic moments (in blue) with the distance between TM atoms in the Fe@Mn@DV.

**Table S17** Barrier for diffusion (in eV) of the Fe–Mn dimer from the vacancy center. The label corresponds to the position of the adsorbed dimer in **Figure S70**.

| system | a | b | c | d | e | f |
|---|---|---|---|---|---|---|
| **Fe@Mn@NSV (a)** | 5.10 | 5.86 | 7.10 | 8.52 | 8.45 | |
| **FeMn@DV (b)** | 7.55 | 5.82 | 7.10 | 5.26 | 5.88 | 8.16 |

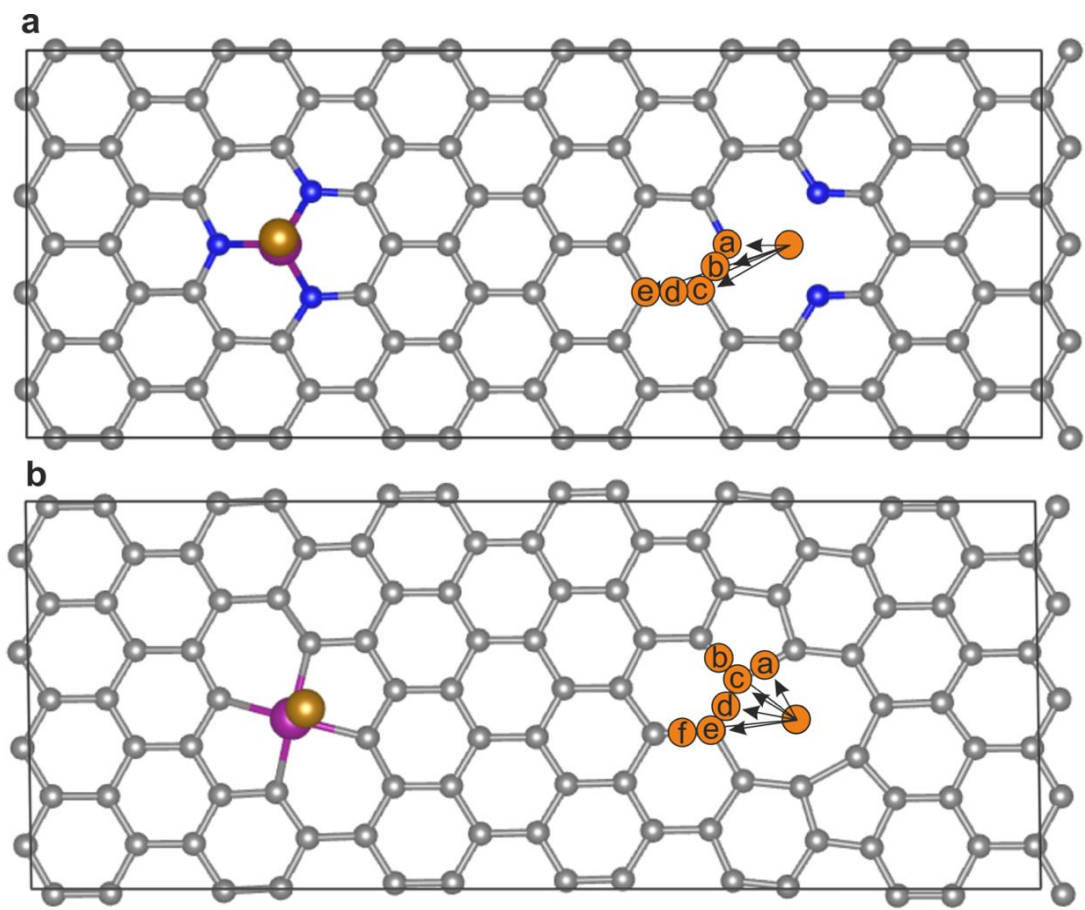

**Figure S70** Scheme showing a diffusion of the Fe–Mn dimer from the vacancy center. The label corresponds to the diffusion energy of the dimer gathered in **Table S17**. **a)** Fe@Mn@NSV and **b)** Fe@Mn@DV.

**Table S18** Last 6 frequencies of the Fe@Mn@DV system revealing imaginary frequencies only below 100 cm$^{-1}$.

| Number of frequency | Wavenumber (cm$^{-1}$) | Energy (meV) |
|:---:|:---:|:---:|
| **139 f** | 52.83 | 6.55 |
| **140 f** | 1.24 | 0.15 |
| **141 f/i** | 0.78 | 0.10 |
| **142 f/i** | 1.32 | 0.16 |
| **143 f/i** | 64.63 | 8.01 |
| **144 f/i** | 81.16 | 10.06 |

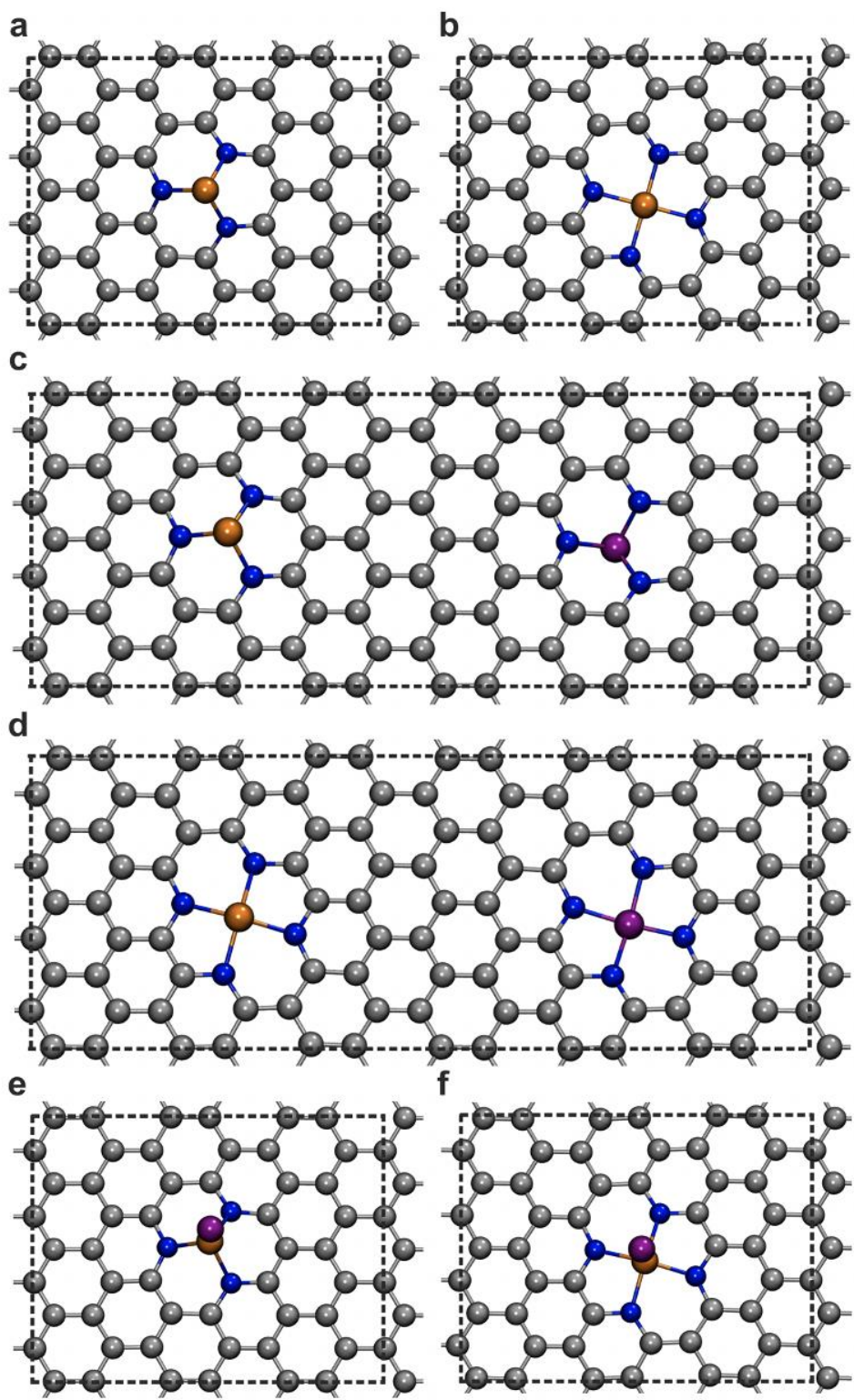

**Figure S71 a)** TM@NSV, **b)** TM@NDV, **c)** $TM_1TM_2$@NSV, **d)** $TM_1TM_2$@NDV, **e)** $TM_2$@$TM_1$@NSV, **f)** $TM_2$@$TM_1$@NDV. Nitrogen atoms are in blue, carbon atoms in grey, TM atoms in orange/purple.

**Table S19** The binding energy $E_{bind}$ (eV), TM–N bond length (Å), total magnetic moment $\mu_{tot}$ ($\mu_B$), magnetic moment of TM $\mu_{TM}$ ($\mu_B$), magnetic moment of N atoms $\mu_N$ ($\mu_B$), Bader charges of TM (e⁻), MAE$_{(TE)}$ and MAE$_{(FT)}$ (both in meV) of TM@NSV and TM@NDV (**Figures S71a,b**).

| system | $E_{bind}$ | TM–N | $\mu_{tot}$ | $\mu_{TM}$ | $\mu_N$ | Bader | MAE$_{(TE)}$ | MAE$_{(FT)}$ |
|---|---|---|---|---|---|---|---|---|
| **Cr@NSV** | -2.74 | 1.82 | 1.39 | 2.16 | -0.06 | 1.20 | 0.10 | -25.17 |
| **Mn@NSV** | -2.40 | 1.79 | 0.97 | 0.90 | -0.01 | 1.04 | -0.30 | -24.82 |
| **Fe@NSV** | -3.33 | 1.76 | 3.00 | 3.02 | 0.05 | 1.01 | 0.62 | 10.88 |
| **Cr@NDV** | -5.92 | 1.95 | 4.00 | 3.53 | -0.05 | 1.25 | 0.10 | -19.11 |
| **Mn@NDV** | -6.05 | 1.91 | 3.00 | 3.07 | -0.04 | 1.29 | 0.84 | 13.87 |
| **Fe@NDV** | -6.95 | 1.89 | 2.00 | 1.97 | -0.02 | 1.17 | -0.28 | 5.57 |

**Table S20** Total spin moments $\mu_{s\_tot}$, local spin moments $\mu_{s\_loc}$ and orbital moments $\mu_l$ (all in $\mu_B$) oriented along the "$x$", and "$z$" magnetization axes and their anisotropy ($\Delta$) for TM@NSV and TM@NDV (**Figures S71a,b**).

| system | $\mu_{s\_tot\,(x)}$ | $\mu_{s\_loc\,(x)}$ | $\mu_{l\,(x)}$ | $\mu_{s\_tot\,(z)}$ | $\mu_{s\_loc\,(z)}$ | $\mu_{l\,(z)}$ | $\mu_{s\_tot\,(\Delta)}$ | $\mu_{s\_loc(\Delta)}$ | $\mu_{l\,(\Delta)}$ |
|---|---|---|---|---|---|---|---|---|---|
| **Cr@NSV** | 0.68 | 0.79 | 0.02 | 0.68 | 0.79 | 0.01 | 0.00 | 0.00 | 0.01 |
| **Mn@NSV** | 0.49 | 0.43 | 0.04 | 0.49 | 0.43 | 0.00 | 0.00 | 0.00 | 0.04 |
| **Fe@NSV** | 1.51 | 1.46 | 0.05 | 1.51 | 1.46 | 0.05 | 0.00 | 0.00 | 0.00 |
| **Cr@NDV** | 0.68 | 0.79 | 0.02 | 0.68 | 0.79 | 0.01 | 0.00 | 0.00 | 0.01 |
| **Mn@NDV** | 1.50 | 1.42 | 0.04 | 1.50 | 1.42 | 0.00 | 0.00 | 0.00 | 0.04 |
| **Fe@NDV** | 1.00 | 0.95 | 0.10 | 1.00 | 0.95 | 0.03 | 0.00 | 0.00 | 0.07 |

**Table S21** The spin-up (spin-down) bandgaps (eV) of TM@NSV, TM@NDV, TM$_1$TM$_2$@NSV, TM$_1$TM$_2$@NDV, TM$_2$@TM$_1$@NSV, TM$_2$@TM$_1$@NDV.

| system | bandgap |
|:---:|:---:|
| **Cr@NSV** | 0.0 (0.0) |
| **Mn@NSV** | 0.0 (0.0) |
| **Fe@NSV** | 0.3 (0.3) |
| **Cr@NDV** | 0.3 (0.1) |
| **Mn@NDV** | 0.4 (0.3) |
| **Fe@NDV** | 0.2 (0.2) |
| **CrFe@NSV** | 0.0 (0.1) |
| **FeMn@NDV** | 0.0 (0.1) |
| **Fe@Mn@NSV** | 0.0 (0.0) |
| **Fe@Fe@NDV** | 0.0 (0.2) |

**Table 22** The binding energy $E_{bind}$ (eV), TM$_1$-TM$_2$ distance (Å), total magnetic moment $\mu_{tot}$ ($\mu_B$), magnetic moments of TM$_1$ and TM$_2$ $\mu_{TM1}$, $\mu_{TM2}$ ($\mu_B$), and MAE$_{(TE)}$ and MAE$_{(FT)}$ (per computational cell, both in meV) of TM$_1$TM$_2$@NSV and TM$_1$TM$_2$@NDV (**Figures S71c,d**).

| Structure | $E_{bind}$ | TM$_1$-TM$_2$ | $\mu_{tot}$ | $\mu_{TM1}, \mu_{TM2}$ | MAE$_{(TE)}$ | MAE$_{(FT)}$ |
|---|---|---|---|---|---|---|
| **CrFe@NSV** | -7.69 | 13.18 | 4.86 | 2.18, 3.09 | 0.73 | 0.43 |
| **CrMn@NSV** | -6.13 | 12.84 | 2.58 | 2.20, 0.91 | -0.21 | -0.17 |
| **FeMn@NSV** | -6.42 | 13.31 | 8.71 | 3.31, 4.14 | 5.18 | 0.57 |
| **CrFe@NDV** | -12.88 | 12.95 | 6.00 | 3.50, 2.15 | 0.36 | -32.10 |
| **CrMn@NDV** | -13.12 | 12.88 | 7.00 | 3.52, 3.10 | 1.28 | 0.98 |
| **FeMn@NDV** | -13.89 | 12.80 | 5.00 | 2.06, 3.09 | 0.87 | 0.81 |

**Table S23** Total spin moments $\mu_{s\_tot}$, local spin moments $\mu_{s\_loc}$ and orbital moments $\mu_l$ (all in $\mu_B$) oriented along the "$x$", and "$z$" magnetization axes and their anisotropy ($\Delta$) for TM$_1$TM$_2$@NSV and TM$_1$TM$_2$@NDV (**Figures S71c,d**).

| system | $\mu_{s\_tot\,(x)}$ | $\mu_{s\_loc\,(x)}$ | $\mu_{l\,(x)}$ | $\mu_{s\_tot\,(z)}$ | $\mu_{s\_loc\,(z)}$ | $\mu_{l\,(z)}$ | $\mu_{s\_tot\,(\Delta)}$ | $\mu_{s\_loc(\Delta)}$ | $\mu_{l\,(\Delta)}$ |
|---|---|---|---|---|---|---|---|---|---|
| **CrFe@NSV** | 2.52 | 2.44 | 0.05 | 2.50 | 2.44 | 0.11 | 0.01 | 0.00 | -0.06 |
| **CrMn@NSV** | 1.24 | 1.25 | 0.01 | 1.23 | 1.25 | 0.01 | 0.01 | 0.00 | 0.00 |
| **FeMn@NSV** | 1.50 | 1.35 | 0.01 | 1.55 | 1.38 | 0.08 | -0.05 | -0.03 | -0.06 |
| **CrFe@NDV** | 3.00 | 2.79 | 0.08 | 3.00 | 2.79 | 0.00 | 0.00 | 0.00 | 0.08 |
| **CrMn@NDV** | 3.50 | 3.20 | 0.02 | 3.50 | 3.20 | 0.02 | 0.00 | 0.00 | 0.00 |
| **FeMn@NDV** | 2.50 | 2.42 | 0.12 | 2.50 | 2.42 | 0.02 | 0.00 | 0.00 | 0.10 |

**Table S24** The binding energy $E_{bind}$ (eV), TM$_1$–TM$_2$ distance (Å), total magnetic moment $\mu_{tot}$ ($\mu_B$), magnetic moments of TM$_1$ and TM$_2$ $\mu_{TM1}$, $\mu_{TM2}$ ($\mu_B$), and MAE$_{(TE)}$ and MAE$_{(FT)}$ (per computational cell, both in meV) of dimers TM$_2$@TM$_1$@NSV and TM$_2$@TM$_1$@NDV (**Figures S71e,f**).

| Structure | $E_{bind}$ | TM$_1$–TM$_2$ | $\mu_{tot}$ | $\mu_{TM1}$, $\mu_{TM2}$ | MAE$_{(TE)}$ | MAE$_{(FT)}$ |
|---|---|---|---|---|---|---|
| **Cr@Cr@NSV** | -0.63 | 2.67 | 3.78 | 2.99, 0.21 | -17.62 | -0.89 |
| **Fe@Cr@NSV** | -2.48 | 2.09 | 0.48 | 3.22, -3.26 | -6.16 | -0.13 |
| **Mn@Cr@NSV** | -0.44 | 2.14 | 7.54 | 2.92, 3.91 | 0.67 | 0.23 |
| **Fe@Fe@NSV** | -1.93 | 2.09 | 2.92 | -0.25, 3.17 | -2.73 | -1.96 |
| **Fe@Mn@NSV** | -2.77 | 2.39 | 1.51 | 3.98, -3.00 | -6.46 | 0.23 |
| **Mn@Mn@NSV** | -1.10 | 2.41 | 5.62 | 1.15, 4.18 | -1.15 | -0.57 |
| **Cr@Cr@NDV** | -0.64 | 1.95 | 4.00 | -0.16, 3.34 | -0.15 | -0.40 |
| **Fe@Cr@NDV** | -1.86 | 2.23 | 6.00 | 2.19, 3.00 | -1.61 | -1.32 |
| **Mn@Cr@NDV** | -1.62 | 2.16 | 2.78 | -2.41, 4.05 | -0.21 | -0.17 |
| **Fe@Fe@NDV** | -1.91 | 2.14 | 4.00 | 0.57, 3.03 | -0.68 | -0.60 |
| **Fe@Mn@NDV** | -1.42 | 2.36 | 5.23 | 2.28, 3.44 | 0.96 | 0.32 |
| **Mn@Mn@NDV** | -1.40 | 2.57 | 8.00 | 2.74, 4.39 | 0.50 | 0.34 |

**Table S25** Total spin moments $\mu_{s\_tot}$, local spin moments $\mu_{s\_loc}$ and orbital moments $\mu_l$ (all in $\mu_B$) oriented along the "$x$", and "$z$" magnetization axes and their anisotropy ($\Delta$) for dimers $TM_2@TM_1@NSV$ and $TM_2@TM_1@NDV$ (**Figures S71e,f**).

| system | $\mu_{s\_tot\,(x)}$ | $\mu_{s\_loc\,(x)}$ | $\mu_{l\,(x)}$ | $\mu_{s\_tot\,(z)}$ | $\mu_{s\_loc\,(z)}$ | $\mu_{l\,(z)}$ | $\mu_{s\_tot\,(\Delta)}$ | $\mu_{s\_loc(\Delta)}$ | $\mu_{l\,(\Delta)}$ |
|---|---|---|---|---|---|---|---|---|---|
| **Cr@Cr@NSV** | 2.04 | 1.84 | 0.18 | 2.10 | 1.87 | 0.09 | -0.05 | -0.03 | 0.10 |
| **Fe@Cr@NSV** | 0.20 | 0.02 | 0.11 | 0.20 | 0.05 | 0.14 | -0.01 | -0.03 | -0.03 |
| **Mn@Cr@NSV** | 3.77 | 3.43 | 0.01 | 3.77 | 3.43 | 0.02 | 0.00 | 0.01 | -0.01 |
| **Fe@Fe@NSV** | 1.45 | 1.56 | 0.12 | 1.45 | 1.56 | 0.05 | 0.00 | 0.00 | 0.07 |
| **Fe@Mn@NSV** | 0.69 | 0.49 | 0.13 | 0.69 | 0.51 | 0.14 | 0.00 | -0.03 | -0.01 |
| **Mn@Mn@NSV** | 2.82 | 2.63 | 0.02 | 2.82 | 2.63 | 0.04 | 0.00 | 0.00 | -0.02 |
| **Cr@Cr@NDV** | 2.00 | 1.70 | 0.01 | 2.00 | 1.70 | 0.02 | 0.00 | 0.00 | -0.01 |
| **Fe@Cr@NDV** | 3.00 | 2.71 | 0.12 | 3.00 | 2.71 | 0.01 | 0.00 | 0.00 | 0.11 |
| **Mn@Cr@NDV** | 1.39 | 1.02 | 0.04 | 1.39 | 1.02 | 0.01 | 0.00 | 0.00 | 0.03 |
| **Fe@Fe@NDV** | 2.00 | 1.87 | 0.12 | 2.00 | 1.87 | 0.11 | 0.00 | 0.00 | 0.01 |
| **Fe@Mn@NDV** | 2.58 | 2.69 | 0.13 | 2.61 | 2.69 | 0.08 | -0.03 | -0.01 | 0.05 |
| **Mn@Mn@NDV** | 4.00 | 3.59 | 0.03 | 4.00 | 3.59 | 0.00 | 0.00 | 0.00 | 0.03 |

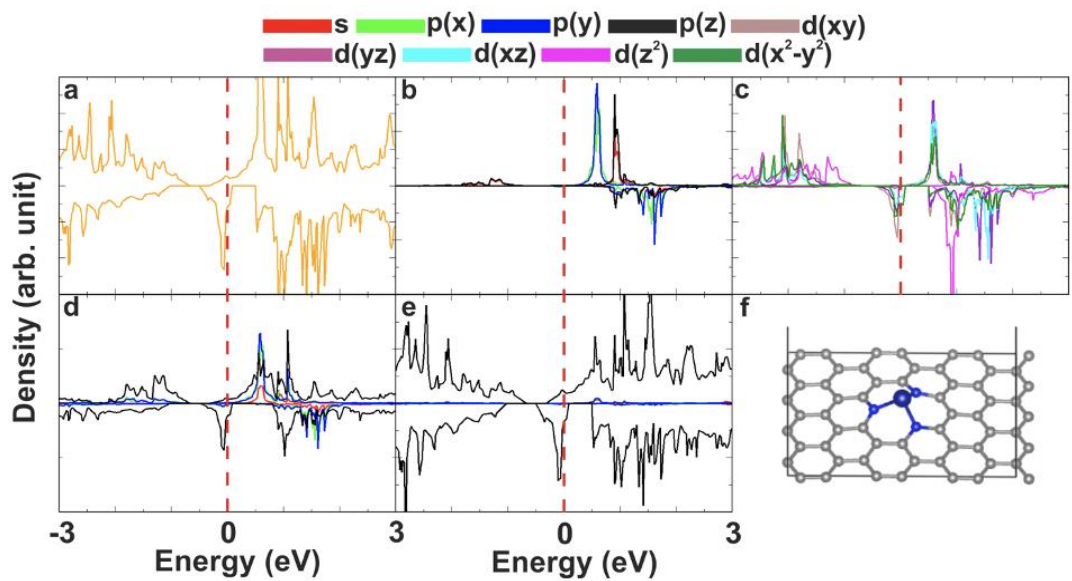

**Figure S72** Atom/orbital decomposed DOS of Cr@NSV. **a)** total DOS, **b,c)** transition metal, **d)** nitrogen atoms, **e)** carbon atoms, **f)** structure.

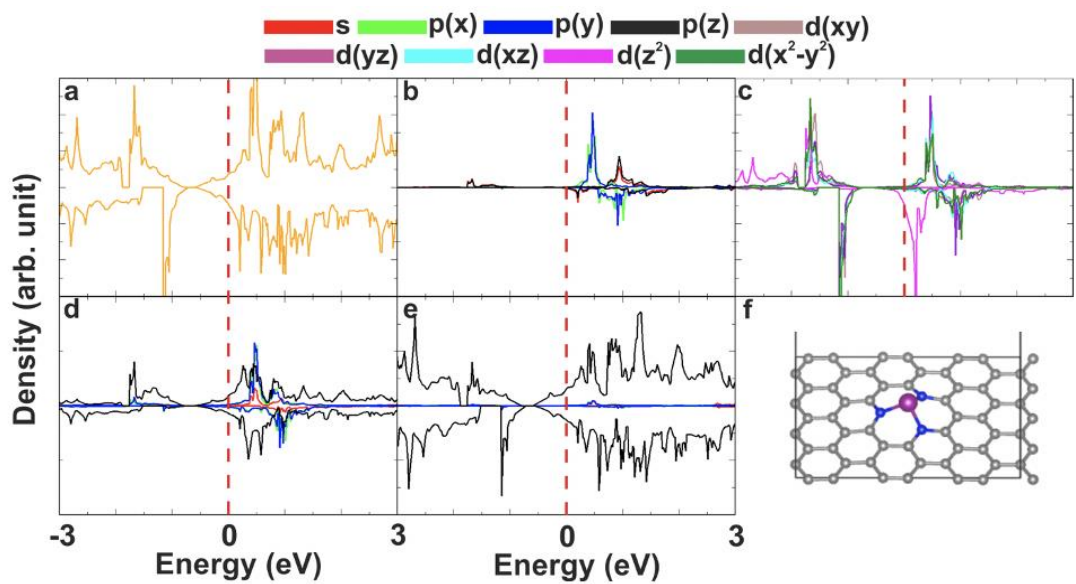

**Figure S73** Atom/orbital decomposed DOS of Mn@NSV. **a)** total DOS, **b,c)** transition metal, **d)** nitrogen atoms, **e)** carbon atoms, **f)** structure.

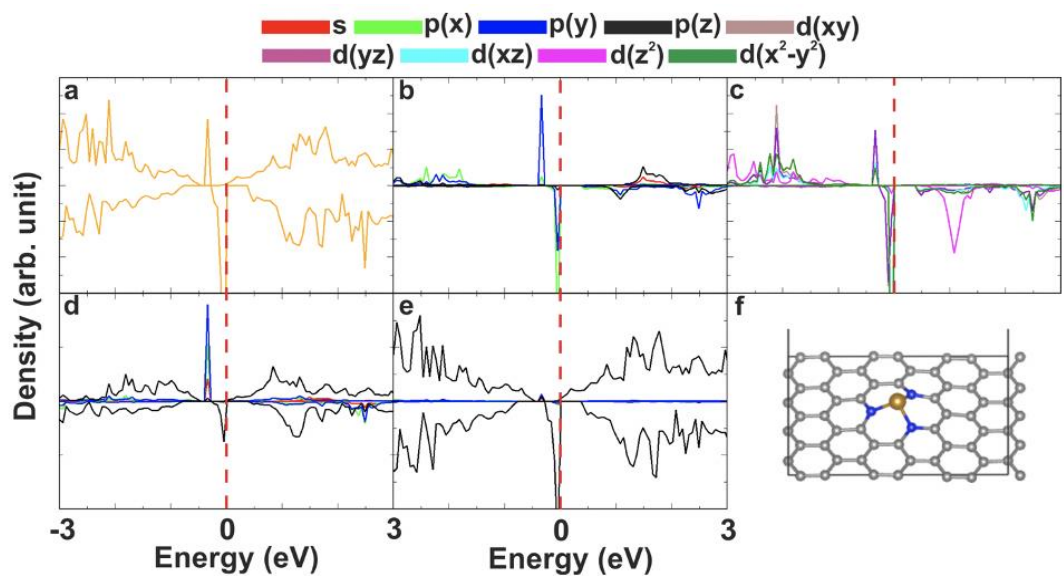

**Figure S74** Atom/orbital decomposed DOS of Fe@NSV. **a)** total DOS, **b,c)** transition metal, **d)** nitrogen atoms, **e)** carbon atoms, **f)** structure.

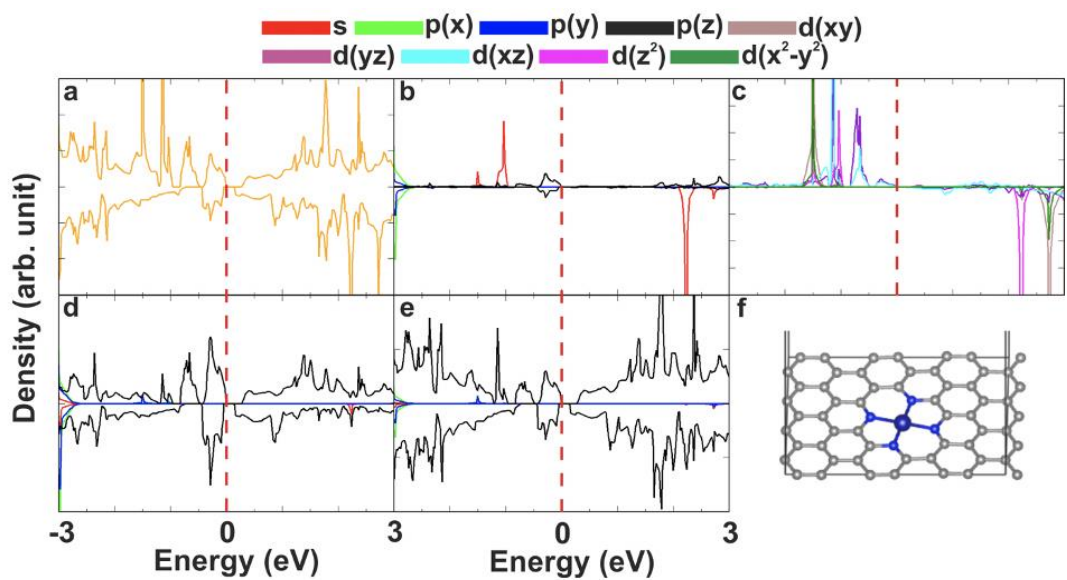

**Figure S75** Atom/orbital decomposed DOS of Cr@NDV. **a)** total DOS, **b,c)** transition metal, **d)** nitrogen atoms, **e)** carbon atoms, **f)** structure.

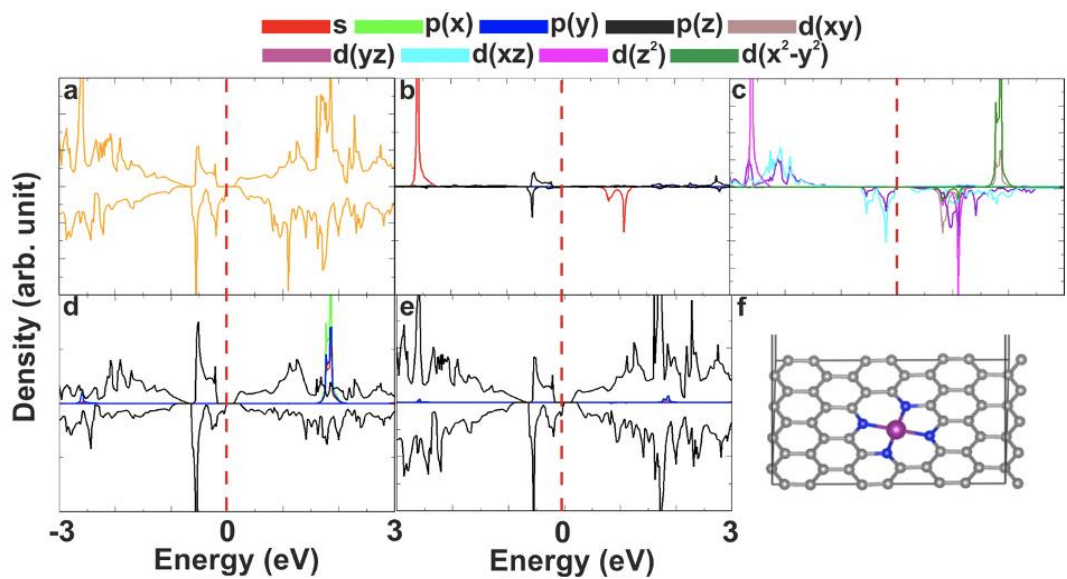

**Figure S76** Atom/orbital decomposed DOS of Mn@NDV. **a)** total DOS, **b,c)** transition metal, **d)** nitrogen atoms, **e)** carbon atoms, **f)** structure.

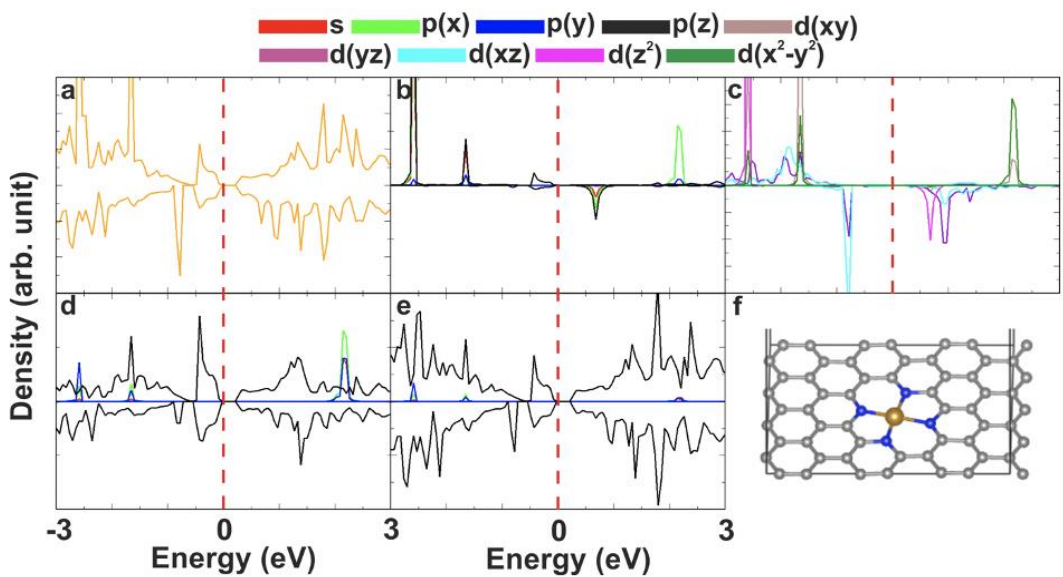

**Figure S77** Atom/orbital decomposed DOS of Fe@NDV. **a)** total DOS, **b,c)** transition metal, **d)** nitrogen atoms, **e)** carbon atoms, **f)** structure.

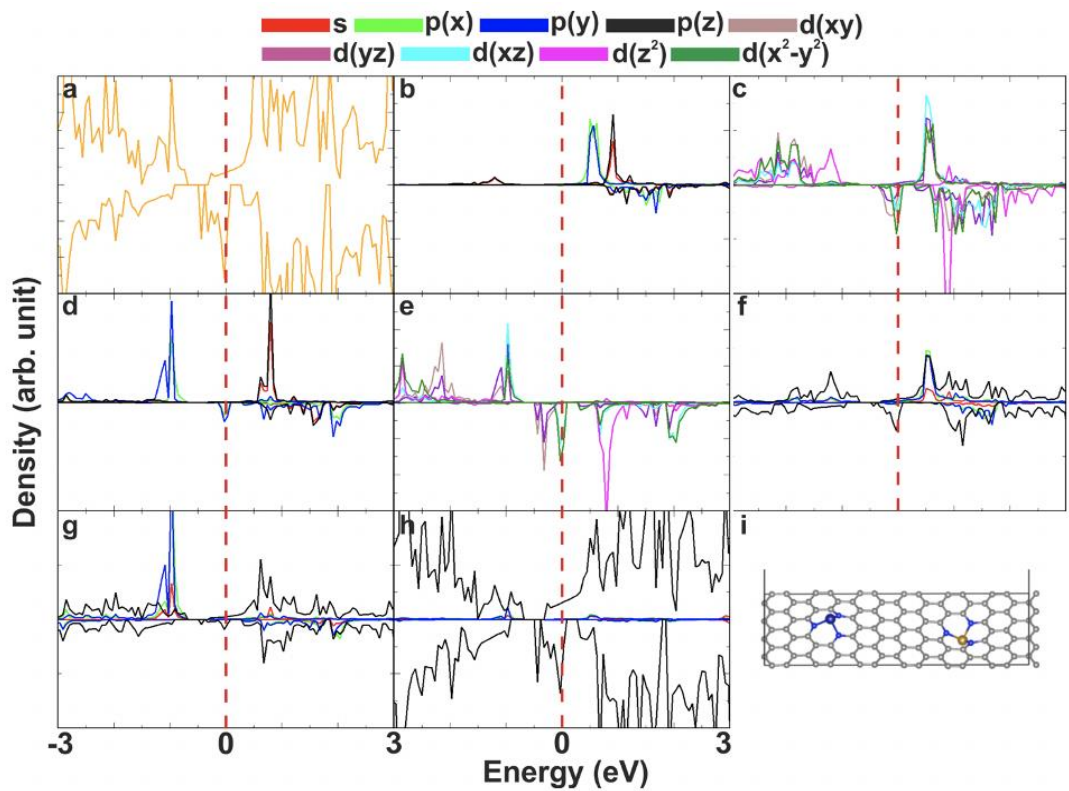

**Figure S78** Atom/orbital decomposed DOS of CrFe@NSV. **a)** total DOS, **b,c)** Cr atom, **d,e)** Fe atom, **f)** nitrogen atoms to Cr, **g)** nitrogen atoms to Fe, **h)** other carbon atoms, **i)** structure.

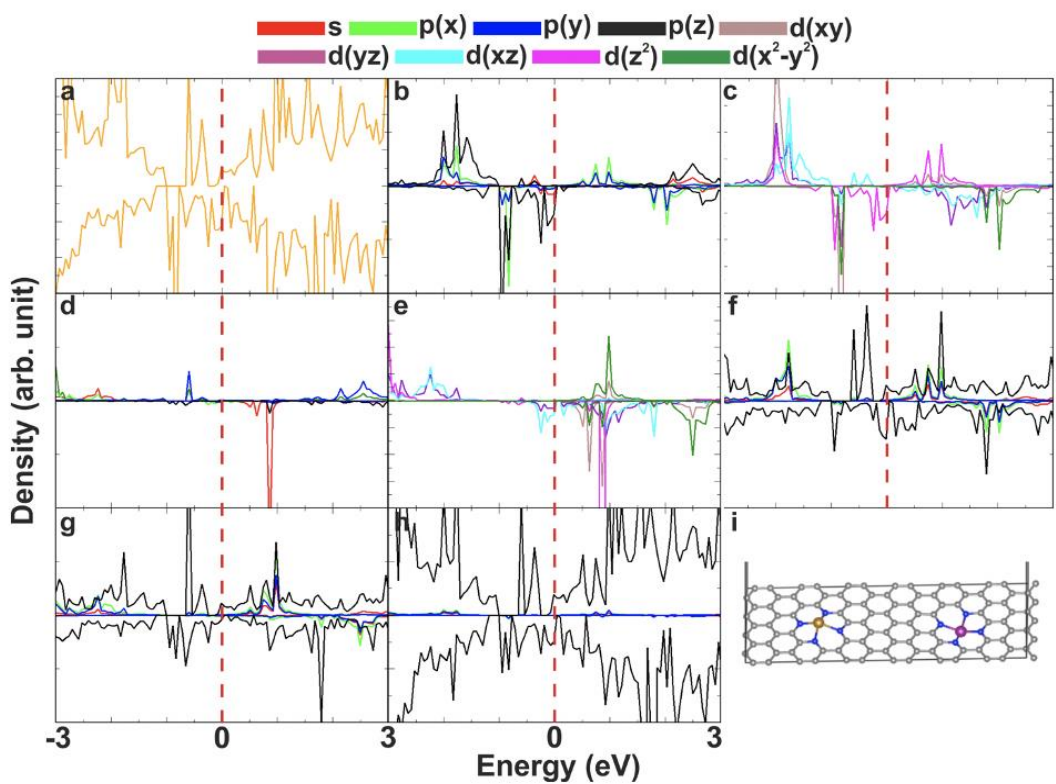

**Figure S79** Atom/orbital decomposed DOS of FeMn@NDV. **a)** total DOS, **b,c)** Cr atom, **d,e)** Fe atom, **f)** nitrogen atoms to Cr, **g)** nitrogen atoms to Fe, **h)** other carbon atoms, **i)** structure.

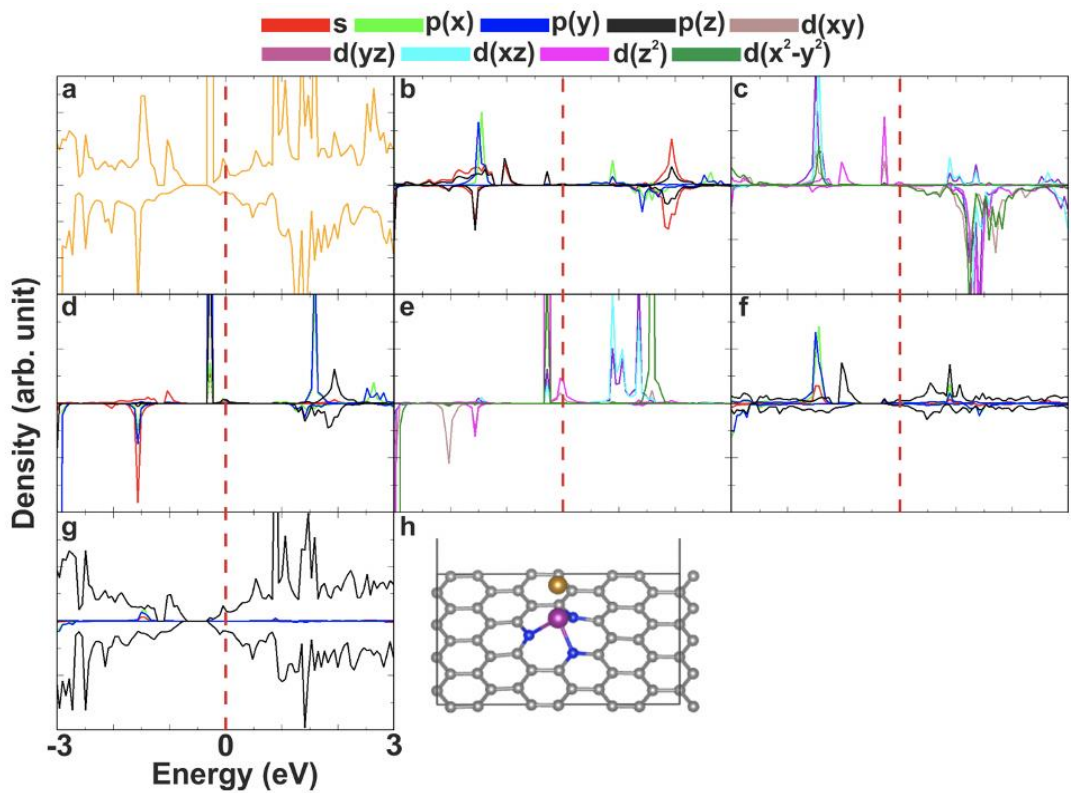

**Figure S80** Atom/orbital decomposed DOS of Fe@Mn@SV. **a)** total DOS, **b,c)** Mn atom, **d,e)** Fe atom, **f)** nitrogen atoms **g)** carbon atoms, **h)** structure.

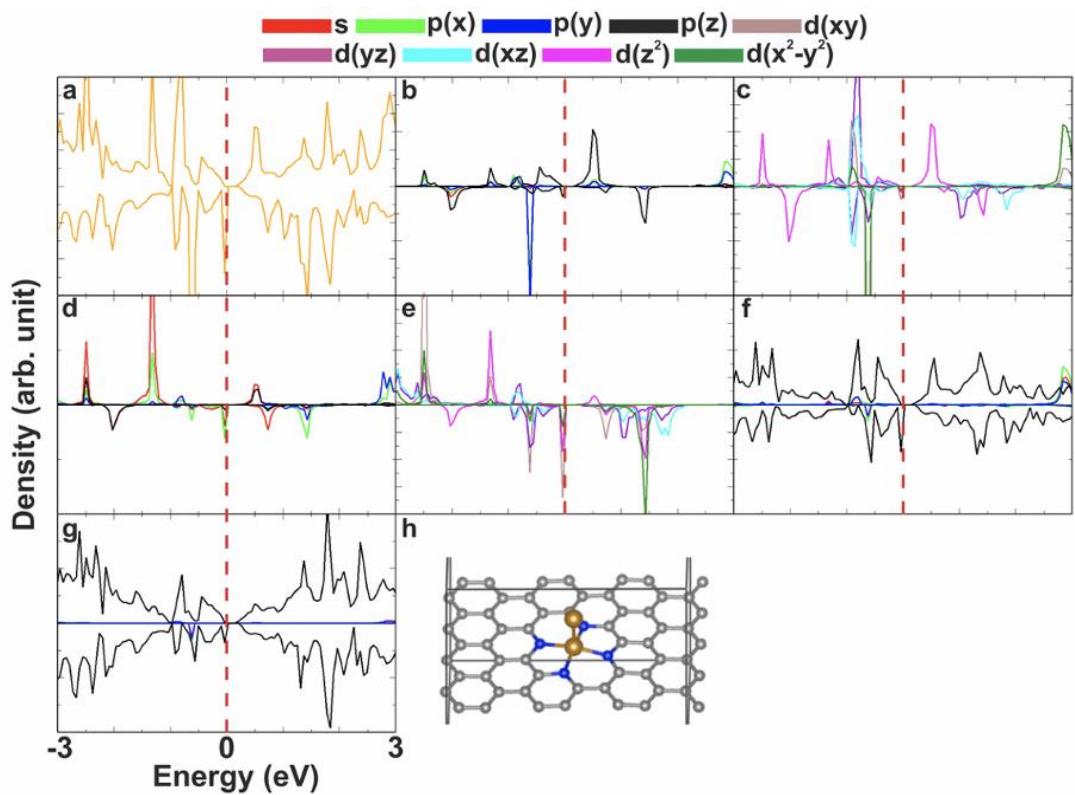

**Figure S81** Atom/orbital decomposed DOS of Fe@Fe@NDV. **a)** total DOS, **b,c)** Fe atom, **d,e)** Fe atom, **f)** nitrogen atoms **g)** carbon atoms, **h)** structure.